\documentclass[11pt,twoside]{article}

\usepackage{amsmath,amssymb,mathrsfs,booktabs,multirow,dcolumn,bigdelim,filecontents,bm,bold-extra}
\usepackage{here,url,fancybox,arydshln,enumitem,sectsty,subfig,stackengine,empheq,cases,braket}
\usepackage[hang,flushmargin]{footmisc}
\usepackage[dvipdfmx]{graphicx,color}
\usepackage[dvipsnames]{xcolor}
\usepackage[T1]{fontenc}
\usepackage{fancyhdr}
\usepackage{xcolor}
\usepackage{pdflscape}
\usepackage{afterpage}
\usepackage{capt-of}
\usepackage{graphbox}
\usepackage[makeroom]{cancel}
\usepackage{mwe}
\usepackage{ulem}
\usepackage{mdframed}
\usepackage[font={small},labelsep=space]{caption}
\DeclareCaptionLabelFormat{vert}{\textbf{{#1} {#2}~|}}
\usepackage{endnotes}

\usepackage[modulo]{lineno}

\DeclarePairedDelimiter\floor{\lfloor}{\rfloor}

\newcommand{\f}[2]{\frac{#1}{#2}}

\newcommand{\ko}[1]{\left( #1 \right)}
\newcommand{\kko}[1]{\left[ #1 \right]}

\newcommand{\bmt}[1]{{{\mbox{\boldmath$ #1 $}}}}
\newcommand{\q}[1]{`#1'}

\renewcommand*{\thepage}{\footnotesize\arabic{page}}
\newcolumntype{d}{D{.}{.}{2.4}}
\newcommand{\RN}[1]{\uppercase\expandafter{\romannumeral #1\relax}}
\newcommand{\dd}{\mathop{}\!d}
\newcommand{\dD}{\mathop{}\!\Delta}

\def\lam{\lambda}
\def\sig{\sigma}
\def\ep{\epsilon}

\renewcommand{\baselinestretch}{1.4}

\usepackage{newtxtext,newtxmath}
\usepackage{etoolbox}

\usepackage[compress,authoryear,round]{natbib}
\setcitestyle{notesep={;},aysep={,},yysep={;}}

\makeatletter
\newcommand*{\myfnsymbol}[1]{\ensuremath{%
\ifcase#1 \or \ast \or \dagger \or \spadesuit \or \diamondsuit \or \clubsuit \or \heartsuit \else \@ctrerr \fi}}
\makeatother

\definecolor{tb}{rgb}{0.24, 0.43, 0.91}
\renewcommand*{\cite}[1]{\textcolor{tb}{\citep{#1}}}
\newcommand*{\citek}[1]{\textcolor{tb}{\citet{#1}}}
\newcommand*{\citec}[3]{\textcolor{tb}{\citep[#1][#2]{#3}}}
\newcommand*{\citea}[2]{\textcolor{tb}{\citealp[#1]{#2}}}

\begin{document}

\quad
\vspace{-1.4cm}
\begin{flushright}
February 2022
\end{flushright}

\vspace{0.5cm}

\begin{center}
\fontsize{15pt}{16pt}\selectfont\bfseries
Scientometric engineering: Exploring citation dynamics via arXiv eprints
\end{center}

\renewcommand*{\thefootnote}{\myfnsymbol{\value{footnote}}}

\vspace*{0.8cm}
\centerline{%
{Keisuke Okamura}\,\footnote{\,{\tt okamura@ifi.u-tokyo.ac.jp}}${}^{,}$%
\footnote{\,{\tt orcid.org/0000-0002-0988-6392}}${}^{;\,1,\,2,\,3}$}

\vspace*{0.6cm}
\small{
\centerline{\textit{
${}^{1}$Institute for Future Initiatives (IFI), The University of Tokyo,}}
\centerline{\textit{
7-3-1 Hongo, Bunkyo-ku, Tokyo 113-0033, Japan.}}
\vspace*{3mm}\centerline{\textit{
${}^{2}$Ministry of Education, Culture, Sports, Science and Technology (MEXT),}}
\centerline{\textit{
3-2-2 Kasumigaseki, Chiyoda-ku, Tokyo 100-8959, Japan.}}
\vspace*{3mm}\centerline{\textit{
${}^{3}$SciREX Center, National Graduate Institute for Policy Studies (GRIPS),}}
\centerline{\textit{
7-22-1 Roppongi, Minato-ku, Tokyo 106-8677, Japan.}}
}
\vspace*{0.5cm}

\vspace{1.5cm}
\noindent\textbf{Abstract.}
\quad
Scholarly communications have been rapidly integrated into digitised and networked open ecosystems, where preprint servers have played a pivotal role in accelerating the knowledge transfer processes.
However, quantitative evidence is scarce regarding how this paradigm shift beyond the traditional journal publication system has affected the dynamics of collective attention on science.
To address this issue, we investigate the citation data of more than 1.5 million eprints on arXiv (\url{https://arxiv.org/}) and analyse the long-term citation trend for each discipline involved.
We find that the typical growth and obsolescence patterns vary across disciplines, reflecting different publication and communication practices.
The results provide unique evidence on the attention dynamics shaped by the research community today, including the dramatic growth and fast obsolescence of Computer Science eprints, 
which has not been captured in previous studies relying on the citation data of journal papers.
Subsequently, we develop a quantitatively-and-temporally normalised citation index with an approximately normal distribution, which is useful for comparing citational attention across disciplines and time periods.
Further, we derive a stochastic model consistent with the observed quantitative and temporal characteristics of citation growth and obsolescence.
The findings and the developed framework open a new avenue for understanding the nature of citation dynamics.

\vspace{0.7cm}

\noindent\textbf{Keywords.}
\quad
Citation dynamics | bibliometrics | arXiv | preprints | research evaluation | stochastic modelling

\vfill

\renewcommand{\UrlFont}{\small\rm}

\thispagestyle{empty}
\setcounter{page}{1}
\setcounter{footnote}{0}
\setcounter{figure}{0}
\setcounter{table}{0}
\setcounter{equation}{0}

\setlength{\skip\footins}{10mm}
\setlength{\footnotesep}{4mm}

\vspace{-1.6cm}

\newpage
\renewcommand{\thefootnote}{\arabic{footnote}}

\setlength{\skip\footins}{10mm}
\setlength{\footnotesep}{4mm}

\urlstyle{same}

\let\oldheadrule\headrule
\renewcommand{\headrule}{\color{VioletRed}\oldheadrule}

\pagestyle{fancy}
\fancyhead[LE,RO]{\textcolor{VioletRed}{\footnotesize{\textsf{\leftmark}}}}
\fancyhead[RE,LO]{}
\fancyfoot[RE,LO]{\color[rgb]{0.04, 0.73, 0.71}{\scriptsize\textsf{arXiv:2106.05027v2~\,|~\,\url{https://doi.org/10.1162/qss_a_00174}}}}
\fancyfoot[LE,RO]{\scriptsize{\textbf{\textsf{\thepage}}}}
\fancyfoot[C]{}

\thispagestyle{empty}

\newpage
\tableofcontents



\vfill

\begin{mdframed}[linecolor=tb]
\begin{tabular}{l}
\begin{minipage}{0.14\hsize}
\includegraphics[width=2.0cm,clip]{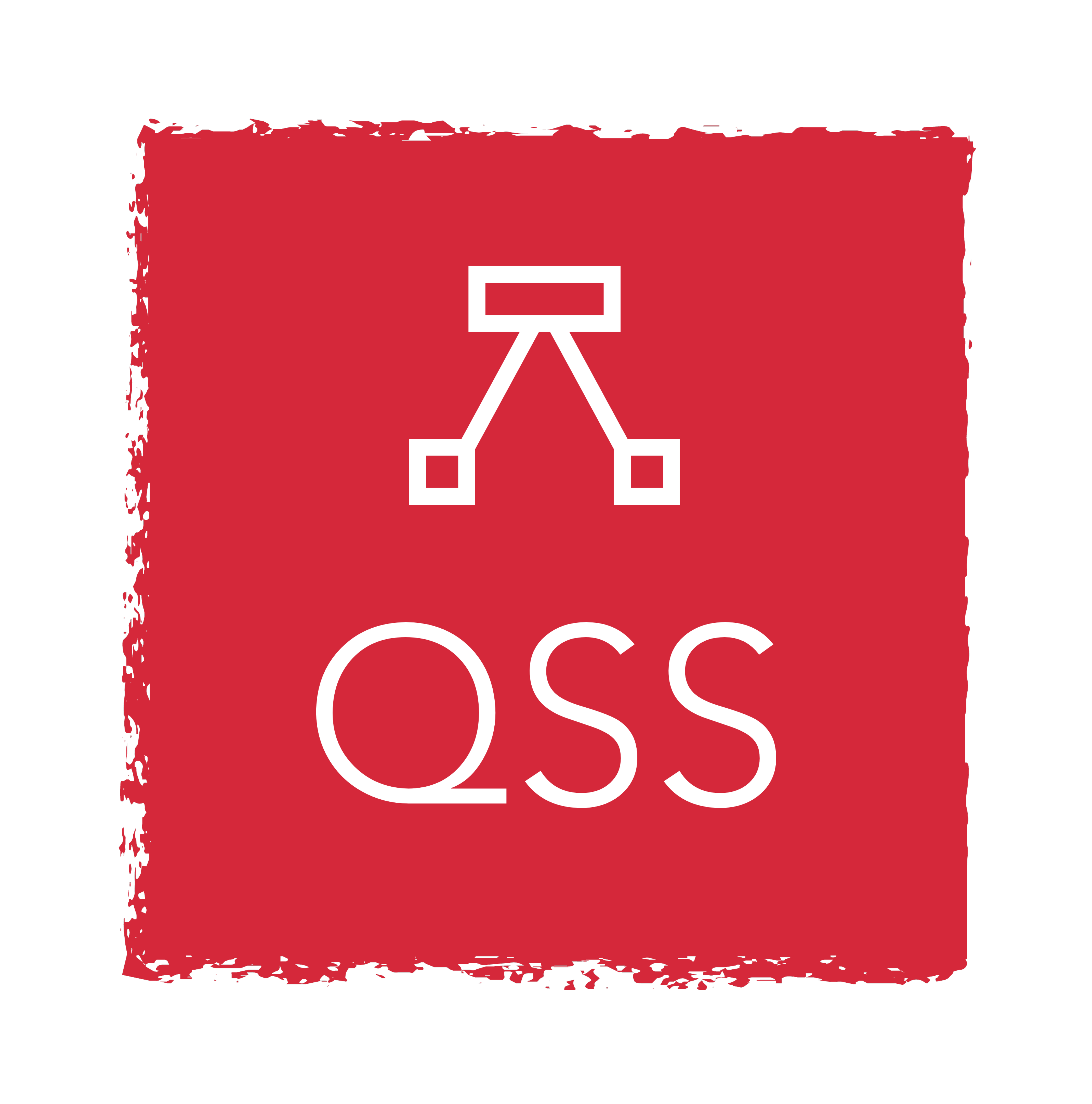}
\end{minipage}
\begin{minipage}{0.83\hsize}
\textcolor{tb}{\textsf{The definitive, peer-reviewed and edited version of this article is published in \textit{Quantitative Science Studies} under a Creative Commons Attribution 4.0 International (CC BY 4.0) license, available at \url{https://doi.org/10.1162/qss _a_00174}, whose contents are the same as the current preprint version.}}
\end{minipage}
\end{tabular}
\end{mdframed}

\newpage
\section{Introduction}

Scientists stand on the shoulders of Giants by citing their predecessors' works in their own works.\endnote{\textit{\q{If I have seen further, it is by standing on the shoulders of Giants}}, Isaac Newton's letter to Robert Hooke, 1675.}
Simultaneously, the scientific community often assumes, either explicitly or implicitly, that the more citations, the higher the scientific value.
This assumption has long made the citation count and its various derivatives influential in research evaluation and administrative decision-making, offering practical, albeit imperfect, proxy metrics of scientific quality or importance \cite{Garfield72,Hirsch05,Waltman16}.
While such use of the citation-based metrics is controversial and often criticised \cite{Garfield79,Line93,Seglen98,Garfield06,Lehmann06,Nicolaisen07,Radicchi17,Aksnes19}, they are nevertheless relevant to quantify the attention of other scientists working on related issues, embodying the scientific knowledge transfer and information flow in the academic sphere.
The life of scientific literature in terms of citations varies, experiencing widely different temporal patterns \cite{Raan04,Redner05,Wang13,Ke15,He18}.
Some papers remain to be continually cited and even become \q{immortal Giants}, while others show fluctuating patterns of citation collection, and, in fact, a significant portion of papers apparently end their lives with little or no citational impact.
Nevertheless, once aggregated and averaged across papers in a given research field, the average citation history curves typically follow a \q{jump--decay} pattern \cite{Barnett89,Glanzel95}; it increases up to a few years after publication to reach a peak and subsequently decreases to some asymptotic value.

The citation pattern has long fascinated researchers and practitioners alike.
They have introduced the concept of literature obsolescence (ageing) \cite{Gross27,Line74,Line93}---the process of becoming no longer cited or less used---to investigate how quickly a scientific publication moves in and out of the focus of researchers' attention.
Here, obsolescence does not necessarily mean deterioration in its intrinsic value but a natural consequence of scientific development or technological advancement over a while.
Scientific discovery presented in a paper and the value it creates are eventually integrated into subsequent papers, thereby contributing to the shoulders to stand on.
Having said that, an interesting question is whether the obsolescence rate has changed over time.
Some studies \cite{Lariviere08,Verstak14,Martin-Martin16} suggested that researchers are increasingly relying on older publications, while others \cite{Evans08,Parolo15} suggested that the obsolescence of scientific knowledge has accelerated over the past decades.
Although results and implications from these previous studies vary without consensus (see also \citea{}{Sinatra15,Zhang17a,Pan18}), a common feature was the use of the citation data on papers published in peer-reviewed journals.
The bibliometric data used were commonly obtained from publishers or major bibliographic databases such as Web of Science \endnote{Web of Science, published by Clarivate analytics or Thomson Reuters, \texttt{https://clarivate.com/webofsciencegroup/solutions/web-of-science/}.} and Scopus.\endnote{Scopus, published by Elsevier, \texttt{https://www.scopus.com}.}

Nowadays, the study of literature growth and obsolescence has embarked on an entirely new phase.
This change is due to the new digital technologies, recourses and online interfaces that enormously increased the speed and volume of scientific knowledge production and consumption.
With the ongoing rapid and large-scale digitisation of information, researchers increasingly rely on digital research documents, or eprints, available on a free and open-access platform.
Those eprints mainly include preprints, yet peer-reviewed and accepted in scientific journals or other traditional means.
Still, they are citable as they are publicly disclosed and accessible on the internet via the preprint servers, which have become an integral part of today's scholarly publishing system.
arXiv,\endnote{arXiv.org [eprint archive], \texttt{https://arxiv.org/}.} the most prominent and oldest eprint archive launched in August 1991, has served as an indispensable research platform (document submission and retrieval system) for physics, mathematics, computer sciences \cite{Ginsparg16,arXiv_stats2020}.
For the past 30 years, it has functioned as a primary source of current and ongoing research, accelerating recognition and dissemination of research findings and facilitating rapid scholarly communication.
The number of eprints posted on arXiv has grown dramatically from 30,601 (2000) to 155,866 (2019), increasing more than five-fold during the past 20 years (Suppl.~Fig.\ \ref{fig:lineplot_comb2}a), and the number of citations received by the arXiv eprints has also snowballed with time (Suppl.~Fig.\ \ref{fig:lineplot_comb2}b).
Previous studies have investigated how arXiv has impacted the traditional publication practices and the researchers' citation activities \cite{Moed07,Gentil-Beccot10,Aman13,Lariviere14,Feldman18,Wang20,Okamura20b}.
For instance, evidence has been provided that arXiv has the advantage to accelerate citation \cite{Moed07,Aman13,Okamura20b} (see also Suppl.~Fig.\ \ref{fig:lineplot_pct_pre}).
Recently, the preprint mode of scholarly communication has seen an exponential rise during the current COVID-19 global pandemic, driven by the vital need for the early and rapid dissemination of the COVID-19-related research results.
Two growing preprint servers are bioRxiv\endnote{bioRxiv [Preprint server], \texttt{https://www.biorxiv.org/}.} (launched in November 2013) and medRxiv\endnote{medRxiv [Preprint server], \texttt{https://www.medrxiv.org/}.} (launched in June 2019), which have offered indispensable platforms for researchers in biomedical disciplines \cite{Abdill19,Fu19,Fraser20,Fraser21,Kirkham20,Sevryugina21}.
Scholarly communications in today's digitalised academic sphere have thus been rapidly shifting from the conventional journal-centred publishing styles towards a new mode of communication via preprint servers and social media (e.g.\ blogging, Facebook and Twitter) per Open Science practices \cite{Gentil-Beccot10,Shuai12,Thelwall13,Berg16,Ginsparg16,Wang20,Fraser21}.

Facing this publishing paradigm shift, what constitutes the growth and obsolescence of scientific knowledge have also been changing and diversifying.
In many research areas and subfields, nowadays, it is quite common that eprints posted on preprint servers or repositories are not yet or never to be published in journals indexed in the commercial bibliometric databases \cite{Lariviere14,Abdill19,Okamura20b,Sevryugina21}.
Even if the preprints are to be eventually published in a database-indexed journal, the citations received during the preprint duration (in their preprint forms) may not be fully covered in major bibliometric databases, leading to bias in the citation analysis.
In addition, there exists a field-dependent bias resulting from the database coverage \cite{Waltman16}.
For instance, computer science has a unique conference-centric publishing culture, in which researchers have a tendency to value top-tier conferences as a publication venue than high-impact journals \cite{Feldman18,Kim19}.
Also, the main research outputs in humanity and social sciences studies have been books rather than journal papers.
For these disciplines, the citational impact of a scientific work tends to be under-represented in many commercial bibliometric databases.
Consequently, research methods and measurement based on the conventional publishing systems may often not be desirable to grasp the whole volume and dynamics of scientific attention, or citations, in this new era of Open Science.

With the aim to obtain a better understanding of the real dynamics underlying the citation network beyond the traditional journal-centred scholarly communication, the present study investigates the citation data of eprints posted on arXiv.
The data contain the information on various types of eprints, regardless of which (or none) journal eventually published (or will publish) them.
A disadvantage of using the arXiv data compared to the commercial databases would be the limited variety in the subject categories; for instance, arXiv's research disciplines do not include biomedical disciplines.
Still, the arXiv data cover various subject categories focusing on physics, mathematics and computer sciences, and the long-term citation data (since 1991) provide a unique testbed for modelling and visualising the distinct disciplinary patterns of scientific attention.
Making full use of the advantages, we investigate two dimensions of citation dynamics: the \q{quantitative} dimension and the \q{temporal} dimension.
The quantitative dimension regards the degree distribution in the citation network \cite{Price65,Price76}.
We consider both the power law model \cite{Price76,Redner98,Peterson10} and the lognormal model \cite{Redner05,Radicchi08,Sheridan18}, and discuss the extent to which each model explains the empirical data.
We conclude that, by and large, the arXiv citation data obeys the Lognormal Law.
The other, temporal dimension regards the time-dependence of citation accumulation at the discipline-average level \cite{Gross27,Barnett89,Glanzel95}.
The entire span of the growth and obsolescence pattern is modelled by a simple nonlinear function of time, which fits the empirical data remarkably well for all disciplines.

The revealed characteristics of the citation dynamics open up some interesting research directions; we discuss two directions in this paper.
The first direction concerns a \q{fair} evaluation of scientific attention cast on individual papers.
Although citation-based metrics have increasingly been applied in research evaluation and science policymaking, various sources of bias have limited their validity and utility \cite{Seglen98,Lehmann06,Waltman16,Aksnes19}.
A particularly significant issue is the disciplinary difference, including the size of the research community, the typical time to publication, the average level of per-paper citations, the growth and obsolescence rates, and the degree to which the preprint (eprint) mode of communication is adopted.\endnote{%
Other biases arising from the motives or limitations on the citer's side include reference copying, self-citations, negative citations, politically motivated flattery, cronyism, limited space for references, homographs, language, nationality, and the bibliometric database coverage, to name but a few. 
See e.g.\ \citek{Seglen98}, \citek{Waltman16} and \citek{Aksnes19} for detailed discussions.} 
The knowledge of citation dynamics revealed in this paper can mitigate some of these biases, both in the quantitative and temporal dimensions, through a new index of citational attention called the $\gamma$-index, as will be demonstrated.
The second direction is more theory-oriented, which is an effort to reveal the underlying mathematical model that consistently explains the observed evolution pattern of citations, both quantity-wise and time-wise.
Unlike the commonly accepted model of preferential attachment \cite{Barabasi99,Newman01,Barabasi02}, which generates networks with the Power Law, our stochastic model naturally reproduces the observed Lognormal Law in the quantitative dimension while also reproducing the observed evolution pattern (\q{jump--decay} plus \q{constant attention}) of the per-paper average citations in the temporal dimension.
We hope this paper serves to provoke further discussions and investigations of the related issues, thereby shedding new light on the nature of collective attention dynamics on scientific knowledge.

\section{The quantitative dimension of citation dynamics}

We first describe the bibliometric data used throughout this study; see \ref{app:data} for the data analysis and visualisation platforms.
Subsequently, we analyse the degree distribution in the arXiv citation network, providing evidence for the Lognormal Law in the quantitative dimension of the citation dynamics.

\subsection{The arXiv data}

Our bibliometric analyses were based on a dataset of 1,589,006 eprints posted on arXiv from its launch in 1991 until the end of 2019---hereafter referred to as \q{the arXiv eprints}---plus the data on their citations.
This raw dataset is the same as that used in \citek{Okamura20b}, which can be found in the Zenodo repository \cite{Okamura21}.
Here, eprints include preprints, conference proceedings, book chapters, data sets and commentary, i.e.\ every electronic material that has been posted on arXiv.
The content and metadata of the arXiv eprints were retrieved from the arXiv API\endnote{arXiv API, \texttt{https://arxiv.org/help/api/}. Accessed January 2020.} as of 21st January 2020, where the metadata included data of the eprint's title, author, abstract, subject category and the arXiv ID, which is the arXiv's original eprint identifier.
In addition, the associated citation data were derived from the Semantic Scholar API\endnote{Semantic Scholar API, \texttt{https://api.semanticscholar.org/}. Accessed January 2020.} from 24th January 2020 to 7th February 2020, containing the citation information in and out of the arXiv eprints and their published versions (if applicable).
Here, whether an eprint has been published in a journal or other means is assumed to be inferrable, albeit indirectly, from the status of the digital object identifier (DOI) assignment.
It is also assumed that if an arXiv eprint received $c_{\mathrm{pre}}$ and $c_{\mathrm{pub}}$ citations until the data retrieval date (7th February 2020) before and after it is assigned a DOI, respectively, then the citation count of this eprint is recorded in the Semantic Scholar dataset as $c_{\mathrm{pre}}+c_{\mathrm{pub}}$.
Both the arXiv API and the Semantic Scholar datasets contained the arXiv ID as metadata, which served as a key variable to merge the two datasets.

Our classification of research disciplines was based on that described in the \citek{arXiv_stats2020} website.
There, the 153 arXiv subject categories are aggregated into several disciplines, of which we restrict our attention to the following six disciplines: Astrophysics (\textit{\q{astro-ph}}), Computer Science (CS) (\textit{\q{comp-sci}}), Condensed Matter Physics (\textit{\q{cond-mat}}), High Energy Physics (HEP) (\textit{\q{hep}}), Mathematics (Math) (\textit{\q{math}}) and Other Physics (\textit{\q{oth-phys}}), which collectively accounted for 98\% of all the eprints.\endnote{%
The rest eprints included those categorised in Electrical Engineering and Systems Science, Statistics, Quantitative Biology and Quantitative Finance.}
Those eprints that are tagged to multiple arXiv disciplines (Suppl.~Fig.\ \ref{fig:overlap}a, b) were counted independently for each discipline.
Due to this overlapping feature, the final dataset contained a cumulative total of 2,011,216 eprints.
See Table \ref{tab:arXiv_disciplines} for the numbers of eprints by discipline (see Suppl.~Table \ref{tab:arXiv_disciplines_de} for a detailed description).
A notable difference was observed between the four physics-based disciplines and the nonphysics disciplines (i.e.\ CS and Math) in the distribution of the time elapsed without being assigned a DOI (Suppl.~Fig.\ \ref{fig:year_lag_all}), in line with the findings of previous studies \cite{Aman13,Lariviere14}.
Specifically, while the majority of physics-based eprints acquire a DOI approximately within two years and exhibit a steep slope down towards the third year, the majority of nonphysics-based eprints are never assigned a DOI (CS: 77\%, Math: 69\%) to the data retrieval date, exhibiting a heavy-tailed histogram.
This is highly suggestive of the empirical fact that journal publication is not necessarily the venue of choice for researchers of CS and Math, and that they heavily rely on arXiv for tracking new research findings \citec{cf.}{}{Kim19,Wang20}.

\begin{table}[t]
\vspace{-0.5cm}
\centering
\caption{\textbf{Classification of the arXiv disciplines.}
The classification scheme is based on that used on the \citek{arXiv_stats2020} website.
(Detailed description is presented in Suppl.~Table \ref{tab:arXiv_disciplines_de}.)}
\label{tab:arXiv_disciplines}
\vspace{-3mm}
{\footnotesize
{\setlength{\tabcolsep}{1.2em}
\begin{tabular}{lll}\\[-2mm] \toprule[1pt] \\[-5.8mm]
\multicolumn{1}{l}{{Discipline}} & \multicolumn{1}{c}{No.\ of eprints} & \multicolumn{1}{c}{{arXiv subject classification}} \\ \midrule
\textit{astro-ph} & \multicolumn{1}{c}{257,864} & Astrophysics \\
\textit{comp-sci} & \multicolumn{1}{c}{386,507} & Computer Science \\
\textit{cond-mat} & \multicolumn{1}{c}{263,506} & Condensed Matter Physics \\
\textit{hep} & \multicolumn{1}{c}{286,840} & High Energy Physics (Theory, Phenomenology, Lattice, Experiment)  \\
\textit{math}& \multicolumn{1}{c}{428,621} & Mathematics (including: Mathematical Physics) \\
\textit{oth-phys} & \multicolumn{1}{c}{387,878} & Other Physics (including: Nuclear Theory and Experiment, General Relativity \\[-1.5mm]
{} & {} & and Quantum Cosmology, Quantum Physics, Nonlinear Sciences) \\[0mm] \bottomrule[1pt] 
\end{tabular}}
}
\vspace{5mm}
\end{table}

\subsection{Degree distribution of the citation network\label{app:lognormlaw}}

\begin{figure}[tp] 
\centering
\vspace{-0.5cm}
\noindent
\includegraphics[align=c, scale=0.7, vmargin=1mm]{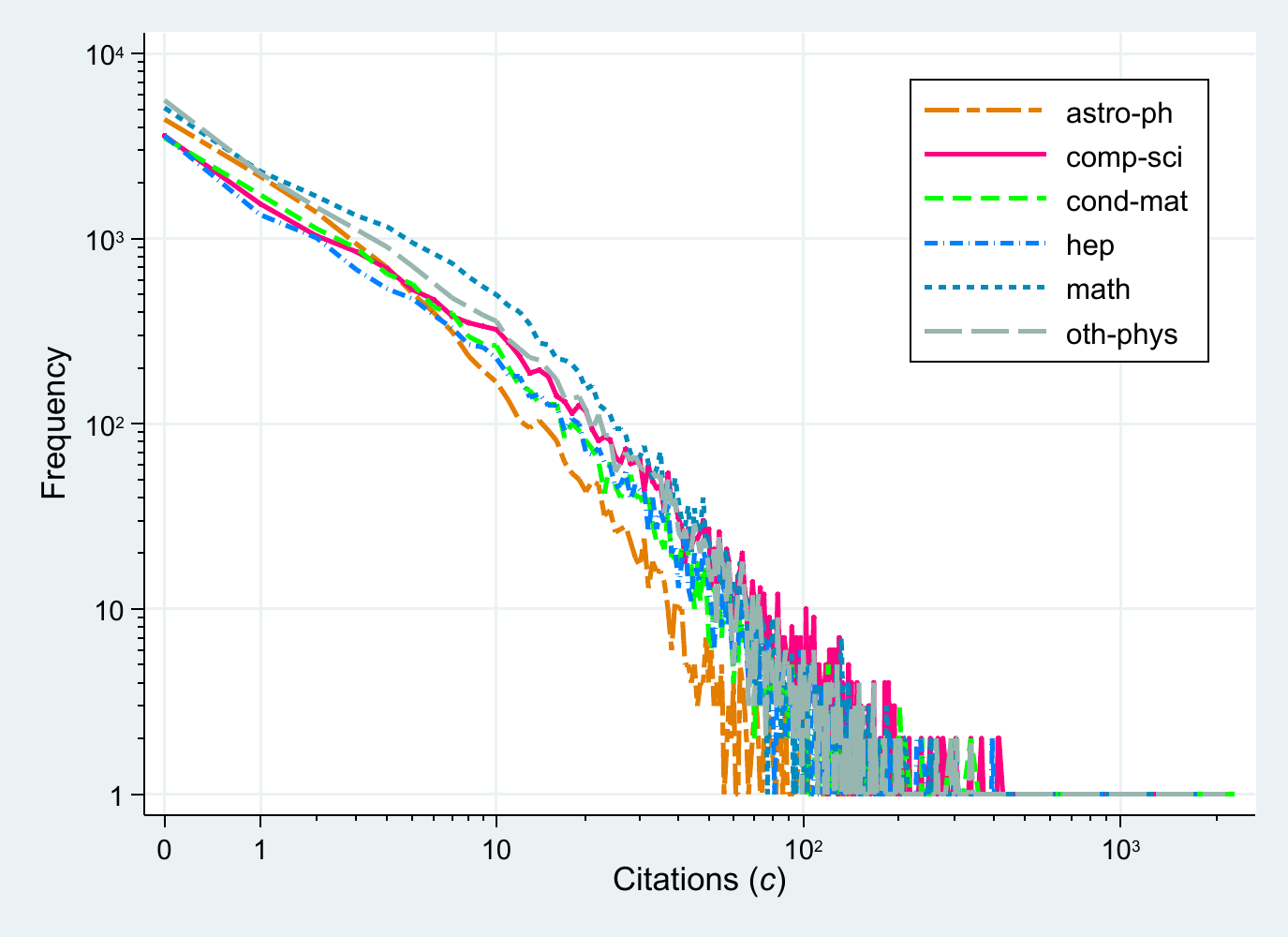}
\caption{\textbf{Distribution of cumulative citations of the arXiv eprints.}
Frequency distributions of citations to the eprints posted on arXiv in 2010 are shown in the log-log scale by discipline.}
\label{fig:lnc_tot_all_1999}
\vspace{5mm}
\end{figure}

We begin with the investigation of the degree distribution in the arXiv citation network.
This field of study has a long history with the pioneering work by \citek{Price65}, which observed that the citation distribution follows a power law.
Following that, previous studies commonly considered either a (variation of) power law \cite{Price76,Redner98,Hajra06,Peterson10,Eom11,Thelwall16} or a lognormal law \cite{Redner05,Radicchi08,Eom11,Thelwall16} to characterise the citation distribution of journal papers.
Here, we investigate the citation data on the arXiv eprints and verify the fitness of each model.
Let $c_{k}\in \mathbb{Z}^{+}_{0}$ be the number of citations that an eprint labelled by $k$ (hereafter referred to as \q{eprint $k$}) accumulates.
As Fig.\ \ref{fig:lnc_tot_all_1999} indicates, the distribution of the raw citations is extremely skewed towards very highly cited eprints (note that the plots are shown in the log-log scale; see also Suppl.~Table \ref{tab:cat_by_pctl}).
Given that, let us define
\begin{equation}
y_{k}\coloneqq \ln(c_{k}+1)
\end{equation}
with the \q{($+1$)-shifted} citation variable \cite{Thelwall16,Sheridan18}.
By this definition, $y$ is a monotonically increasing function of the raw citation variable, $c$, with the zero-citation ($c=0$) case being mapped to $y=0$.
Also, let $q_{k}$ represent the quantile rank of $c_{k}$ in the citation distribution, i.e.\ $q_{k}=P(c\leq c_{k})$.
For the power law (\q{\textsf{PL}}) model with exponent $a>1$,
\begin{equation}\label{powerlaw2}
P_{\mathsf{PL}}(c)\sim (a-1)(c+1)^{-a}\,,\quad 
\text{i.e.}\quad
y_{\mathsf{PL}}\sim \f{\ln(1-q)}{1-a}\,,
\end{equation}
while for the lognormal (\q{\textsf{LN}}) model with mean $b$ and variance $m^{2}$,
\begin{equation}\label{lognorm2}
P_{\mathsf{LN}}(c)\sim \f{1}{(c+1){m}\sqrt{2\pi}}\exp\kko{-\f{\big(\ln (c+1)-{b}\big)^{2}}{2{m}^{2}}}\,,\quad 
\text{i.e.}\quad
y_{\mathsf{LN}}\sim b+m\Phi^{-1}(q)\,,
\end{equation}
where $\Phi(\cdot)$ is the standard normal cumulative distribution function defined by $\Phi(x)=(2\pi)^{-1/2}\int_{-\infty}^{x} e^{-t^{2}/2}\dd t$.

\begin{figure}[tp]
\centering
\vspace{-0.5cm}
    \begin{tabular}{l}
\begin{minipage}{0.5\hsize}
\begin{flushleft}
\raisebox{-\height}{\includegraphics[align=c, scale=0.6, vmargin=0mm]{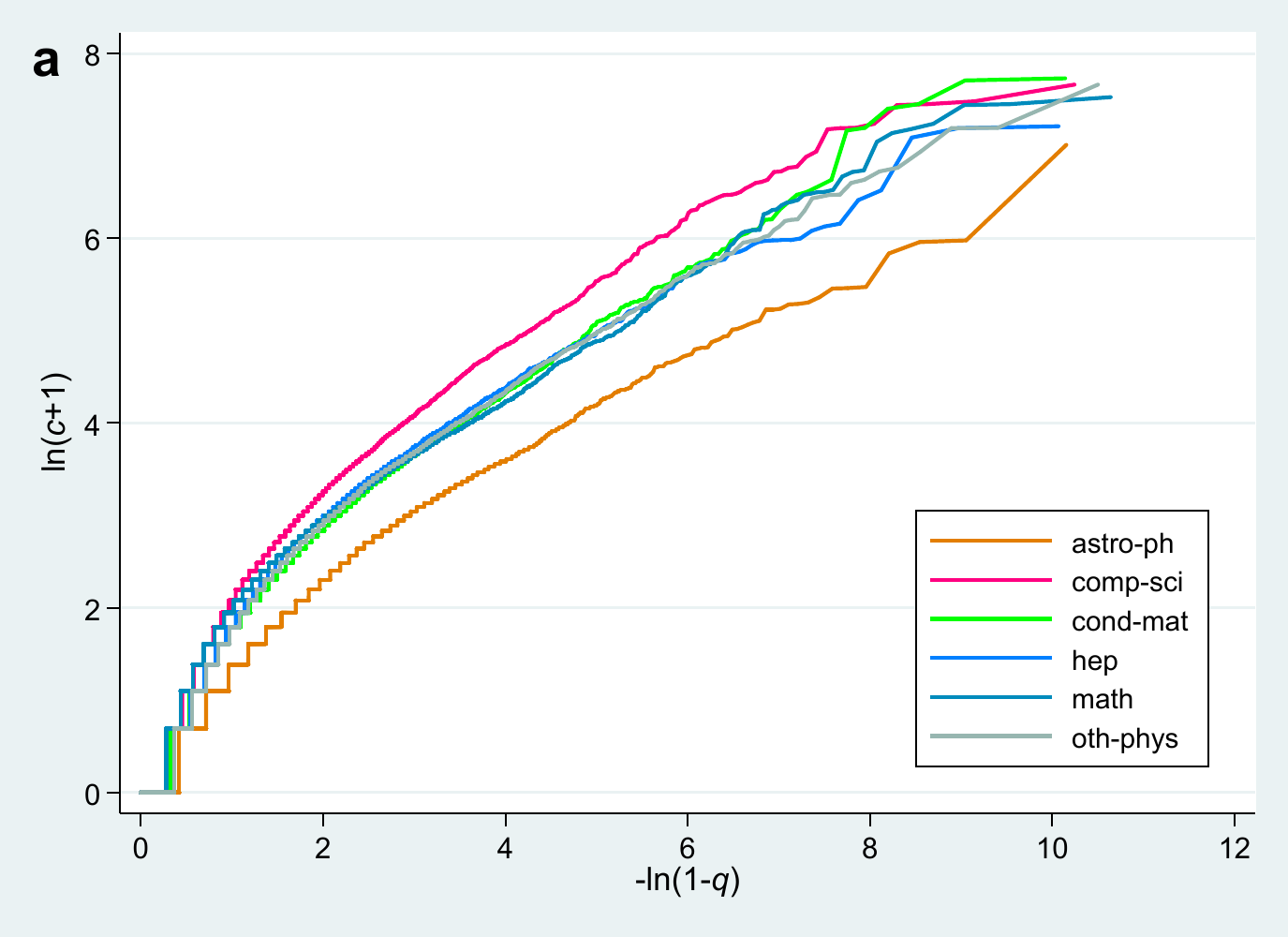}}
\end{flushleft}
      \end{minipage}
\begin{minipage}{0.5\hsize}
\begin{flushleft}
\raisebox{-\height}{\includegraphics[align=c, scale=0.6, vmargin=0mm]{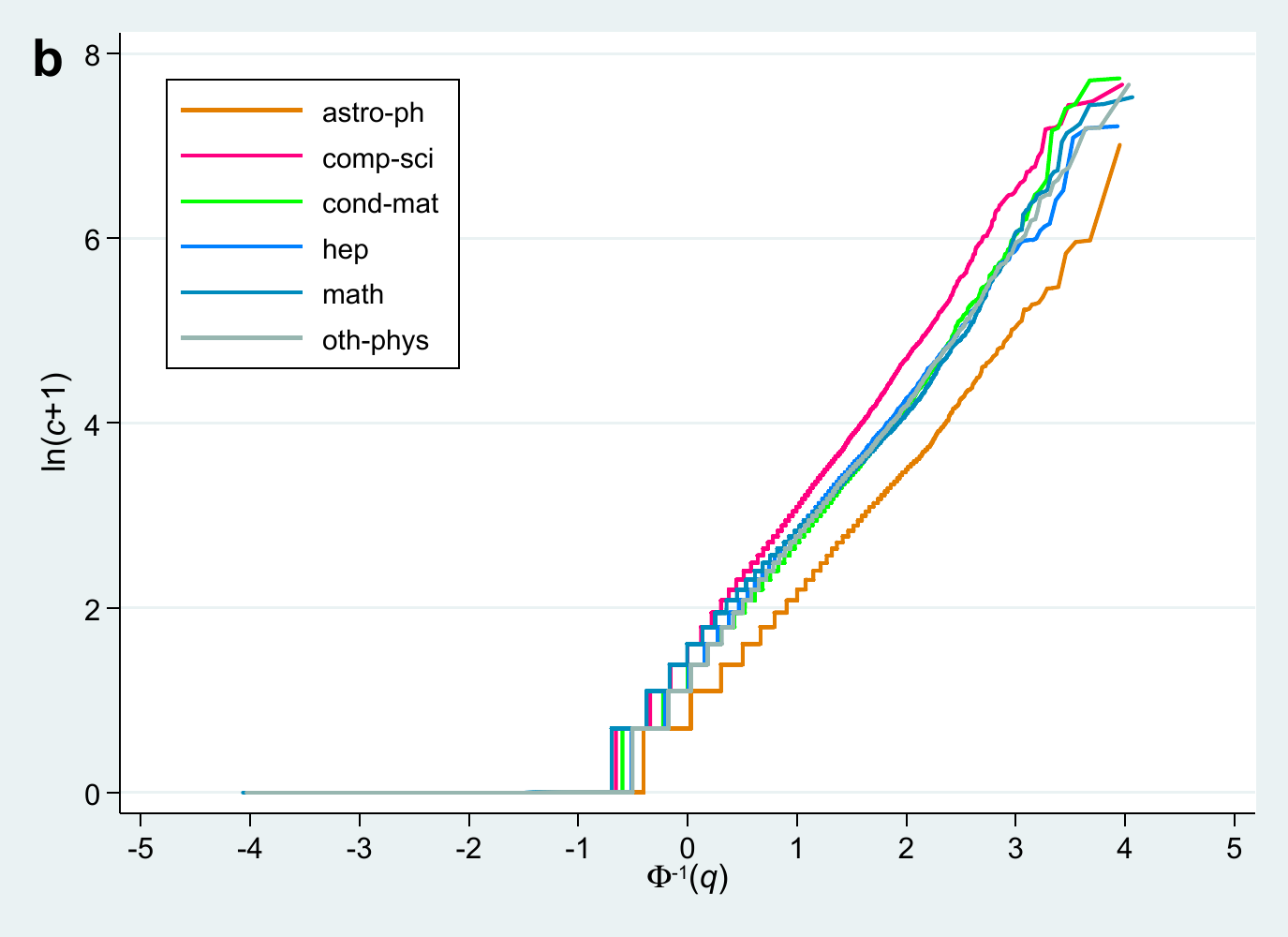}}
\end{flushleft}
          \end{minipage}
    \end{tabular}
\caption{\textbf{Degree distribution of the arXiv citation network: \q{Power Law vs.\ Lognormal Law}.}
\textsf{\textbf{a}}, plot of $y_{k}=\ln(c_{k}+1)$ against $-\ln(1-q_{k})$, where $c_{k}$ is the number of citations eprint $k$ accumulates, and $q_{k}$ is the quantile rank of $c_{k}$ in the citation distribution of each discipline;
\textsf{\textbf{b}}, plot of $y_{k}$ against $\Phi^{-1}(q_{k})$, where $\Phi(\cdot)$ represents the standard normal cumulative density function.
The plots are based on the citation data of eprints posted on arXiv in 2010.}
\label{fig:degreedist}
\vspace{0.5cm}
\end{figure}

To investigate which (or both, or neither) of the models ($y_{\mathsf{PL}}$ and $y_{\mathsf{LN}}$) explains the empirical distribution of the arXiv citation data better, we created the quantile plot of $y_{k}$ against each model, using the sub-datasets of eprints classified according to the year of first submission (1991--2019).
The fitness of each model was checked by visual inspection of the linearity between $y_{k}$ and $\ln(1-q_{k})$ (power law; Suppl.~Fig.\ \ref{fig:qplot_comb_years_1}), and between $y_{k}$ and $\Phi^{-1}(q_{k})$ (lognormal; Suppl.~Fig.\ \ref{fig:qplot_comb_years_2}).
First, let us look into the results for the power law model; see Fig.\ \ref{fig:degreedist}a for the quantile plot based on the year-2010 dataset.
An approximately linear relationship can be seen in the region $-\ln(1-q)\gtrsim 4$, or $q\gtrsim 0.98$ in terms of the quantile rank.
However, this model does not fit well with the vast majority of the observations (i.e.\ below approximately the 98th percentile eprints).
To cure this situation, we also considered a shifted power law model \cite{Eom11,Thelwall16}, $P_{\mathsf{sPL}}(c)\sim \big((a-1)/\theta\big)(c/\theta+1)^{-a}$, by introducing a shift parameter $\theta>0$ (Suppl.~Fig.\ \ref{fig:shiftedpl}).
Indeed, the approximately linear region could be expanded by fine-tuning the level of $\theta$ (see, e.g.\ the plot with $\theta=10$).
However, without any justification for its dynamical origin, the shift parameter ($\theta$) represented nothing more than an arbitrary threshold set by hand.
Also, the value of $\theta$ was quite sensitive to the slope of the linear region.
Consequently, the estimated value of the exponent ($a$) was crucially dependent on the choice of the arbitrary parameter.

By contrast, the lognormal model was shown to fit the empirical data remarkably well; see Fig.\ \ref{fig:degreedist}b for the quantile plot based on the year-2010 dataset.
An approximately linear relationship between $y_{k}$ and $\Phi^{-1}(q_{k})$ is observed for the entire region of $c_{k}>0$, without any arbitrary thresholds as in the shifted power law model case.
Applying an ordinary least squares (OLS) regression model to the nonzero-citation data, the two lognormal parameters ($b$, $m$) were estimated with a very high coefficient of determination ($\text{the adjusted $R$-squared}>0.99$; Suppl.~Table \ref{tab:qplot_2010}), with which the degree distribution was estimated as $y\sim \mathcal{N}(\hat{b},\hat{m}^{2})$.
This feature was also true for each dataset year and each discipline (Suppl.~Fig.\ \ref{fig:qplot_comb_years_2}).
These results lead us to conclude that the arXiv citation distribution is better explained by the lognormal model than the power law model.\endnote{%
Other skewed distributions such as the gamma and Weibull distributions, including the exponential distribution as their special case, were also considered.
However, the lognormal distribution outperformed the others in describing the empirical arXiv citation data.}
Note that for both the power law and the lognormal models, the empirical data deviate from the law in the region $1-q\lesssim \mathcal{O}(10^{-5})$, i.e. exceptionally highly cited eprints.

\section{The temporal dimension of citation dynamics}

Next, we move on to the temporal dimension of the citation dynamics.
As the following argument applies equally to journal papers and eprints, we use the term \q{paper} instead of \q{eprint} for the moment.

\subsection{Discipline-average citation history curves}

There have been two approaches known in the literature to investigate the scientific literature growth and obsolescence.
The first approach is called the retrospective (or \q{synchronous}, \q{backwards-looking}, \q{citations from}) approach \cite{Price65,Nakamoto88,Barnett89,Burrell02,Glanzel04,Redner05,Lariviere08,Verstak14,Yin17,Zhang17a,Pan18}, which looks at the distribution of the age of papers cited by papers published in a given year.
The second approach is called the prospective (or \q{diachronous}, \q{forward-looking}, \q{citations to}) approach \cite{Nakamoto88,Burrell02,Glanzel04,Redner05,Bouabid11,Bouabid13,Wang13,Parolo15,Yin17,Pan18}, which looks at the distribution of citations acquired over time by papers published in a given year.
Our methodology was based on the latter, prospective approach, investigating the citation history of papers until the data retrieval date.
Also, our analyses here are conducted at the discipline-average level, in which the citation dynamics is driven by collective scientific attention.
The discipline-average trajectory of citation accumulation---\textit{average citation history curve}---is indicative of the typical growth and obsolescence patterns of papers associated with the research discipline.

We conceptualise and model the per-paper average citation history curve as the superposition of the three components:
\begin{equation}\label{KeyEq}
\text{average citation history}~\approx~(i)~\text{\q{jump--decay}}~+~ (ii)~\text{\q{constant attention}}~+~(iii)~\text{\q{anomalies}}\,.
\end{equation}
The first component, $(i)$, represents the pattern in which the average citation count increases with time to its peak value and then decreases as additional time passes, shaping a positively skewed curve.
The second component, $(ii)$, is assumed to derive from those papers that continue to acquire citations even after a paper's typical lifespan \cite{Redner05,Bouabid11,Bouabid13}.
The cumulative advantage \cite{Price76,Newman01,Barabasi99,Barabasi02,Redner05,Sheridan18}, i.e.\ a tendency that highly cited papers are more likely to accumulate additional citations than papers with fewer citations, would contribute to this component.
Here, although the classic preferential attachment model \cite{Barabasi99,Newman01,Barabasi02} is not necessarily assumed in its original form, a variation of the Matthew effect (\textit{\q{the rich get richer}}) \cite{Merton68} at the paper level is expected to be in operation.
The third component, $(iii)$, arise from unexpected or unpredictable citation events, accounting for deviations from the \q{regular} part of the curve, $(i)+(ii)$.
This anomalous component includes cases of multiple humps and the Sleeping Beauties in science \cite{Redner05,Raan04,Ke15,He18}, whose citation history exhibits a long unrecognised period followed by sudden and intense attention.

As the number of papers increases, the contributions from this \q{anomalous} component, $(iii)$, tend to be suppressed, and the average citation history curve will become approximately the \q{regular} part, $(i)+(ii)$.
However, cases of very highly cited papers, including some giant Sleeping Beauties, could still be impactful enough to deform the average citation history curve from the \q{regular} pattern.
In fact, previous studies have found that highly-cited publications tend to follow varied citation history curves \cite{Redner05,Wang13}.
To mitigate such anomalous effects on our analysis of the average citation history curve, we set the threshold percentile of the citation distribution as the 99th percentile, and focus on the eprints below this threshold (hereafter referred to as \q{the below-99th percentile eprints} or \q{the below-99th citation data}).\endnote{%
More generally, we refer to the eprints at and below the $p$th percentile of the citation distribution as \q{the below-$p$th percentile eprints} or \q{the below-$p$th citation data}.}
The rest, top 1\% highest cited eprints, were removed from analyses, as the average citation history curve of such a highest cited cluster of eprints no longer follows the regular time-dependence of citation accumulation (Suppl.~Fig.\ \ref{fig:lognormfit_ctp_top1}a, b).

\subsection{Regression model fitting}

Now we are in a position to construct a regression model for the time distribution of citations.
Let $u_{i}$ denote the average citations of eprints of $i$ years old ($i\in \mathbb{Z}_{0}^{+}$).
It is defined as the total citation counts acquired by eprints posted on arXiv $i$ years ago, divided by the total number of eprints posted on arXiv $i$ years ago. 
Note that, in general, the mean value is not regarded as a good measure of central tendency for a highly skewed distribution, in which case the use of the median is usually preferred.
However, regarding the current arXiv data, the median yielded the values of one or two for all disciplines due to the extremely skewed distribution of citations (Suppl.~Table \ref{tab:cat_by_pctl}), which were not useful for comparative analysis.
It then turned out that the mean value served as a useful indicator to investigate the time profile of the citations, allowing for a comparative analytical approach.
Moreover, the use of the mean value also dovetailed with the stochastic modelling of citation dynamics discussed later (Section \ref{sec:Discussion2}).
Let also $t_{i}$ denote the age of eprints posted on arXiv $i$ years ago, i.e.\ $t_{i}\coloneqq i$.
Then, our regression equation can be stated as
\begin{equation}\label{reg_model}
u_{i}=A f(t_{i}+1;\mu,\sigma)+B g(t_{i};\lambda)+\ep_{i}\,,
\end{equation}
where $A>0$ and $B>0$ are the overall scaling factors for parametric functions $f$ and $g$ (see below for the explicit expressions), respectively, and $\ep_{i}$ is the error term.
This model function comprises two parts.
The first component,
\begin{equation}\label{lognorm.distribution}
f(t;\mu,\sigma)=\f{1}{t\sigma\sqrt{2\pi}}\exp\kko{-\f{(\ln t-\mu)^{2}}{2\sigma^{2}}}\,,
\end{equation}
captures the \q{jump--decay} pattern, $(i)$, of Eq.\ (\ref{KeyEq}).
It is represented by a lognormal probability distribution function so that $\ln t\sim \mathcal{N}(\mu,\sig^{2})$.
Here, we note that previous studies have considered a variety of functional forms to model the obsolescence (ageing) process in citations, including power law \cite{Hajra05,Parolo15}, exponential \cite{Hajra06,Parolo15}, polynomial \cite{Bouabid11,Bouabid13} and lognormal \cite{Wang13,Yin17,He18}, of which some are practical or heuristic, and some are theory-based.
Of all these alternative specifications of the model function, we employed the lognormal function not only for its intuitive clarity and neatness but also for the empirical and theoretical soundness (see \ref{app:lognormal} in the Supplementary Materials for a theoretical validation of the lognormal time distribution).
Some basic statistical parameters characterising this distribution are obtained from the two parameters, $\mu$ and $\sigma$, as:
$\text{Mean}=e^{\mu+\sigma^{2}/2}$, $\text{Median}=e^{\mu}$, $\text{Mode}=e^{\mu-\sigma^{2}}$ and $\text{Variance}=\big(e^{\sigma^{2}}-1\big)e^{2\mu+\sigma^{2}}$.
Note that the argument of $f$ in Eq.\ (\ref{reg_model}) is set as $t_{i}+1$, rather than $t_{i}$, so that the contribution of the lognormal component to the total citation outcome at $i=0$ becomes $A f(t_{i}+1;{\mu},{\sigma})\big|_{i=0}=A\big(\sqrt{2\pi}{\sigma}\big)^{-1}e^{-{\mu}^{2}/(2{\sigma}^{2})}$.
The second component in Eq.\ (\ref{reg_model}), 
\begin{equation}\label{step-like}
g(t;\lambda)=\tanh(\lambda t)\,,
\end{equation}
represents a monotonous increasing, sigmoid function with parameter $\lam>0$ so that $g(t=0;\lam)=0$ and $g(t;\lam)\big|_{t\gg 1}=1$.
This component accounts for the \q{constant attention} component, $(ii)$, of Eq.\ (\ref{KeyEq}).
Here, we employed the hyperbolic tangent as the model function simply for its simplicity while acknowledging other sigmoid-type alternatives.
The rationale of using a \q{soft} sigmoid function rather than a \q{hard} Heaviside step function---the large-$\lam$ limit of (\ref{step-like})---is that the rate of rising can vary among disciplines \cite{Wang13b}.
Specifically, for some disciplines, the \q{constant attention} comes into effect with an almost instant rise, while for others, it can take a few years before reaching a plateau.
Note that the argument of $g$ in Eq.\ (\ref{reg_model}) is set as $t_{i}$ rather than $t_{i}+1$ since the long-lasting citation effect can only come into effect after an eprint is posted on arXiv, so that $g(t_{i};{\lam})\big|_{i=0}=0$ for arbitrary ${\lam}$.
The three parameters ($\mu$, $\sig$ and $\lam$) and the two overall factors ($A$ and $B$) were estimated by applying the nonlinear regression model, Eq.\ (\ref{reg_model}), to the yearly citation data for each discipline.\endnote{The nonlinear regression analyses were conducted using STATA/IC software (version 13; StataCorp LP, Texas, USA) via the --\texttt{nl}-- command, which fits an arbitrary nonlinear function by least squares.
\texttt{https://www.stata.com/manuals13/rnl.pdf}.}
Thus, in contrast to most previous studies that treat the growth and the obsolescence phases separately, often focusing on the latter, we quantify the discipline-average characteristics of the entire citation history.

\subsection{Revealed quantitative and temporal characteristics of citation evolution\label{sec:citation_curves}}

\begin{figure}[tp] 
\centering
\vspace{-0.5cm}
\noindent
\includegraphics[align=c, scale=1.17, vmargin=1mm]{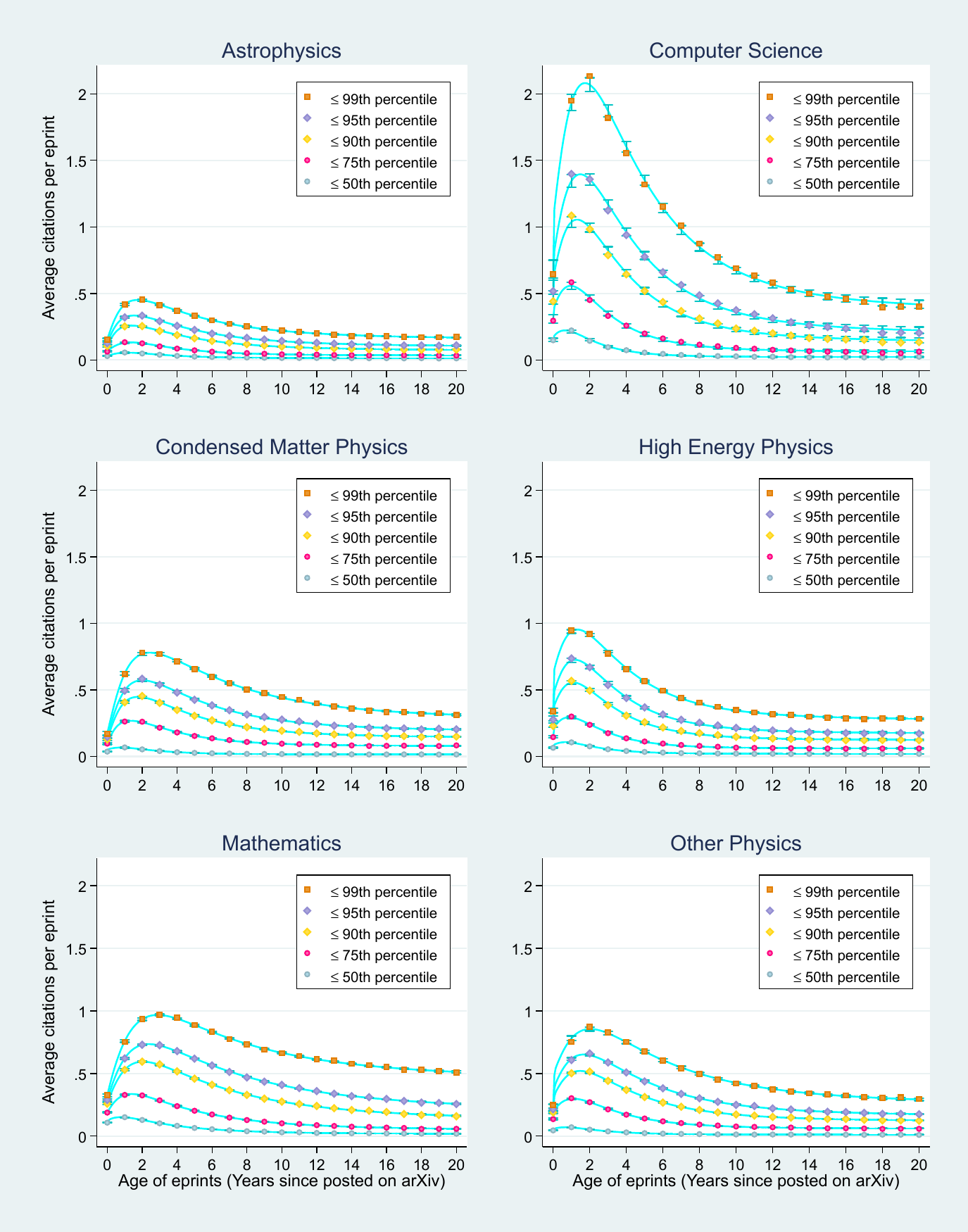}
\caption{\textbf{The citation distribution and the fitted regression curves by discipline.}
Symbols indicate the observations in the arXiv data at each percentile of the citation distribution.
The curves represent the model function of Eq.\ (\ref{reg_model}) with the estimated parameters given in Table \ref{tab:lognorm_fitting_p99} (for the below-99th citation data) and Suppl.~Table \ref{tab:lognorm_fitting_p50-95} (for the citation data with the lower percentile thresholds).
The 99\% confidence intervals are indicated by capped vertical lines.}
\label{fig:citation_curves_cat}
\end{figure}

\begin{figure}[tp] 
\centering
\vspace{-0.5cm}
\noindent
\includegraphics[align=c, scale=0.7, vmargin=1mm]{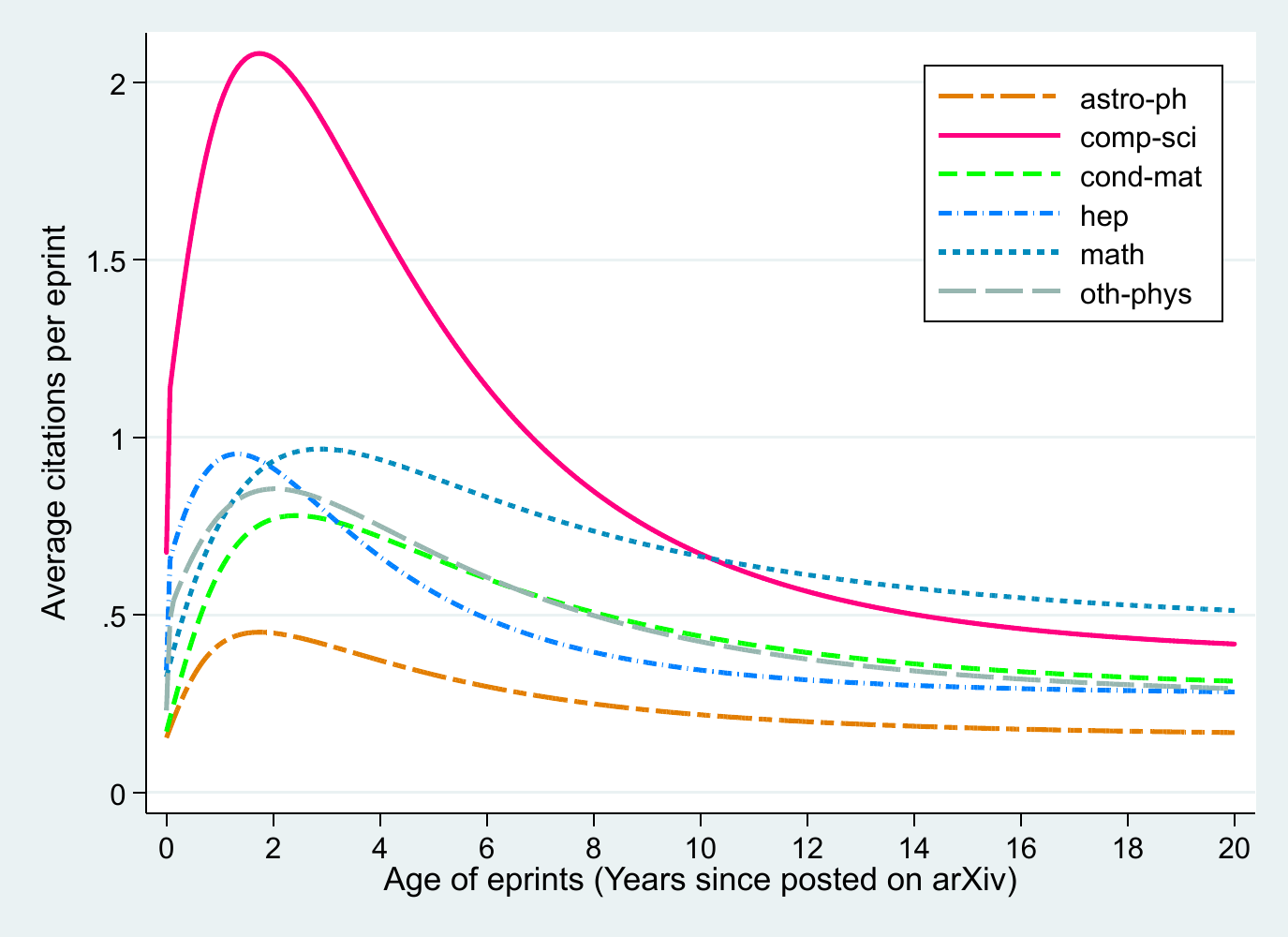}
\caption{\textbf{Discipline-average citation history curves for the arXiv disciplines.}
Fitted curves are based on the regression results for the below-99th percentile eprints.
See Table \ref{tab:lognorm_fitting_p99} for the summary statistics.}
\label{fig:citation_curves_p99}
\vspace{5mm}
\end{figure}

Figure \ref{fig:citation_curves_cat} shows the time distribution of the empirical data, $\{(t_{i},u_{i})\,|\,i=0,\,\dots,\,20\}$, overlaid by the fitted regression curves based on Eq.\ (\ref{reg_model}), i.e.\ $\hat{u}(t)=\hat{A}f(t;\hat{\mu},\hat{\sigma})+\hat{B}g(t;\hat{\lambda})$, $0\leq t\leq 20$, for each arXiv discipline and for the below-$p$th percentile eprints ($p=99,\,95,\,90,\,75,\,50$).
Here, the estimates of parameters are indicated by a \q{hat} symbol.
The coefficient of determination (the adjusted $R$-squared) was higher than 0.99 for all disciplines.
Note that, considering the extremely skewed distribution of citations (Fig.\ \ref{fig:lnc_tot_all_1999}), the vertical axis may be better thought of as shown by an arbitrary scale, which nevertheless is useful for a cross-discipline comparative analysis.
A wide variety of differences can be seen among the fitted curves both across disciplines and percentile sections.
For each discipline panel, the vertical gaps between the regression curves reflect the variance of the quantitative citation distribution (cf.\ Suppl.~Fig.\ \ref{fig:qplot_slope_lnc_tot_all}).
Regarding the percentile section, focusing on the below-99th citation data, Table \ref{tab:lognorm_fitting_p99} summarises the estimated regression coefficients with the derivative metrics, and Fig.\ \ref{fig:citation_curves_p99} shows the predicted discipline-average citation history curves (see Suppl.~Table \ref{tab:lognorm_fitting_p50-95} and Suppl.~Fig.\ \ref{fig:citation_curves_pctl_rest} for the results regarding the lower percentile data).
As can be seen, CS exhibits the highest peak in a short time interval, indicating the highest growth rate of citation counts; its peak value ($\hat{u}_{\mathrm{p}}=2.08$) is more than double that of any other discipline.
Also, Math exhibits the heaviest-tailed curve profile with the highest constant attention component ($\hat{B}=0.452$).\endnote{%
A caveat in interpreting the result for Math is that around 10--20\% of eprints (submission year: 2010--2019) categorised in the Math discipline was also cross-listed in the CS discipline (cf.\ Suppl.~Fig.\ \ref{fig:overlap}b), which has grown exponentially in the past decades.
Therefore, the quantitative characteristics of Math discussed in this study may not be directly compared to the findings of other works based on different data sources \citec{e.g.}{}{Lariviere14,Wang20}.}
These characteristics are in sharp contrast to the rest, four physics-based disciplines.
Differences among the physics-based disciplines are also evident concerning the peak height, the mode year and the degree of stretching in the time direction.
For example, Astrophysics has the lowest peak value ($\hat{u}_{\mathrm{p}}=0.450$) and the lowest constant attention ($\hat{B}=0.158$).
Also, the \q{jump--decay} phase of HEP occurs in a much shorter time interval than the other physics-based disciplines.
All these characteristics reflect different citation, publication or/and communication practices intrinsic of each research discipline \cite{Schubert96,Zitt05,Bjork13}.

\begin{table}[t]
\vspace{-0.5cm}
\centering
\caption{\textbf{Estimated regression coefficients with the derived metrics.}
Results are shown for the below-99th citation data.
The adjusted $R$-squared was higher than 0.99 for all disciplines.
In addition to the estimated regression coefficients (left half), the peak value ($\hat{u}_{\mathrm{p}}$), the typical time interval of the growth phase ($\delta_{1}$) and the obsolescence phase ($\delta_{2}$), the internal obsolescence rate ($\mathcal{S}=\delta_{1}/\delta_{2}$) and the retention rate ($\mathcal{R}=\hat{B}/\hat{u}_{\mathrm{p}}$) (right half) are also shown by discipline.
The value of $\hat{\lam}$ larger than ten is displayed as \q{$\gg 1$}.
(Results regarding the lower percentile data are presented in Suppl.~Table \ref{tab:lognorm_fitting_p50-95}.)}
\label{tab:lognorm_fitting_p99}
{\footnotesize
{\setlength{\tabcolsep}{0.6em}
\begin{tabular}{l@{\quad}dddddd@{\hspace{-8mm}}dddddd}\\[-2mm] \toprule[1pt] \\[-6.0mm]
Category & \multicolumn{1}{c}{$\hat{A}$}        & \multicolumn{1}{c}{$\hat{\mu}$} & \multicolumn{1}{c}{$\hat{\sigma}$} & \multicolumn{1}{c}{$\hat{B}$}  & \multicolumn{1}{c}{$\hat{\lam}$}  &{} & \multicolumn{1}{c}{$\hat{u}_{\mathrm{p}}$} & \multicolumn{1}{c}{$\delta_{1}$} & \multicolumn{1}{c}{$\delta_{2}$} & \multicolumn{1}{c}{$\mathcal{S}$} & \multicolumn{1}{c}{$\mathcal{R}$} \\ \cline{1-6}\cline{8-13}\\[-5.0mm] 
\textit{astro-ph} & \multicolumn{1}{d}{2.19} & \multicolumn{1}{d}{1.61} & \multicolumn{1}{d}{0.817} & \multicolumn{1}{d}{0.158} & \multicolumn{1}{d}{1.21} &{} & \multicolumn{1}{d}{0.450} & \multicolumn{1}{d}{2.56} & \multicolumn{1}{d}{2.43} & \multicolumn{1}{d}{1.05} & \multicolumn{1}{d}{0.351}   \\
\textit{comp-sci} & \multicolumn{1}{d}{1.14\times 10} & \multicolumn{1}{d}{1.56} & \multicolumn{1}{d}{0.741} & \multicolumn{1}{d}{0.379} & \multicolumn{1}{d}{\gg 1}\hspace{-8mm} &{} & \multicolumn{1}{d}{2.08} & \multicolumn{1}{d}{2.74} & \multicolumn{1}{d}{2.00} & \multicolumn{1}{d}{1.37}  & \multicolumn{1}{d}{0.182} \\
\textit{cond-mat} & \multicolumn{1}{d}{4.60} & \multicolumn{1}{d}{1.83} & \multicolumn{1}{d}{0.802} & \multicolumn{1}{d}{0.279} & \multicolumn{1}{d}{0.916} &{} & \multicolumn{1}{d}{0.779} & \multicolumn{1}{d}{3.26} & \multicolumn{1}{d}{2.95} & \multicolumn{1}{d}{1.11}  & \multicolumn{1}{d}{0.359} \\
\textit{hep} & \multicolumn{1}{d}{3.71} & \multicolumn{1}{d}{1.37} & \multicolumn{1}{d}{0.725} & \multicolumn{1}{d}{0.277} & \multicolumn{1}{d}{\gg 1}\hspace{-8mm} &{} & \multicolumn{1}{d}{0.953} & \multicolumn{1}{d}{2.32} & \multicolumn{1}{d}{1.61} & \multicolumn{1}{d}{1.45}  & \multicolumn{1}{d}{0.290} \\
\textit{math} & \multicolumn{1}{d}{6.25} & \multicolumn{1}{d}{1.91} & \multicolumn{1}{d}{0.927} & \multicolumn{1}{d}{0.452} & \multicolumn{1}{d}{0.439} &{} & \multicolumn{1}{d}{0.918} & \multicolumn{1}{d}{2.85} & \multicolumn{1}{d}{3.88} & \multicolumn{1}{d}{0.735} & \multicolumn{1}{d}{0.493} \\
\textit{oth-phys} & \multicolumn{1}{d}{5.04} & \multicolumn{1}{d}{1.76} & \multicolumn{1}{d}{0.805} & \multicolumn{1}{d}{0.259} & \multicolumn{1}{d}{\gg 1}\hspace{-8mm} &{} & \multicolumn{1}{d}{0.855} & \multicolumn{1}{d}{3.03} & \multicolumn{1}{d}{2.77} & \multicolumn{1}{d}{1.10} & \multicolumn{1}{d}{0.303} \\[0mm] \bottomrule[1pt] \\
\end{tabular}}
}
\vspace{5mm}
\end{table}

Besides, the timing in which the \q{constant attention} component---the second term in Eq.\ (\ref{reg_model})---comes into effect also varies across disciplines, from an almost instant rise (e.g.\ CS and HEP) to a gradual rise (Math).
To elaborate on this point, let us decompose each predicted citation history curve into its components, i.e.\ the \q{jump--decay} (lognormal) component and the \q{constant attention} (sigmoid) component.
Let $\rho(T)$ be the ratio of the citation contribution from the lognormal part to the total citations during the period $[0,T]$.
Introducing the collective shorthand notations, ${{\omega}}_{1}=\{{A},{\mu},{\sig}\}$, ${{\omega}}_{2}=\{{B},{\lam}\}$ and ${{\Omega}}=\{{{\omega}}_{1},{{\omega}}_{2}\}$, it is evaluated by
\begin{equation}\label{cumulative}
\rho(T;\hat{{\Omega}})\coloneqq\f{F(T;\hat{{\omega}}_{1})}{{H}(T;\hat{{\Omega}})}
=1-\f{G(T-1;\hat{{\omega}}_{2})}{{H}(T;\hat{{\Omega}})}\,,
\end{equation}
where $F$, $G$ and $H$ are the cumulative functions defined by
\begin{equation}
F(T;{{\omega}}_{1})=A\int_{0}^{T}f(t;{\mu},{\sigma})\dd t
=A\Phi\ko{\f{\ln T-{\mu}}{{\sig}}}\,,\quad
G(T;{{\omega}}_{2})=B\int_{0}^{T}g(t;{\lam})\dd t
=\f{B}{\lam}\ln\big(\cosh(\lam T)\big)\,,
\end{equation}
and $H(T;{{\Omega}})=F(T;{{\omega}}_{1})+G(T-1;{{\omega}}_{2})$, respectively.
The time behaviour of $\rho(T)$ varies across disciplines (Suppl.~Figs.\ Figs.\ \ref{fig:fg_balance_by_pctl_rest}, \ref{fig:fg_balance_years}).
The CS, HEP and Other Physics curves exhibit a cusp soon after the posting on arXiv, reflecting the rapid saturation (i.e.\ $\hat\lambda\gg 1$).
As the age increases, $\rho$ tends to be independent of the shape parameters $(\hat{\mu},\,\hat{\sig},\,\hat{\lam})$ and becomes dependent only on the overall factors $(\hat{A},\,\hat{B})$.
Specifically, $\rho\sim \big(1+(\hat{B}/\hat{A})T\big)^{-1}$ as $T\gg 1$.
Throughout the period (except for the first few years of the cusp period), $\rho_{\mathsf{CS}}$ remains at a high level $(\gtrsim 60\%)$ for all percentile sections, reflecting the largest overall factor for the lognormal component ($\hat{A}=11.4$).
By contrast, $\rho_{\mathsf{Math}}$ is the lowest for the below-99th citation data, but the highest for the below-50th citation data for $T\in [5,10]$; the lower the percentile threshold, the higher the level of $\rho_{\mathsf{Math}}$.

\subsection{Speed of obsolescence and memory of science}

Having obtained the average citation history curve with quantitative characteristics, we now conduct a cross-discipline comparative analysis of the obsolescence rate.
To characterise the obsolescence phenomenon, previous studies commonly used measures such as \q{half-life} (which often also refers to the median citation age) \cite{Burton60,Lariviere08,Parolo15,Zhang17a}, average citation age \cite{Redner05,Lariviere08,Zhang17a}, \q{life-expectancy} (or \q{life-time}) \cite{Bouabid11,Bouabid13} and Price Index \cite{Lariviere08,Zhang17a}.\endnote{%
The \q{half-life} (median citation age) and the average citation age are defined either in a retrospective/synchronous or a prospective/diachronous manner, depending on the context.}
An alternative way to conceptualise the obsolescence rate may be to use the magnitude of the down-slope (gradient) of the citation history curve after its peak, which involves both the quantitative and temporal dimensions.
Whichever of the above indicators is used, it is subject to discipline-specific biases arising from the disciplinarily different publication practices, systems and processes \cite{Schubert96,Zitt05,Bjork13}.
For instance, regarding the quantitative bias, ten citations received by a Life Science paper would not represent the same citational impact as that received by a Math paper (e.g.\ \citea{}{Schubert96}, Table 1; \citea{}{Okamura19}, Suppl.~Table S2).
This means that the drop in the raw number of yearly citations would not be a well-suited indicator of obsolescence.
Also, regarding the temporal bias, what a year means to the academic life-cycle can be very different between disciplines.
For instance, some disciplines have a high turnover rate of citation with a fast-paced research environment compared to other disciplines, as already evidenced in our arXiv data (Fig.\ \ref{fig:citation_curves_p99}, Suppl.~Fig.\ \ref{fig:citation_curves_pctl_rest}).
Indeed, an investigation of the predicted cumulative citation history curves revealed that, on average, more than half of the citations throughout the life-to-date of a HEP eprint has already been received within the first three years since posted on arXiv, whereas the corresponding figure for a Math eprint is below 30\%.
Besides, publishing delay also varies across disciplines \cite{Aman13,Bjork13}.
Therefore, a comparative analysis of the obsolescence rates can be conducted only after first gauging the relevant bibliometric coordinates against a quantitatively and temporally normalised reference standard.

To achieve this goal, we consider the quantitative and temporal aspects of obsolescence separately and introduce two kinds of obsolescence metrics defined for each dimension.
First, we investigate the time-wise obsolescence by focusing on the temporal characteristics of the lognormal (\q{jump--decay}) component of the average citation history curve.
We introduce a discipline-specific time-scale, i.e.\ a typical unit time associated with each discipline.
Just as the biological age---measured by a biological clock---is more relevant than the chronological age in some biological studies, here, \q{bibliodynamical} age---measured by a \q{bibliodynamical} clock---is more relevant to measure the obsolescence perceived by the researchers of each discipline.
Let $\delta_{1}$ and $\delta_{2}$ denote the typical time interval of the growth phase and the obsolescence phase associated with a citation history curve.
Among several possible ways to identify these characteristics in the lognormal distribution,\endnote{%
Alternative ways to quantify $\delta_{2}$, based on the lognormal characterisation, include $\delta_{2}'=\text{Mean}-\text{Mode}$ and $\delta_{2}''=\text{Mean}-\text{Median}$, the resulting internal obsolescence rate simply being Eq.\ (\ref{ratio}) with the term $\sig^{2}$ replaced with $3\sig^{2}/2$ and $\sig^{2}/2$, respectively.} 
we adopt the definitions that $\delta_{1}\coloneqq \text{Mode}$ and $\delta_{2}\coloneqq \text{Median}-\text{Mode}$.
These metrics can be directly obtained from the estimated regression parameters of the lognormal distribution, yielding $\delta_{1}=e^{\hat{\mu}-\hat{\sigma}^{2}}$ and $\delta_{2}=e^{\hat{\mu}}(1-e^{-\hat{\sigma}^{2}})$.
Subsequently, we define the time-adjusted rate of temporal obsolescence as the ratio of $\delta_{1}$ to $\delta_{2}$,
\begin{equation}\label{ratio}
\mathcal{S}\coloneqq \f{\delta_{1}}{\delta_{2}}=\f{1}{\exp\big(\hat{\sig}^{2}\big)-1}\,,
\end{equation}
which we call the \textit{internal obsolescence rate}.
Literally, it represents how long it takes to reach the peak time (Mode) after being posted on arXiv, divided by how long it takes to reach the typical point in time lying in the skirts of the curve (Median) after the peak time.
This quantity is dimensionless, depending only on the shape parameter, $\hat{\sig}$, without dependence on the scale parameter, $\hat{\mu}$, of the lognormal distribution.
Also, it behaves as $\mathcal{S}(\hat{\sig})\sim \hat{\sigma}^{-2}$ for $\hat{\sigma}\ll 1$, indicating a very fast (almost instant) obsolescence, and as $\mathcal{S}(\hat{\sig})\sim e^{-\hat{\sigma}^{2}}$ for $\hat{\sigma}\gg 1$, indicating a very slow (no) obsolescence.

\begin{figure}[tp]
\centering
\vspace{-0.5cm}
    \begin{tabular}{l}
\begin{minipage}{0.5\hsize}
\begin{flushleft}
\raisebox{-\height}{\includegraphics[align=c, scale=0.6, vmargin=0mm]{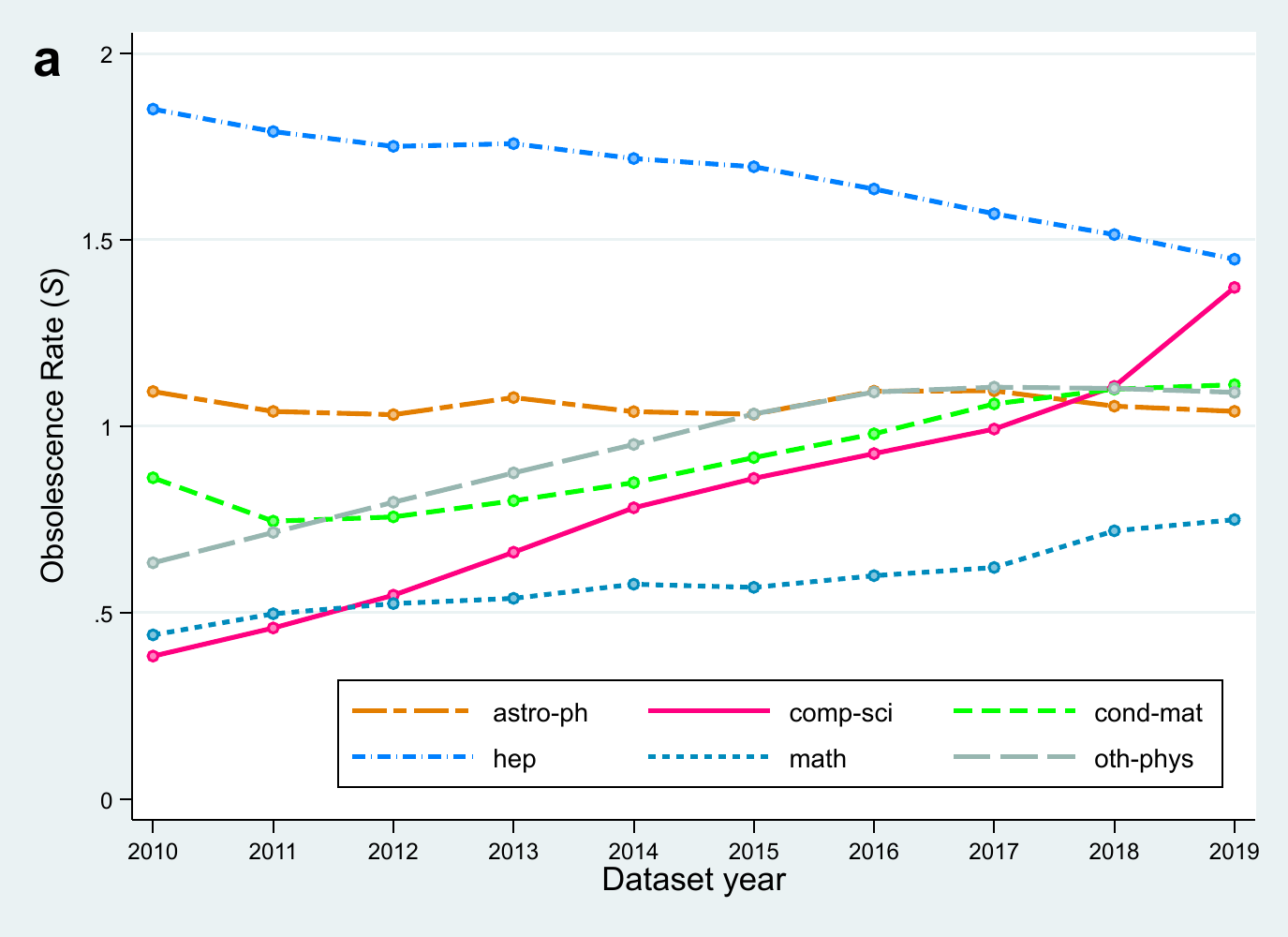}}
\end{flushleft}
      \end{minipage}
\begin{minipage}{0.5\hsize}
\begin{flushleft}
\raisebox{-\height}{\includegraphics[align=c, scale=0.6, vmargin=0mm]{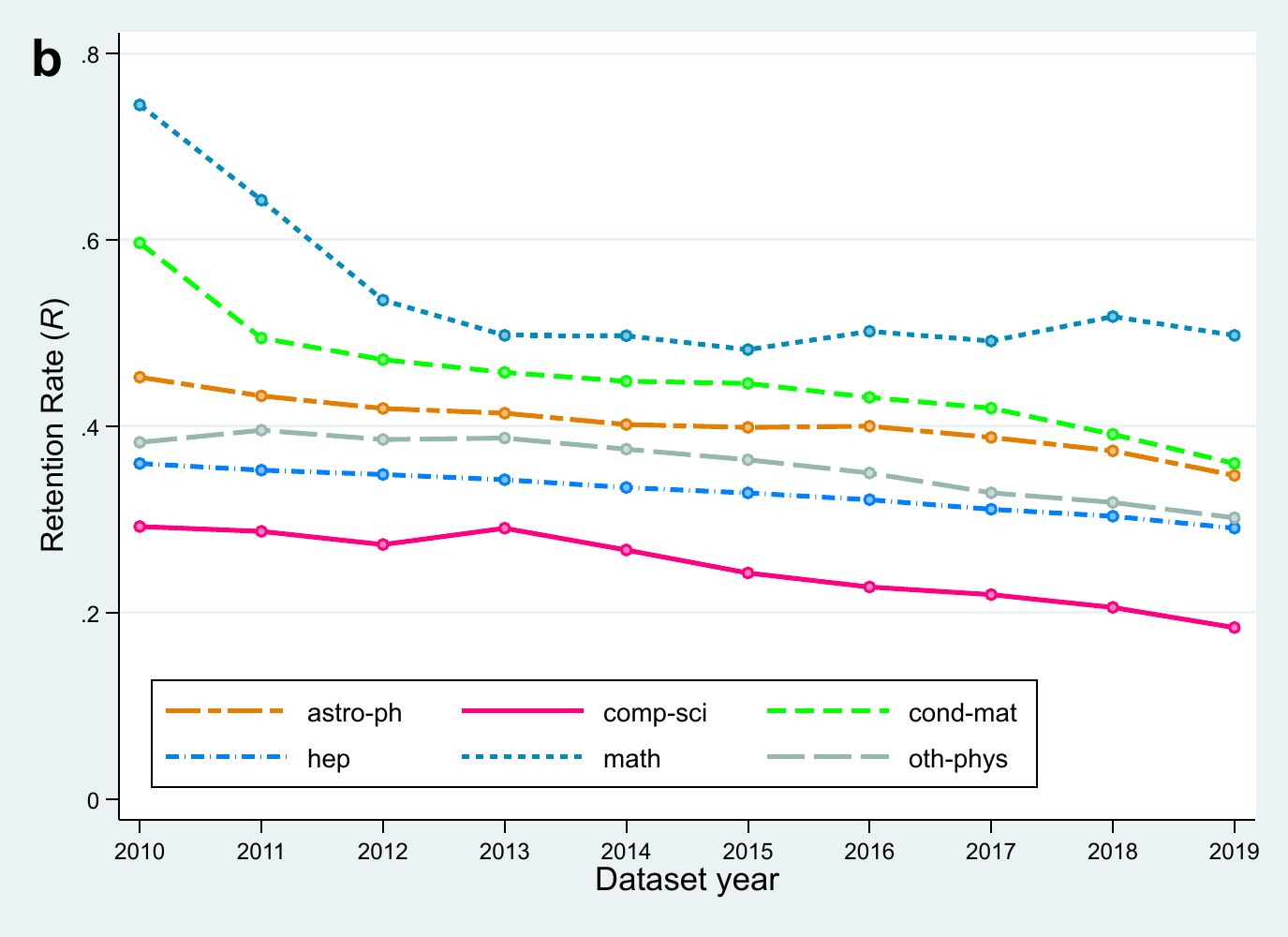}}
\end{flushleft}
          \end{minipage}
    \end{tabular}
\caption{\textbf{Trends in the internal obsolescence rate ($\bmt{\mathcal{S}}$) and the retention rate ($\bmt{\mathcal{R}}$).}
The analyses are based on the below-99th percentile eprints posted on arXiv during 2010--2019. 
\textsf{\textbf{a}}, trends in the internal obsolescence rate defined by $\mathcal{S}=\delta_{1}/\delta_{2}$, where $\delta_{1}$ and $\delta_{2}$ respectively represent the typical time interval of the growth phase and the obsolescence phase identified in the average citation history curve; 
\textsf{\textbf{b}}, trends in the retention rate defined by $\mathcal{R}=\hat{B}/\hat{u}_{\mathrm{p}}$, where $\hat{B}$ and $\hat{u}_{\mathrm{p}}$ respectively represent the asymptotic value and the peak value of the average citation history curve.}
\label{fig:obso_rate_99}
\vspace{0.5cm}
\end{figure}

It is intriguing to know how the obsolescence rate has changed over time \cite{Evans08,Lariviere08,Verstak14,Parolo15,Martin-Martin16,Zhang17a,Pan18}.
The findings from the previous studies have been inconclusive or inconsistent, as they use different sources of the citation data with different field categorisations and time windows.
We revisited this issue by using the arXiv data with the proposed internal obsolescence rate of Eq.\ (\ref{ratio}).
We constructed a series of ten consecutive sub-datasets from the entire arXiv dataset such that the first, the second and the $n$th sub-dataset respectively contain the eprints posted on arXiv before and including the year 2019, 2018 and $2020-n$, with $n$ running from 1 to 10.
Subsequently, the regression model of Eq.\ (\ref{reg_model}) was applied to each sub-dataset.
The estimated regression coefficients were used to obtain $\delta_{1}$ and $\delta_{2}$ for each dataset year (2010--2019), with which the internal obsolescence rate ($\mathcal{S}$) is obtained by the formula (\ref{ratio}).
Figure \ref{fig:obso_rate_99} shows the resultant trends in $\mathcal{S}$ by discipline.
During the past decade, the most remarkable increase is observed for CS, for which $\mathcal{S}$ has increased from 0.383 (2010; the lowest of all) to 1.37 (2019; close to the highest).
It implies that for CS researchers today, the knowledge at the research front has increasingly become obsolete with the exponential growth in the CS eprints (Suppl.~Fig.\ \ref{fig:lineplot_comb2}).
Condensed Matter Physics and Other Physics also follow an increasing trend with relatively similar levels of $\mathcal{S}$. 
Math shows a less steep increase from 0.440 (2010) to 0.749 (2019), remaining at the lowest level since 2012.
Astrophysics stays relatively flat around slightly above $\mathcal{S}=1$, suggesting that the speed of obsolescence does not vary much over time.
By contrast, HEP exhibits a decreasing trend, for which $\mathcal{S}$ has steadily dropped from 1.85 (2010) to 1.45 (2019).
Still, HEP kept at the highest level throughout the period, suggesting a highly competitive situation among researchers.

In contrast to the time-wise obsolescence discussed above, the quantity-wise obsolescence can be quantified as the degree to which the early peak level is retained after a while to reach the \q{constant attention} level.
The corresponding metric is defined by
\begin{equation}\label{retention}
\mathcal{R}\coloneqq \f{\hat{B}}{\hat{u}_{\mathrm{p}}}\,,
\end{equation}
which we call the \textit{retention rate}.
By definition, it depends only on the (yearly) citation peak value and the asymptotic value, without dependence on the process or trajectory of the citation evolution.
Using the same datasets as used for the internal obsolescence rates, the retention rates for the past years and the trend therein were also analysed (Fig.\ \ref{fig:obso_rate_99}b).
The ranking of the six arXiv disciplines has not changed over the period (2010--2019).
CS, the lowest throughout, shows a steady decreasing trend, and the four physics-based disciplines also show decreasing trend, indicating the gradual progress of the quantity-wise obsolescence. 
Math, the highest throughout, shows a significant drop during 2010--2013, but remains in a relatively flat trend around $\mathcal{R}=0.5$ after 2013, suggesting that rather \q{old} Math eprints have continued to have a presence as a knowledge source.
Finally, we note that the retention rate is reciprocally related to what could be called the \textit{inflation rate}, defined by the ratio of the peak value, $\hat{u}_{\mathrm{p}}$, to the baseline level, $\hat{B}$.
Put differently, it measures the degree to which the early growth trend is inflationary compared to the stationary level reached after a while (see Suppl.~Fig.\ \ref{fig:inflation_99}).

\section{Discussion}

Based on the revealed knowledge of the quantitative and temporal dimensions of citation dynamics, this section extends our results in two directions.
First, we develop a new citation-based index that can be used to compare the citational impact of papers published in different disciplines at different times.
Second, we provide a mathematical (stochastic) model of citation dynamics that reproduces both the quantitative and temporal patterns observed in the previous sections.
The first part will be of interest to practitioners of research evaluation and science policymakers who wish to improve the informed peer-review process, while the second part is mainly for interested researchers working on the intersection of scientometrics and mathematical physics/economics.

\subsection[Quantitatively-and-temporally normalised index of citational attention (\q{$\gamma$-index})]{Quantitatively-and-temporally normalised index of citational attention (\q{$\bmt{\gamma}$-index})\label{sec:Discussion1}}

Previous studies have developed a number of citation-based indices to quantify the research impact of scientific publications.
These indices are often \q{field-normalised}, reflecting the fact that the average citations per paper vary widely between fields \cite{Schubert96,Zitt05,Waltman16}.
However, the normalisation scales used there have mostly been based on the bibliometric data on journal papers.
Considering that scholarly communications, including citations, have already been or are being heavily dependent on eprint (preprint) systems in many research disciplines (see Introduction), the normalisation scales are necessarily biased in both the quantitative and temporal dimensions without consideration of nonjournal-based citations.\endnote{%
As an illustration, Suppl.~Table S2 of \citek{Okamura19} reported the per-paper average citations (10-year average) for \q{Computer Science and Mathematics}, \q{Physics \& Space Sciences} and \q{Basic Life Sciences} to be 5.26, 11.9 and 14.7, respectively.
A more detailed study (unpublished) by the same author also revealed that the corresponding figure for \q{CS}, \q{Math} and \q{Physics} are 6.76, 4.37 and 11.2, respectively.
These results were obtained from the Essential Science Indicators (ESI) database, published by Clarivate Analytics, which only covers journal-published papers.
By contrast, the current analysis based on the arXiv data predicts that the per-eprint average citations for CS and Math as $H_{\mathsf{CS}}\big|_{T=10}=18.7$ and $H_{\mathsf{Math}}\big|_{T=10}=14.1$, respectively, which are far above those for the four physics-based disciplines, in sharp contrast to the journal-based findings.
These differences apparently stem from the fact that research outputs in CS and Math are significantly underrepresented in the ESI database.}
In addition, an appropriate citation window varies between disciplines due to the different \q{bibliodynamic clock} discussed earlier.
Towards a less biased evaluation and comparative analysis of the scientific attention, we propose a new, quantitatively-and-temporally normalised citation index based on the revealed characteristics of the average citation history curve.

\begin{figure}[tp] 
\centering
\vspace{-0.5cm}
\noindent
\includegraphics[align=c, scale=0.7, vmargin=1mm]{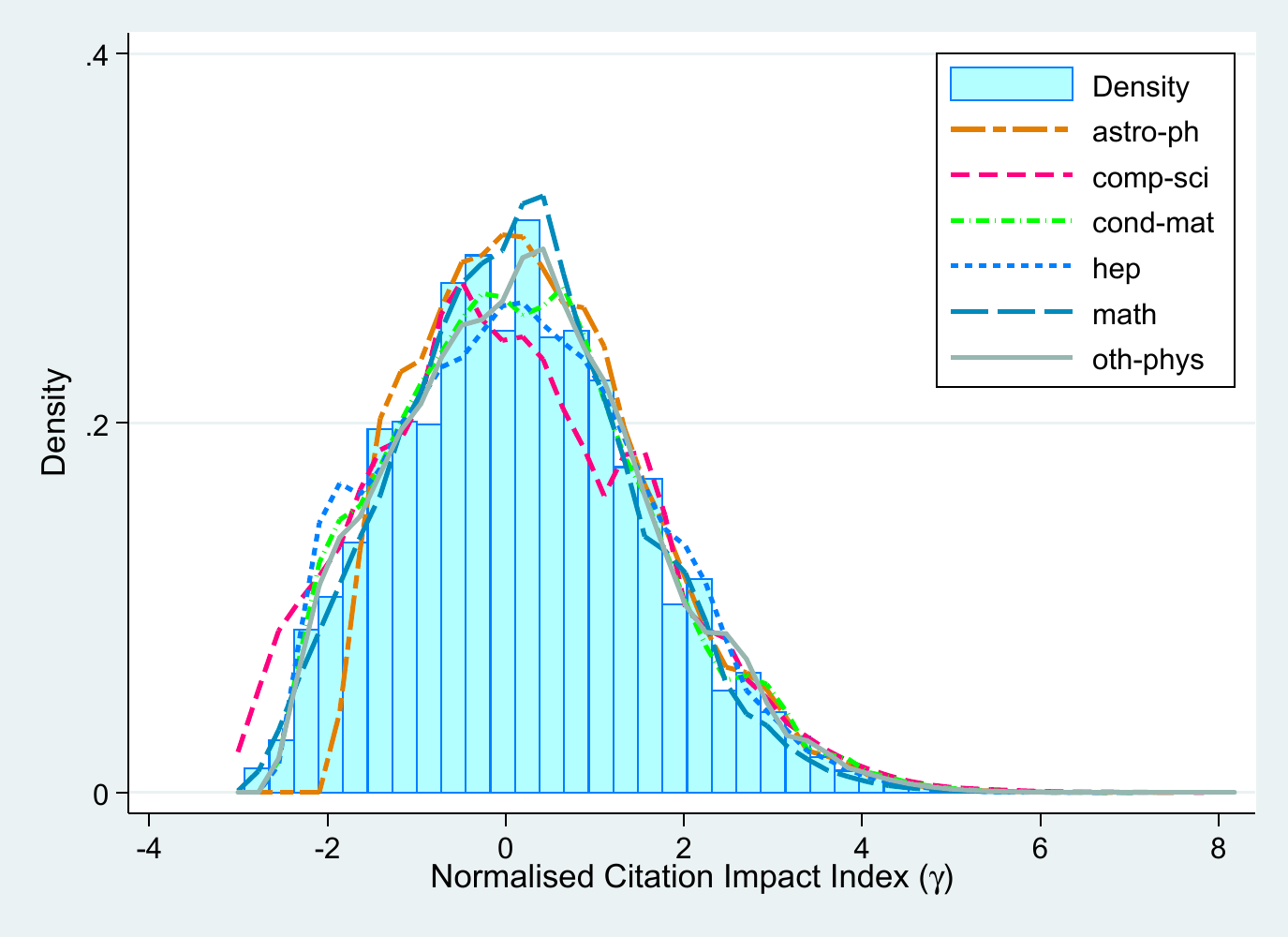}
\caption{\textbf{Distribution of the quantitatively-and-temporally normalised citation index ($\bmt{\gamma}$).}
Results are shown for the below-99th percentile eprints with nonzero citations.
Shown is the frequency distribution of the $\gamma$-index, overlaid by the graphs of the kernel density estimation ($\textrm{half-width}=0.2$) by discipline.}
\label{fig:impact}
\vspace{5mm}
\end{figure}

Let $c_{k}(T)$ denote the number of cumulative citations of eprint $k$ at some time point $t=T>0$, posted on arXiv at $t=0$.
The above consideration regarding the quantitative and temporal biases leads us to introduce a proper normalisation factor, which is precisely given by the cumulative density function introduced before (see the definition below Eq.\ (\ref{cumulative})), $H_{\mathsf{s}(k)}(T)\coloneqq H(T;\hat{{\Omega}}_{\mathsf{s}(k)})$, evaluated for discipline $\mathsf{s}(k)$ to which eprint $k$ belongs.
The quantitatively-and-temporally normalised citation index for eprint $k$ at time $T$ is then defined by
\begin{equation}\label{impact}
\gamma_{k}(T)\coloneqq\ln\ko{\f{c_{k}(T)}{H_{\mathsf{s}(k)}(T)}}\,,
\end{equation}
which we call the \textit{$\gamma$-index}.
By this definition, $\gamma=0$, $\gamma>0$ and $\gamma<0$ correspond to, respectively, a citation equal to, more than and less than that expected by the discipline-average citation history curve.
Here, we remind that $H_{\mathsf{s}(k)}(T)$ has been estimated based on the mean values of the citation distribution, which is extremely skewed.
In this sense, the $\gamma$-index may also be better thought of as quantified by an arbitrary scale.
In what follows, we focus on the eprints with nonzero citations (hereafter referred to as \q{the nonzero-citation data}; $N=1,206,997$).
Below, we restrict our analysis to those eprints posted on arXiv after and including the year 1999, which is the same set of eprints as used in the regression analysis before.
Figure \ref{fig:impact} shows the frequency distribution of the $\gamma$-index for the nonzero-citation data, overlaid by the graphs of the associated kernel density estimation by discipline (see also Suppl.~Table \ref{tab:impact_nlx}).
Overall, the $\gamma$-index tends to be distributed normally.
However, the Gaussian approximation becomes invalid for the left tail of the distribution around $\gamma\lesssim -2$.
This fact can also be verified through the lognormal quantile plot of $\gamma$ (Suppl.~Fig.\ \ref{fig:gamma_star}a).
Also, the one-way analysis-of-variance (ANOVA) and the posthoc Bonferroni test for inter-discipline comparison showed that the mean values of the distribution of the $\gamma$-index were statistically different among disciplines ($p<0.01$ for all pairs except $p=0.169$ between Condensed Matter Physics and HEP).

To further refine the index, we also considered the associated standardised index defined by
\begin{equation}\label{gamma}
\gamma^{*}_{k}(T)\coloneqq\Phi^{-1}\big(Q_{k}(T)\big)\,,
\end{equation}
where $Q_{k}(T)$ represents the quantile rank of $\gamma_{k}$ at time $T$, restricted to the discipline $\mathsf{s}(k)$.
This index, called the \textit{$\gamma^{*}$-index}, has a distribution very close to the standard normal distribution (Suppl.~Fig.\ \ref{fig:gamma_star}b, Suppl.~Table \ref{tab:impact_normal}).
This time the one-way ANOVA and the posthoc Bonferroni test for inter-discipline comparison did not reject the null hypothesis that there was no difference among the disciplines ($p=1.00$ for all pairs).
This result supports the validity of the $\gamma^{*}$-index as the \emph{standard} citation index, which can serve as a further less biased alternative to the existent citation-based impact measures.
In practice, however, it is not always possible to calculate the quantile ranks due to limitations on the data availability.
In that case, we can still use the $\gamma$-index of Eq.\ (\ref{impact}) as it can be obtained with only the knowledge of pre-estimated average citation history curves without the need of calculating the quantile ranks.
Indeed, a high correlation between $\gamma$ and $\gamma^{*}$ was verified (Suppl.~Fig.\ \ref{fig:gamma_star}a, Suppl.~Table \ref{tab:pwcorr}; the Pearson's correlation coefficient between the two indices was greater than 0.98 for all disciplines at the 0.1\% significance level), suggesting that the $\gamma$-index is a good proxy of the $\gamma^{*}$-index.

For easy reference of users, Table \ref{tab:hayamihyou99} presents a \q{ready reckoner} for the $\gamma$-index pre-calculated based on the below-99th citation data.\endnote{%
The numerics presented in Table \ref{tab:hayamihyou99} are based on the below-99th citation data, and therefore, provide a \q{harsher} evaluation of the citational impact than those numerics calculated with the lower percentile thresholds.
For comparison, Suppl.~Table \ref{tab:hayamihyou90} presents another ready reckoner for the $\gamma$-index pre-calculated based on the below-90th citation data.}
Due to the monotonically increasing property of the cumulative density function, $H(T)$, the $\gamma$-index for each citation count monotonically decreases as the elapsed years ($T$) after being posted on arXiv becomes longer.
Note that we have omitted the negative values of $\gamma$ from the table since, as discussed, the $\gamma$-index for the below-99th citation data could be conservatively valid for $\gamma\gtrsim 0$.
As an illustration, a total of 10 citations at $T=2$ in the CS discipline ($\gamma=1.73$) is comparable to a total of 50 citations at $T=6$ in the same discipline ($\gamma=1.71$).
As another illustration, a total of 10 citations at $T=3$ in the HEP discipline ($\gamma=1.68$) is seen as more impactful than the same number of citations at the same elapsed time in the CS discipline ($\gamma=0.96$).
Also, a total of 50 citations at $T=9$ in the Astrophysics discipline ($\gamma=2.87$) is comparable to the same number of citations at $T=4$ in the HEP discipline ($\gamma=2.91$), or a total of 100 citations at $T=4$ in the CS discipline ($\gamma=2.85$).

We have demonstrated how the idea of the quantitatively-and-temporally normalised index of the citational impact works by applying the $\gamma$-index to the real world (arXiv) data.
However, it should be kept in mind that this index would still produce a significantly biased result if applied without an appropriate classification of research disciplines or subfields.
For instance, even within the same discipline of {HEP}, papers on theoretical HEP and experimental HEP could show quite distinct characteristics regarding the typical citation history profile.
The same situation can also occur for the {Math} discipline, e.g., between papers on pure mathematics and applied mathematics.
We also note that, depending on the research discipline, the $\gamma$-index may not be a beneficial indicator of the citational impact for particularly \q{young} papers produced within a couple of years \citec{cf.}{}{Wang13b}.
With all these caveats, the $\gamma$-index (or the $\gamma^{*}$-index) can offer a useful, albeit approximate and imperfect, measure to evaluate and compare the citational attention on scientific publications across disciplines and time periods.

\begin{table}[tp]
\renewcommand{\arraystretch}{0.7}
\vspace{-0.5cm}
\centering
\caption{\textbf{\q{Ready reckoner} for the normalised citation index ($\bmt{\gamma}$).}
Pre-calculated values are shown by discipline, based on the below-99th citation data.
Negative $\gamma$-values are omitted and shown as a dash (--).}
\label{tab:hayamihyou99}
{\scriptsize
{\setlength{\tabcolsep}{0.7em}
\begin{tabular}{ld@{\hspace{0cm}}d@{\hspace{0cm}}ddddddddd}\\[-2mm] 
\toprule[1pt] \\[-3.5mm]
{} & {} &{} & \multicolumn{1}{c}{\hspace{-4mm}$T=2$}    & \multicolumn{1}{c}{\hspace{-4mm}$T=3$}     & \multicolumn{1}{c}{\hspace{-4mm}$T=4$}     & \multicolumn{1}{c}{\hspace{-4mm}$T=5$}     & \multicolumn{1}{c}{\hspace{-4mm}$T=6$}     & \multicolumn{1}{c}{\hspace{-4mm}$T=7$}     & \multicolumn{1}{c}{\hspace{-4mm}$T=8$}     & \multicolumn{1}{c}{\hspace{-4mm}$T=9$}     & \multicolumn{1}{c}{\hspace{-4mm}$T=10$}    \\ \cline{4-12}\\[-2.5mm]
{\footnotesize{\textit{astro-ph}}} & \multicolumn{1}{l}{~$c={}$} & 5         & 2.61 & 1.82 & 1.39  & 1.12  & 0.92  & 0.77  & 0.66  & 0.56  & 0.48  \\
{} & \multicolumn{1}{l}{~$c={}$} & 10        & 3.31 & 2.51 & 2.08  & 1.81  & 1.61  & 1.47  & 1.35  & 1.26  & 1.17  \\
{} & \multicolumn{1}{l}{~$c={}$} & 50        & 4.92 & 4.12 & 3.69  & 3.42  & 3.22  & 3.08  & 2.96  & 2.87  & 2.78  \\
{} & \multicolumn{1}{l}{~$c={}$} & 100       & 5.61 & 4.82 & 4.39  & 4.11  & 3.92  & 3.77  & 3.65  & 3.56  & 3.48  \\[0mm] \hline \\[-2.5mm]
{\footnotesize{\textit{comp-sci}}} & \multicolumn{1}{l}{~$c={}$} & 5        & 1.04  & 0.27  & \multicolumn{1}{c}{---} & \multicolumn{1}{c}{---} & \multicolumn{1}{c}{---} & \multicolumn{1}{c}{---} & \multicolumn{1}{c}{---} & \multicolumn{1}{c}{---} & \multicolumn{1}{c}{---} \\
{} & \multicolumn{1}{l}{~$c={}$} & 10        & 1.73  & 0.96  & 0.54  & 0.28  & 0.10  & \multicolumn{1}{c}{---} & \multicolumn{1}{c}{---} & \multicolumn{1}{c}{---} & \multicolumn{1}{c}{---} \\
{} & \multicolumn{1}{l}{~$c={}$} & 50        & 3.34  & 2.57  & 2.15  & 1.89  & 1.71  & 1.58  & 1.49  & 1.41  & 1.35  \\
{} & \multicolumn{1}{l}{~$c={}$} & 100       & 4.03  & 3.27  & 2.85  & 2.59  & 2.41  & 2.28  & 2.18  & 2.10  & 2.04  \\[0mm] \hline \\[-2.5mm]
{\footnotesize{\textit{cond-mat}}} & \multicolumn{1}{l}{~$c={}$} & 5         & 2.35 & 1.43  & 0.93  & 0.61  & 0.38  & 0.21  & 0.08  & \multicolumn{1}{c}{---} & \multicolumn{1}{c}{---} \\
{} & \multicolumn{1}{l}{~$c={}$} & 10        & 3.04 & 2.13  & 1.62  & 1.30  & 1.08  & 0.91  & 0.77  & 0.67  & 0.57  \\
{} & \multicolumn{1}{l}{~$c={}$} & 50        & 4.65 & 3.74  & 3.23  & 2.91  & 2.69  & 2.52  & 2.38  & 2.27  & 2.18  \\
{} & \multicolumn{1}{l}{~$c={}$} & 100       & 5.35 & 4.43  & 3.93  & 3.61  & 3.38  & 3.21  & 3.08  & 2.97  & 2.88  \\[0mm] \hline \\[-2.5mm]
{\footnotesize{\textit{hep}}} & \multicolumn{1}{l}{~$c={}$} & 5         & 1.68 & 0.98  & 0.61  & 0.37  & 0.21  & 0.09  & \multicolumn{1}{c}{---} & \multicolumn{1}{c}{---} & \multicolumn{1}{c}{---} \\
{} & \multicolumn{1}{l}{~$c={}$} & 10        & 2.38 & 1.68  & 1.30  & 1.06  & 0.90  & 0.78  & 0.68  & 0.61  & 0.54  \\
{} & \multicolumn{1}{l}{~$c={}$} & 50        & 3.99 & 3.29  & 2.91  & 2.67  & 2.51  & 2.39  & 2.29  & 2.21  & 2.15  \\
{} & \multicolumn{1}{l}{~$c={}$} & 100       & 4.68 & 3.98  & 3.60  & 3.37  & 3.20  & 3.08  & 2.99  & 2.91  & 2.84  \\[0mm] \hline \\[-2.5mm]
{\footnotesize{\textit{math}}} & \multicolumn{1}{l}{~$c={}$} & 5         & 1.98 & 1.17  & 0.69  & 0.37  & 0.13  & \multicolumn{1}{c}{---} & \multicolumn{1}{c}{---} & \multicolumn{1}{c}{---} & \multicolumn{1}{c}{---} \\
{} & \multicolumn{1}{l}{~$c={}$} & 10        & 2.67 & 1.86  & 1.38  & 1.06  & 0.83  & 0.65  & 0.50  & 0.39  & 0.29  \\
{} & \multicolumn{1}{l}{~$c={}$} & 50        & 4.28 & 3.47  & 2.99  & 2.67  & 2.43  & 2.26  & 2.11  & 1.99  & 1.89  \\
{} & \multicolumn{1}{l}{~$c={}$} & 100       & 4.97 & 4.16  & 3.68  & 3.36  & 3.13  & 2.95  & 2.81  & 2.69  & 2.59  \\[0mm] \hline \\[-2.5mm]
{\footnotesize{\textit{oth-phys}}} & \multicolumn{1}{l}{~$c={}$} & 5         & 1.94 & 1.17  & 0.74  & 0.45  & 0.25  & 0.10  & \multicolumn{1}{c}{---} & \multicolumn{1}{c}{---} & \multicolumn{1}{c}{---} \\
{} & \multicolumn{1}{l}{~$c={}$} & 10        & 2.63 & 1.86  & 1.43  & 1.15  & 0.94  & 0.79  & 0.67  & 0.57  & 0.49  \\
{} & \multicolumn{1}{l}{~$c={}$} & 50        & 4.24 & 3.47  & 3.04  & 2.75  & 2.55  & 2.40  & 2.28  & 2.18  & 2.10  \\
{} & \multicolumn{1}{l}{~$c={}$} & 100       & 4.93 & 4.17  & 3.73  & 3.45  & 3.25  & 3.09  & 2.97  & 2.88  & 2.80  \\[-0.3mm] \bottomrule[1pt] \\
\end{tabular}}
}
\vspace{5mm}
\end{table}

\subsection{Stochastic modelling of fundamental citation dynamics\label{sec:Discussion2}}

So far, we have investigated the macroscopic, discipline-average picture of citation distribution and evolution pattern.
Indeed, the regression model of Eq.\ (\ref{reg_model}) achieved a remarkable fit to the empirical arXiv data.
However, the underlying microscopic mechanism that governs the quantitative and temporal dimensions of citation dynamics at the level of individual papers, consistent with the revealed macroscopic picture, is yet to be uncovered.
To address this issue, herein, we provide mathematical modelling of the citation dynamics through a stochastic differential equation (SDE).
A key observation is that the Lognormal Law observed for the degree distribution is related to geometric Brownian motion.
This stochastic model plays a central role in many mathematical modelling, most notably in Black--Scholes option pricing \cite{Black-Scholes73,Merton73} known in the financial engineering literature.
Recall that the Black--Scholes model involves the deterministic and stochastic parts, reminding us of some structural similarity to the citation dynamics.
In the real world of scholarly communication via citations, when an author of scientific papers determine which literature to cite, the author's attitudes and behaviours are often influenced---whether conscious or unconscious---by the recent citation performance of the literature.
However, that does not solely determine the paper's reference list; the author's finite cognitive capacity, resource limitations, time constraints and pursuit of novelty introduce some random, nondeterministic factors in the citation phenomenon.
Other random events would include the discovery of unexpected relationships to other works and the subsequent development of research.
These observations lead us to speculate that the citation dynamics is also described by a deterministic plus stochastic model, as in the Black--Scholes model.
Then, intuition suggests that the deterministic part is encoded in the average citation history function, whereas the stochastic part introduces the statistical variability in the trajectory distribution of citations.
It will be an extended Black--Scholes model, where the \q{drift} and the \q{volatility} parameters are now time-dependent, encoding the nontrivial temporal characteristics of the citation dynamics.
Below, we show that it is indeed the case.

Let $\Delta c(t_{i},t_{i+1})$ be the number of citations a paper receives during the time interval $[t_{i},\,t_{i+1}]$ with $t_{0}=0$ and $\Delta t=t_{i+1}-t_{i}$, $i\in \mathbb{Z}_{0}^{+}$.
We interpret $\Delta c(t_{i},t_{i+1})$ as a discretised realisation of a continuous latent attention function, $X(t)$, evaluated at $t_{i}$, such that $\Delta c(t_{i},t_{i+1})=\floor*{X(t_{i})\dD t}$, where the floor function is defined by $\floor*{x}\coloneqq \max\{n\in\mathbb{Z}\,|\,n\leq x\}$ for $x\in\mathbb{R}$.
The latent attention function, $X(t)$, is assumed to be strictly positive for all $t\in \mathbb{R}^{+}_{0}$.
The cumulative citation count at time $t=T=N\dD t$, denoted as $c(T)$, is related to $X(t)$ as $c(T)=\sum_{i=0}^{N-1}\Delta c(t_{i},t_{i+1})=\sum_{i=0}^{N-1}\floor*{X(t_{i})\dD t}\approx \int_{0}^{T}X(t)\dd t$ for sufficiently small $\Delta t$, or large $N$.\endnote{%
More precisely, the relation $c(T)\leq \floor*{\int_{0}^{T}X(t)\dd t}\leq c(T)+1$ follows from the property of the floor function.}
Then, the problem of finding the underlying SDE boils down to obtaining $X(t)$ satisfying the following four properties:
\begin{enumerate}[noitemsep]
\item \label{cond1} $X(t)$ is continuous and positive for for all $t\in \mathbb{R}^{+}_{0}$.
\item \label{cond2} $X(t)$ consists of a deterministic part and a stochastic part.
\item \label{cond3} The mean value of $X(t)$ reproduces the observed average citation history curve.
\item \label{cond4} The distribution of cumulative citations for a large ensemble of papers follows the observed Lognormal Law.
\end{enumerate}
Skipping all the details of the derivation (see \ref{app:stochastic} in the Supplementary Materials), the SDE whose solution satisfies Properties 1--4 is given by
\begin{equation}\label{gGBM0}
dX(t)=X(t)\big[d\ln u(t)+\beta(t)\dd W(t)\big]\,.
\end{equation}
Here, $\{ W(t)\,|\, t\in \mathbb{R}^{+}_{0} \}$ is the standard Brownian motion (or the Wiener process) with $W(t+\Delta t)-W(t)\sim \mathcal{N}(0,\Delta t)$, $u(t)\in\mathbb{R}^{+}$ is the average citation history function as before, and $\beta(t)\in \mathbb{R}^{+}$ controls the random fluctuations of $X(t)$ during the time interval $dt$ as a response to external random events discussed above.
The SDE (\ref{gGBM0}) can be solved straightforwardly, in the It\^{o}'s sense, yielding
\begin{align}\label{solution}
X(t)=u(t)\exp\ko{-\f{1}{2}\int_{0}^{t}\beta(\tau)^{2}\dd\tau +\int_{0}^{t}\beta(\tau) \dd W(\tau)}
=\f{u(t)}{w(t)}\exp\ko{\int_{0}^{t}\beta(\tau) \dd W(\tau)}\,,
\end{align}
where we introduced $w(t)\coloneqq \exp\big(\f{1}{2}\int_{0}^{t}\beta(\tau)^{2}\dd\tau\big)$.
The probability density function for $X(t)$ is obtained by solving the corresponding Fokker--Planck equation.
Here we only present the final result (see \ref{app:stochastic} in the Supplementary Materials for the derivation):
\begin{equation}\label{GBM-PDF0}
p(x,t)=\f{1}{x\sqrt{4\pi\ln w(t)}}\exp\kko{-\f{\ln^{2}\big(xw(t)/u(t)\big)}{4\ln w(t)}}\,.
\end{equation}
We can check that Eqs.\ (\ref{solution}, \ref{GBM-PDF0}) readily satisfy Properties 1--4.
Property \ref{cond1} is automatically satisfied with a continuous positive function $u(t)$.
Property \ref{cond2} is already ensured at the level of the SDE (\ref{gGBM0}), which is also manifest in the solution (\ref{solution}).
Property \ref{cond3} can be directly checked by calculating the expectation value on both sides of Eq.\ (\ref{solution}), yielding $\mathrm{E}[X(t)]=u(t)$.
Consequently, if we choose the citation history function specifically to be $u(t;\hat{\Omega})=\hat{A} f(t+1;\hat{\mu},\hat{\sigma})+\hat{B} g(t;\hat{\lambda})$ with $f$ and $g$ defined in (\ref{lognorm.distribution}) and (\ref{step-like}), respectively, then the solution (\ref{solution}) precisely reproduces the evolution pattern observed for the arXiv citation data.
Finally, regarding Property 4, the probability density function (\ref{GBM-PDF0}) reproduces the lognormal distribution Eq.\ (\ref{lognorm2}) under the identification of $\hat{b}\sim \ln \big(u(T)/w(T)\big)$ and $\hat{m}\sim \sqrt{2\ln w(T)}$.
Via the latter identification, $w(t)$ can be constructed through the information extracted from $\hat{m}$ (Suppl.~Table \ref{tab:reg_beta}).
Also, $u(t)$ can be obtained from the estimated regression parameters presented in Table \ref{tab:lognorm_fitting_p99}.
With the so obtained $w(t)$ and $u(t)$, the resulting probability density function (\ref{GBM-PDF0}) is shown to exhibit an extremely skewed lognormal distribution at each time section, where most of the probability density is concentrated in the region $x\ll 1$.
Therefore, the distribution of $c(T)\approx \int_{0}^{T}X(t)\dd t$ is also approximated by a lognormal distribution, reproducing the empirical Lognormal Law; thus, Property \ref{cond4} is satisfied.
It is noteworthy that, as is clear from the above construction, the proposed stochastic model of citation dynamics is not only applicable to the typical \q{jump--decay} plus \q{constant attention} citation pattern but also various \q{atypical} patterns \cite{Raan04,Ke15,He18}.
Moreover, it would also be applicable to a wide range of studies concerning popularity and collective attention \cite{Simon71,Lorenz-Spreen19}.

\section{Summary and conclusions}

This study investigated the citations of more than 1.5 million eprints on arXiv to model and analyse the quantitative and temporal characteristics of collective attention on scientific knowledge.
The eprints included not only papers published in journals but also their preprint versions and other open electronic materials.
The developed conceptual and technical framework allowed us to explore the citation dynamics beyond the traditional journal-centred scholarly communication, making the previously invisible visible.
By applying a nonlinear regression model to the long-term discipline-average citation data, some interesting facts were revealed regarding how quickly and how many, on average, an eprint in each discipline acquires citations and how quickly it becomes obsolete.
Apart from very highly cited papers, it was uncovered that while the \q{jump--decay} plus \q{constant attention} patterns were consistent across disciplines, the quantitative characteristics such as the average peak height, the average time to reach the peak and the curve's skewness varied among the disciplines.
In particular, CS exhibited the steepest gradient with the highest peak in the growth phase, while Math exhibited the heaviest-tailed curve profile in the obsolescence phase.
The regression results were used to analyse the trends in the obsolescence rate, both in the quantitative and temporal dimensions.
The temporal obsolescence rate, or the internal obsolescence rate, was quantified by $\mathcal{S}=\delta_{1}/\delta_{2}$, where $\delta_{1}$ and $\delta_{2}$ represented the typical time interval of the growth phase and the obsolescence phase, respectively, identified in the average citation history curve.
The quantity-wise obsolescence rate, or the retention rate, was quantified by $\mathcal{R}=\hat{B}/\hat{u}_{\mathrm{p}}$, where $\hat{B}$ and $\hat{u}_{\mathrm{p}}$ represented the asymptotic value and the peak value of the average citation history curve, respectively.
Both obsolescence metrics exhibited varied characteristics across disciplines, reflecting the different publication and citation practices.
Further, the revealed characteristics of the discipline-average citation history curves were used to develop a new citation index, called the $\gamma$-index, which was normalised across disciplines and time periods.
When applied to the arXiv citation data, the distribution of $\gamma$ was fairly close to the standard normal distribution (except its left tail).
This fact suggested that the $\gamma$-index could be used as an improved, less biased alternative to those citation-based impact measures widely used in the current academic or government practice.
Moreover, a stochastic modelling of the citation dynamics is presented, which successfully reproduced both the observed Lognormal Law for the cumulative citation distribution---the quantitative dimension---and the observed time behaviour of the average citation history curve---the temporal dimension---in a unified formalism.

The new conceptual and methodological framework developed in this paper to explore the dynamics of collective attention on science would be of interest to a wide range of research communities, including academic researchers, practitioners and policymakers.
However, as with any bibliometric research, this study also faced various limitations that may have impacted the general validity of the findings.
First, papers not being posted on arXiv, including those directly submitted to journals, were not considered due to the data availability.
Although a growing number of research outputs in the fields of physical, mathematical, and computer sciences have relied on arXiv (Suppl.~Fig.\ \ref{fig:lineplot_comb2}a, b), those citations outside the current arXiv data could have impacted some quantitative aspects of the findings.
Second, the results should be assessed with caution as they are likely to be highly dependent on the research discipline classification scheme \cite{Schubert96,Zitt05,Waltman16}.
Although this study relied on the same classification as that used on the \citek{arXiv_stats2020} website, different specifications based on different subject areas or specialities could have also been applied, leading to quantitatively different implications on the citation dynamics.
The validity and utility of the $\gamma$-index also wholly depends on the discipline classification scheme. 
Third, the citation data would have been incomplete due to the limitation of the external source and its connectivity to the arXiv system.
There have not been consistent tools to measure citations in the sphere of eprints, and arXiv's citation-tracking algorithm may also be imperfect.
If a significant number of papers or related citations have not been correctly recorded in the Semantic Scholar API data, then the quantitative results could have been different from---albeit qualitatively similar to---the findings of this paper.
Fourth and finally, as also emphasised in the discussion on the $\gamma$-index, it should be noted that there are inherent limitations in using citation-based methods to evaluate the research impact.
Citation is not necessarily a measure of the absolute quality or value of scientific works; its primary criterion is utility in research \cite{Garfield79}.
Therefore, our quantitative approaches must not solely be used to evaluate individual research outcomes and must always be combined with expert judgements or trustworthy peer-reviews.

There are many directions in which this work may be extended or applied.
For instance, the quantitative information on the discipline-average citation history curve can be used to adjust the citation window to calculate various popular citation-based metrics, including the journal impact factor \cite{Garfield72,Garfield06} and the $h$-index \cite{Hirsch05}, alleviating the problem of biased assessment.
Such knowledge will also be helpful to better detect \q{atypical} citation patterns for individual papers, including the cases of multiple humps and Sleeping Beauties \cite{Raan04,Ke15,He18}.
It would also be interesting to investigate the determinants of various bibliometric indicators considered in this paper, including the internal obsolescence rate, the retention/inflation rate and the $\gamma$-index ($\gamma^{*}$-index).
Our compiled arXiv dataset \cite{Okamura21} contained the information of the arXiv ID, which can be used to identify various information about each eprint, including title, author names, affiliations and the number of authors, institutions to which the authors belong, researcher ID, journal titles and DOI (if applicable).
The researcher ID can also be used to draw information about the authors' research records and various achievements.
By integrating these information resources into a more comprehensive dataset, we will be able to conduct causal inference with both quantitative and qualitative methods for a deeper understanding of the citation dynamics.
In so doing, it is important to extend the research area beyond the arXiv disciplines, including the areas of not only other natural sciences and biomedical sciences but also humanities and social sciences.
The yet-to-be-seen landscape of interdisciplinary research driven by open eprints would be different from that seen through the lens of journal papers (e.g.\ \citea{}{Okamura19}).
Finally, it would be interesting to investigate further how the networks of citations \cite{Barabasi02,Golosovsky17,Wang13,Pan18}, or more broadly, collective attention \cite{Lorenz-Spreen19} is formed and disseminated through time and social space, based on the framework developed in this paper.
The increasing availability of large-scale data will enable us to explore new frontiers in scientometrics than ever.

We close this paper with a remark on what \q{citation} would represent in the coming new era of Open Science.
As ever more scientific knowledge is posted, published and made available to the public through various open digital materials and platforms, those who \q{cite} the source of knowledge will be diversified across the stakeholders of science and society.
Simultaneously, the reasons and ways for \q{citation} will also be diversified.
The scientific landscape and the scenery, which researchers have long seen over {the shoulder of Giants}, will necessarily change as Open Science provides a new way to see the world from high above {the Data Clouds}.
It will enable people to access and experience collaborative knowledge production and consumption processes through open, transparent and shared platforms.
Accordingly, what the term \q{citation}---or scientific attention, attitudes and behaviour---has meant and means today in scientific communities will not be the same tomorrow.
A deeper understanding of the \emph{who}, \emph{what}, \emph{why} and \emph{how} of \q{citation} in a much broader context will be required to build a balanced and sustainable relationship between science and the future society.


\vspace{0.0cm}
\paragraph{\textbf{Acknowledgement.}}
The author would like to thank Ryo Suzuki and two anonymous reviewers for their valuable comments on the manuscript.
The views and conclusions contained herein are those of the author and should not be interpreted as necessarily representing the official policies or endorsements, either expressed or implied, of any of the organisations to which the author is affiliated.

\paragraph{\textbf{Author Contributions.}}
Keisuke Okamura: Conceptualisation, Methodology, Validation, Formal analysis, Investigation, Writing, Visualisation, Project administration.

\vspace{-0.3cm}
\paragraph{\textbf{Competing Interests.}}
The author has no competing interests.

\vspace{-0.3cm}
\paragraph{\textbf{Funding Information.}}
The author did not receive any funding for this research.

\vspace{-0.3cm}
\paragraph{\textbf{Data Availability.}}
The datasets generated and/or analysed during this study can be found in the Zenodo repository at \url{https://doi.org/10.5281/zenodo.5803962}.

\vspace{-0.2cm}
\afterpage{\clearpage%
\vspace{-0.8cm}
\theendnotes
\addcontentsline{toc}{section}{Notes}
}

\afterpage{\clearpage%
\bibliographystyle{apalike}
\addcontentsline{toc}{section}{References}

\renewcommand{\headrule}{\color{VioletRed}\oldheadrule}

\renewcommand{\baselinestretch}{1.2}

}

\afterpage{\clearpage%
\quad\\[-1cm]
\begin{center}
\textcolor{gray}{\large\textsf{[THIS PAGE INTENTIONALLY LEFT BLANK]}}
\end{center}
\pagestyle{fancy}
\fancyhead[LE,RO,RE,LO]{}
\fancyfoot[RE,LO]{\color[rgb]{0.04, 0.73, 0.71}{\scriptsize\textsf{arXiv:2106.05027v2~\,|~\,\url{https://doi.org/10.1162/qss_a_00174}}}}
\fancyfoot[LE,RO]{\scriptsize{\textbf{\textsf{\thepage}}}}
\fancyfoot[C]{}
}



\renewcommand{\thesubsection}{Appendix \Alph{subsection}}

\afterpage{\clearpage%

\pagestyle{fancy}
\fancyhead[LE,RO]{\textcolor{orange}{\footnotesize{\textsf{SUPPLEMENTARY MATERIALS}}}}
\fancyhead[RE,LO]{}
\fancyfoot[RE,LO]{\color[rgb]{0.04, 0.73, 0.71}{\scriptsize\textsf{arXiv:2106.05027v2~\,|~\,\url{https://doi.org/10.1162/qss_a_00174}}}}
\fancyfoot[LE,RO]{\scriptsize{\textbf{\textsf{\thepage}}}}
\fancyfoot[C]{}

\addcontentsline{toc}{section}{Supplementary Materials}

\quad
\vspace{-0.5cm}
\begin{center}
\fontsize{15pt}{16pt}\selectfont\bfseries
Supplementary Materials
\end{center}

\vspace{-0.5cm}
\begin{center}
{for \textit{\q{Scientometric engineering: Exploring citation dynamics via arXiv eprints}} by K.\ Okamura (2022).}
\end{center}
}
\vspace{3.0cm}

\renewcommand{\figurename}{Suppl.~Figure}
\renewcommand{\tablename}{Suppl.~Table}
\renewcommand{\thefigure}{S\arabic{figure}}
\renewcommand{\thetable}{S\arabic{table}}
\renewcommand{\theequation}{S\arabic{equation}}
\setcounter{section}{0}
\setcounter{figure}{0}
\setcounter{table}{0}
\setcounter{equation}{0}




\subsection[{\hspace{3eM}} Data analysis and visualisation]{Data analysis and visualisation\label{app:data}}

All data were analysed with STATA/IC software (version 13; StataCorp LP, Texas, USA) throughout the paper.
Frequency tables, cross-tabulations, histograms, scatter plots, quantile plots, kernel density estimations and multiple regression analyses were conducted.
The chord diagrams of Fig.\ \ref{fig:overlap} were produced by using R software (version 3.3.3; R Core Team).
The datasets generated and/or analysed during this study can be found in the Zenodo repository \cite{Okamura21}.

\renewcommand{\headrule}{\color{orange}\oldheadrule}

\subsection[{\hspace{3eM}} Theoretical validation of the lognormal time distribution of citations]{Theoretical validation of the lognormal time distribution of citations\label{app:lognormal}}

This appendix section provides a brief theoretical validation of the lognormal time distribution of citations given in Eq.\ (\ref{lognorm.distribution}) in the main text, used to model the \q{jump--decay} temporal pattern of the citation dynamics.
The key observation for the validation is that the lognormal distribution generally arises when the value of a variable is determined by a multiplication of many random factors.

Consider a typical paper in a given field with nonzero citations.
For simplicity, assume that all those who cite the paper (hereinafter \q{citers}) have a similar tendency for citation pattern, including the timing of citation.
Let $t_{0}$ be the inherently \q{right} timing of citing the paper, shared among the potential citers in the field.
The slight differences in the behaviours of the citers produce a certain distribution of the actual timing of citation, denoted here as $t_{n}$ with $n$ the total number of the citing events.
We assume that $t_{n}$ is collectively shaped in the research community by a multiplicative process of independent random factors associated with the individual citation events.
The randomness may arise for various reasons, including the citers' cognitive capacity, resource limitations and time constraints.
Subsequently, the actual timing of citation is given by
\begin{equation}\label{t_n}
t_{n}=\prod_{j=1}^{n}\ko{1+\ep_{j}}t_{0}\,,
\end{equation}
where $\ep_{j}$ represents a small perturbation of the citation timing introduced by the citing event labelled by $j$.
By taking the logarithm of the both sides of Eq.\ (\ref{t_n}) and using the approximation $\ln (1+\ep)\approx \ep$ for small $\ep$, we have
\begin{equation}\label{ln_t_n}
\ln t_{n}\approx \sum_{j=1}^{n}\ep_{j}+\ln t_{0}\,.
\end{equation}
If the number of the citing events, $n$, is sufficiently large, this relation implies that the probability distribution of $\ln t_{n}$ is closely approximated by a normal distribution regardless of the distribution of $\ep$ by virtue of the central limit theorem.
Equivalently, the probability distribution of $t_{n}$ is closely approximated by a lognormal distribution with $n\gg 1$.
This line of reasoning provides a possible explanation as to why the \q{jump--decay} component of the average citation history curve is well-fitted by the lognormal probability distribution function, as shown in the main text.

\subsection[{\hspace{3eM}} Derivation of the stochastic model of citation dynamics]{Derivation of the stochastic model of citation dynamics\label{app:stochastic}}

This appendix section provides the derivation of Eqs.\ (\ref{gGBM0}--\ref{GBM-PDF0}) in the main text.
We assume that the continuous version of Gibrat's law \cite{Saichev10} (also known as the law of proportionate growth) is in operation, as in the case with the Black--Scholes option pricing \cite{Black-Scholes73,Merton73}.
Namely, the growth rate of attention is independent of attention.
This implies a stochastic differential equation (SDE) of the form:
\begin{equation}\label{gGBM}
dX(t)=X(t)\big[\alpha(t)\dd t+\beta(t) \dd W(t)\big]\,,\quad 
X(0)\eqqcolon x_{0}>0\,,
\end{equation}
where $\alpha(t)\in\mathbb{R}$ and $\beta(t)\in \mathbb{R}^{+}$ are the \q{drift} and the \q{volatility} parameters, respectively, and $\{ W(t)\,|\, t\in \mathbb{R}^{+}_{0} \}$ is the standard Brownian motion (or the Wiener process).
In solving this equation for $X(t)$, it is convenient to introduce the new stochastic process defined by $Y(t)\coloneqq \ln X(t)$.
By applying the It\^{o}'s lemma to $Y(t)$, we obtain the SDE for $Y(t)$ as
\begin{equation}\label{gGBM2}
dY(t)=D(t)\dd t+\beta(t) \dd W(t)\,,\quad 
D(t)\coloneqq\alpha(t)-\f{\beta(t)^{2}}{2}\,.
\end{equation}
Taking the exponential of this equation, we obtain
\begin{equation}\label{M(t)}
X(t)=x_{0}e^{Y(t)}=x_{0}\exp\kko{\int_{0}^{t}D(\tau)\dd\tau+\int_{0}^{t}\beta(\tau) \dd W(\tau)}\,.
\end{equation}
The probability density function for our generalised geometric Brownian motion, denoted as $p(x,t)$, can be obtained by solving the corresponding Fokker--Planck equation.
To make life easier, instead of solving the Fokker--Planck equation for $p(x,t)$, we consider the Fokker--Planck equation for $q(y,t)$ with $y=\ln x$, which is given by 
\begin{equation}\label{FP_eq2}
\f{\partial}{\partial t}q(y,t)=-D(t)\f{\partial}{\partial y} q(y,t)+\f{\beta(t)^{2}}{2}\f{\partial^{2}}{\partial y^{2}} q(y,t)\,,\quad 
q(y,0)=\delta(y-y_{0})\,.
\end{equation}
We first look for solutions in Fourier space, in which Eq.\ (\ref{FP_eq2}) becomes
\begin{equation}
\f{\partial}{\partial t}\tilde q(k,t)=\ko{ikD(t)-\f{k^{2}}{2}\beta(t)^{2}}\tilde q(k,t)\,,\quad 
\tilde q(k,t)\coloneqq \f{1}{2\pi}\int_{-\infty}^{\infty}e^{iky}q(y,t)\dd y\,.
\end{equation}
This equation can be straightforwardly solved as
\begin{equation}
\tilde q(k,t)=\exp\kko{ik\ko{y_{0}+\int_{0}^{t}D(\tau)\dd\tau}-\f{k^{2}}{2}\int_{0}^{t}\beta(\tau)^{2}\dd\tau}\,.
\end{equation}
By Fourier-transforming the above equation back into the space domain, we obtain
\begin{equation}\label{GBM-PDF2}
q(y,t)=\f{1}{2\pi}\int_{-\infty}^{\infty}e^{-iky}\tilde q(k,t)\dd k
=\f{1}{\sqrt{2\pi\int_{0}^{t}\beta(\tau)^{2}\dd\tau}}\exp\kko{-\f{\ko{y-y_{0}-\int_{0}^{t}D(\tau)\dd\tau}^{2}}{2\int_{0}^{t}\beta(\tau)^{2}\dd\tau}}\,,
\end{equation}
which follows a normal distribution with the expectation value and the variance given by, respectively,
\begin{equation}\label{E[y]}
\mathrm{E}[Y(t)]=y_{0}+\int_{0}^{t}D(\tau)\dd\tau\quad 
\text{and}\quad
\mathrm{Var}[Y(t)]=\int_{0}^{t}\beta(\tau)^{2}\dd\tau\,.
\end{equation}
Finally, the probability density function for $X(t)$ is given by, using that $\int_{-\infty}^{\infty}q(y,t)\dd y=\int_{0}^{\infty}p(x,t)\dd x=1$ and $dy=dx/x$,
\begin{equation}\label{GBM-PDF}
p(x,t)=\f{1}{x\sqrt{2\pi\int_{0}^{t}\beta(\tau)^{2}\dd\tau}}\exp\kko{-\f{\big(\ln (x/x_{0})-\int_{0}^{t}D(\tau)\dd\tau\big)^{2}}{2\int_{0}^{t}\beta(\tau)^{2}\dd\tau}}
\end{equation}
with the expectation value and the variance given by, respectively,
\begin{equation}\label{E[x]}
\mathrm{E}[X(t)]=x_{0}\exp\ko{\int_{0}^{t}\alpha(\tau)\dd\tau}\quad 
\text{and}\quad 
\mathrm{Var}[X(t)]=\kko{x_{0}\exp\ko{\int_{0}^{t}\alpha(\tau)\dd\tau}}^{2}\kko{\exp\ko{\int_{0}^{t}\beta(\tau)^{2}\dd\tau}-1}.
\end{equation}
One can check that this solution in fact solves the SDE (\ref{gGBM}). 
Its cumulative density function is given by 
\begin{equation}
\int_{0}^{x}p(x',t)\dd x'=\Phi\ko{\f{\ln (x/x_{0})-\int_{0}^{t}D(\tau)\dd\tau}{\sqrt{\int_{0}^{t}\beta(\tau)^{2}\dd\tau}}}
\end{equation}
with $\Phi(\cdot)$ the standard normal distribution function.

\subsubsection*{Application to the arXiv citation data}
As we discussed in the main text, the probability density function (\ref{GBM-PDF}) reproduces the lognormal distribution of Eq.\ (\ref{lognorm2}) under the identification of $\hat{b}\sim \ln \big(u(t)/w(t)\big)$ and $\hat{m}\sim \sqrt{2\ln w(t)}$ with $u(t)=\mathrm{E}[X(t)]$ and $w(t)=\exp\big(\f{1}{2}\int_{0}^{t}\beta(\tau)^{2}\dd\tau\big)$.
Note that $\hat{b}$ and $\hat{m}$ can be obtained from the regression analysis, using the equation of the form $y_{k}=b+m\Phi^{-1}(q_{k})+\ep_{k}$ with $\ep_{k}$ the error term.
Suppl.~Fig.\ \ref{fig:qplot_slope_lnc_tot_all} shows the time trends in the estimate $\hat{m}$, indicating that $\hat{m}$ has grown with the age of the dataset.
By fitting a regression function of the form $\hat{m}_{i}=\sqrt{s_{2}\ln(t_{i}/s_{1}+1)}+\ep_{i}$ with $s_{1}$ and $s_{2}$ regression coefficients and $\ep_{i}$ the error term, the estimates $\hat{s}_{1}$ and $\hat{s}_{2}$ are obtained.
Suppl.~Table \ref{tab:reg_beta} summarises the regression results, with the adjusted $R$-squared nearly one for all disciplines.
This regression equation, together with the above relation between $\hat{m}$ and $w$, leads to the following form for the \q{volatility} term,
\begin{equation}\label{eq:reg_beta}
\beta_{\star}(t;\hat{s}_{1},\hat{s}_{2})=\sqrt{\f{\hat{s}_{2}}{t+\hat{s}_{1}}}\,,
\end{equation}
which behaves as $\beta_{\star}\sim \sqrt{\hat{s}_{2}/\hat{s}_{1}}\big(1-t/(2\hat{s}_{1})\big)$ for $t\ll 1$ and $\beta_{\star}\sim \sqrt{\hat{s}_{2}/t}$ for $t \gg 1$.
In addition, the \q{jump--decay} pattern of citation evolution observed for the arXiv citation data can be reproduced by identifying $\alpha(t)$ with
\begin{equation}\label{LN-alpha}
\alpha_{\star}(t;{{\Omega}})\coloneqq 
\f{d}{dt}\ln u_{\star}(t;\Omega)\,,\quad 
u_{\star}(t;\Omega)=A f(t+1;\mu,\sigma)+B g(t;\lambda)\,,
\end{equation}
where $f$ and $g$ given by (\ref{lognorm.distribution}) and (\ref{step-like}), respectively, and $\Omega=\{{{\omega}}_{1},{{\omega}}_{2}\}=\{\{A,\mu,\sig\},\{B,\lam\}\}$ is the collective shorthand notation for the regression coefficients introduced before.
By this construction, the time behaviour of $X(t)$ precisely agrees with the movement we saw for the arXiv citation data. 
This in turn identifies the initial value of $X(t)$ to be ${x_{0}}=u_{\star}(0)=A\big(\sqrt{2\pi}\sigma\big)^{-1}e^{-{\mu}^{2}/(2{\sigma}^{2})}$.
Thus, the average citation history curve, $u(t)$, is governed solely by the deterministic \q{drift} component of the stochastic motion, $\alpha(t)$.

In summary, the SDE associated with the arXiv data is given by
\begin{equation}
dX(t)=X(t)\big[\alpha_{\star}(t;\hat{{\Omega}})\dd t+\beta_{\star}(t;\hat{s}_{1},\hat{s}_{2}) \dd W(t)\big]\,,
\end{equation}
where the hatted variables refer to the estimated quantities as given in Table \ref{tab:lognorm_fitting_p99} ($\hat{\Omega}$) and Suppl.~Table \ref{tab:reg_beta} ($\hat{s}_{1},\,\hat{s}_{2}$) for the below-99th citation data.
More generally, the SDE associated with the lognormal distribution in the quantitative dimension and arbitrary time behaviour $u_{\star}(t)$ in the temporal dimension can be modelled as $dX(t)=X(t)\big[d\ln u_{\star}(t)+{\beta}_{\star}(t) \dd W(t)\big]$.
Here, the time-dependent \q{drift} parameter is given by $\alpha_{\star}(t)=d\ln u_{\star}(t)/dt$.
This form is the same as Eq.\ (\ref{gGBM0}) in the main text.
Eqs.\ (\ref{solution}, \ref{GBM-PDF0}) also follow from Eqs.\ (\ref{gGBM2}, \ref{M(t)}, \ref{GBM-PDF}, \ref{E[x]}).


\addcontentsline{toc}{subsection}{Supplementary Figures}

\afterpage{\clearpage%
\begin{figure}[htp]
\centering
\vspace{-0.5cm}
    \begin{tabular}{c}
\begin{minipage}{0.5\hsize}
\begin{flushleft}
\raisebox{-0.5cm}{\textsf{\textbf{a}}\quad }
\raisebox{-\height}{\includegraphics[align=c, scale=0.6, vmargin=0mm]{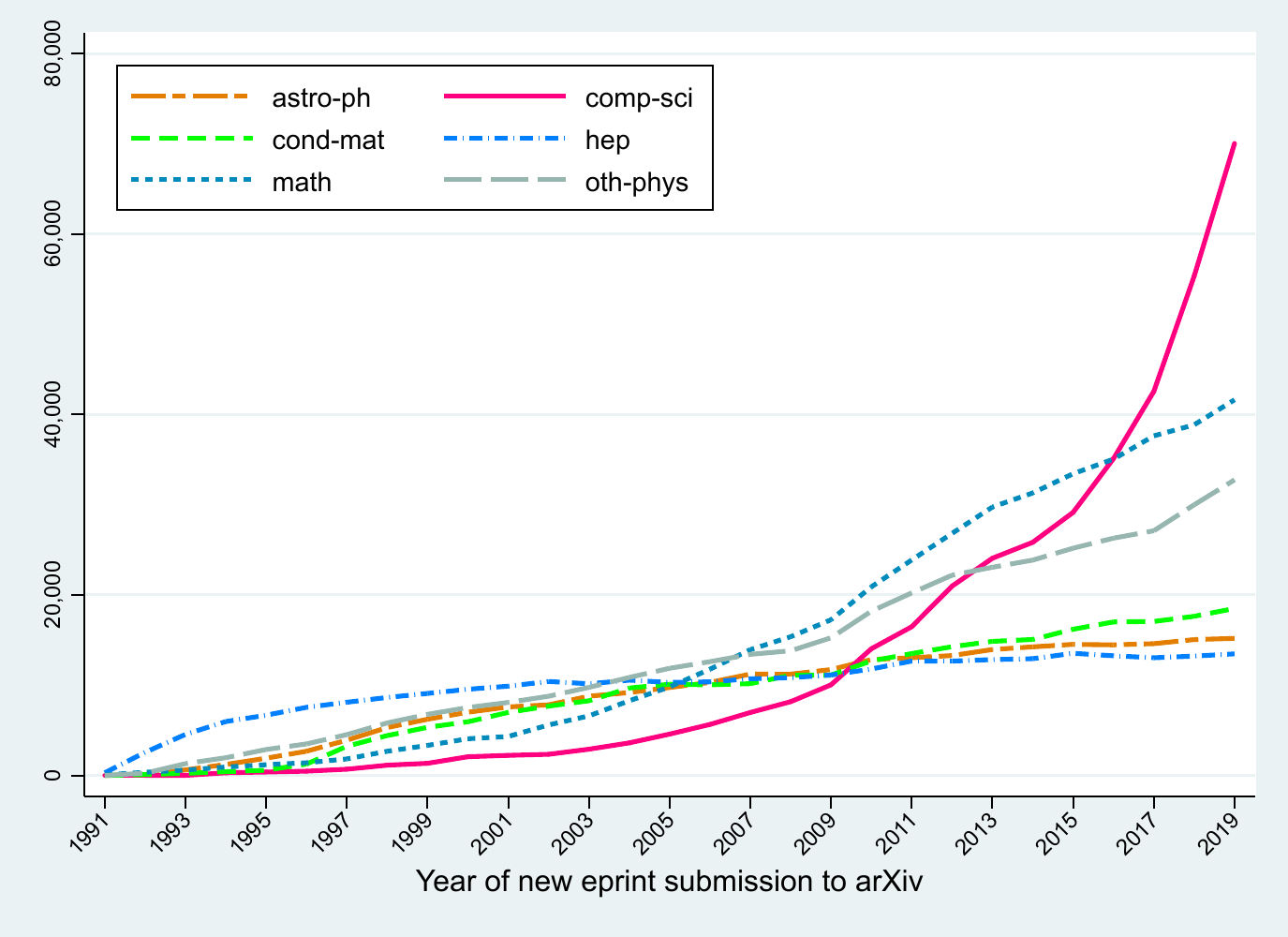}}
\end{flushleft}
      \end{minipage}
\begin{minipage}{0.5\hsize}
\begin{flushleft}
\raisebox{-0.5cm}{\textsf{\textbf{b}}\quad }
\raisebox{-\height}{\includegraphics[align=c, scale=0.6, vmargin=0mm]{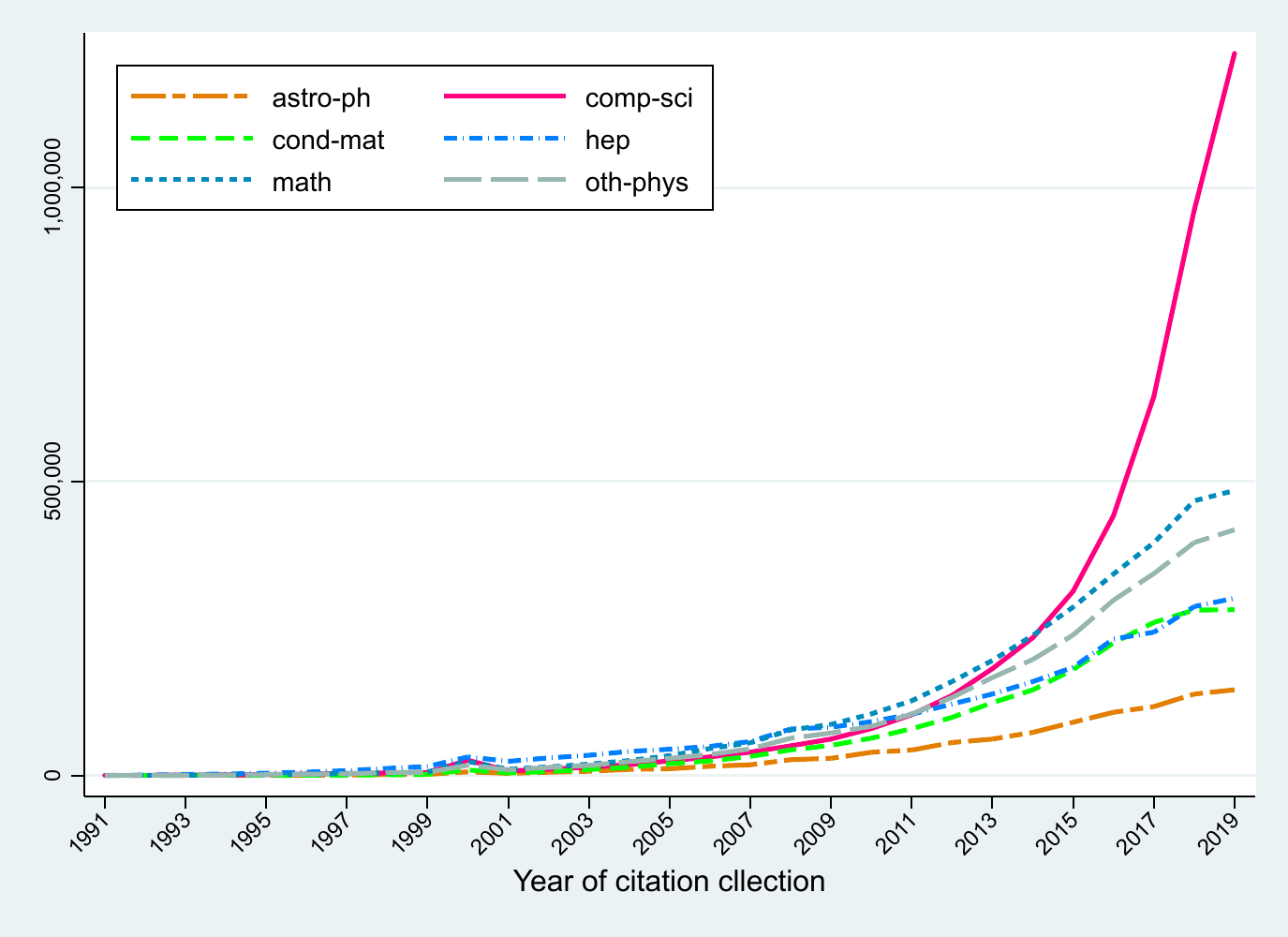}}
\end{flushleft}
          \end{minipage}
    \end{tabular}
\caption{\textbf{The rise of arXiv eprints by discipline.}
The graphs are excerpts, translated from \citek{Okamura20b}.
\textbf{a}, trends in the number of new eprint submissions to arXiv; \textbf{b}, trends in the number of newly collected citations by the arXiv eprints.
Those eprints tagged to multiple arXiv disciplines are counted independently for each discipline.
Both graphs manifest the progressive rise of the arXiv eprints for the past decades.
In particular, the number of CS eprints increased from 14,041 (2010) to 70,029 (2019), increasing approximately five-fold during the past decade, which is in line with the rise of modern deep learning.}
\label{fig:lineplot_comb2}
\end{figure}
\quad\\
\vfill
}

\afterpage{\clearpage%
\begin{figure}[tp] 
\centering
\vspace{-0.5cm}
\noindent
\includegraphics[align=c, scale=0.7, vmargin=1mm]{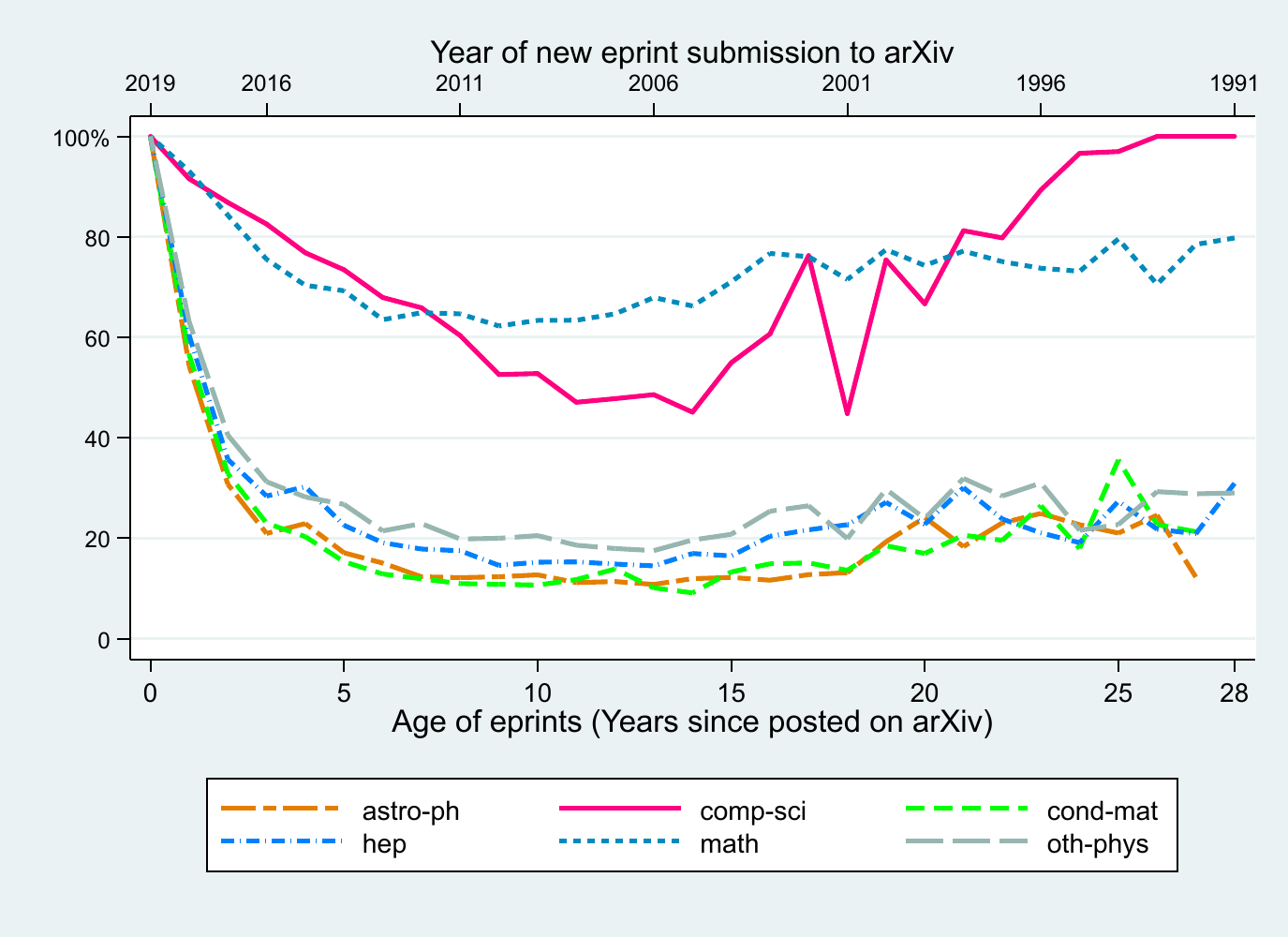}
\caption{\textbf{Percentage of citations made to arXiv eprints before a DOI is assigned.}
The graph is an excerpt, translated from \citek{Okamura20b}.
The number of eprints for each discipline is provided in Table \ref{tab:arXiv_disciplines} in the main text.
Those eprints tagged to multiple arXiv disciplines are counted independently for each discipline.
On average, approximately 15--30\% of the citations to an eprint posted on the physics-based disciplines (submission year: 1991--2015) are made before it is assigned a DOI.}
\label{fig:lineplot_pct_pre}
\vspace{5mm}
\end{figure}
\quad\\
\vfill
}

\afterpage{\clearpage%
\begin{figure}[htp]
\centering
\vspace{-0.5cm}
    \begin{tabular}{c}
\begin{minipage}{0.5\hsize}
\begin{flushleft}
\raisebox{-0.5cm}{\textsf{\textbf{a}\quad {\small 2000-2009}}\quad }
\raisebox{-\height}{\includegraphics[align=c, scale=0.57, vmargin=-5mm]{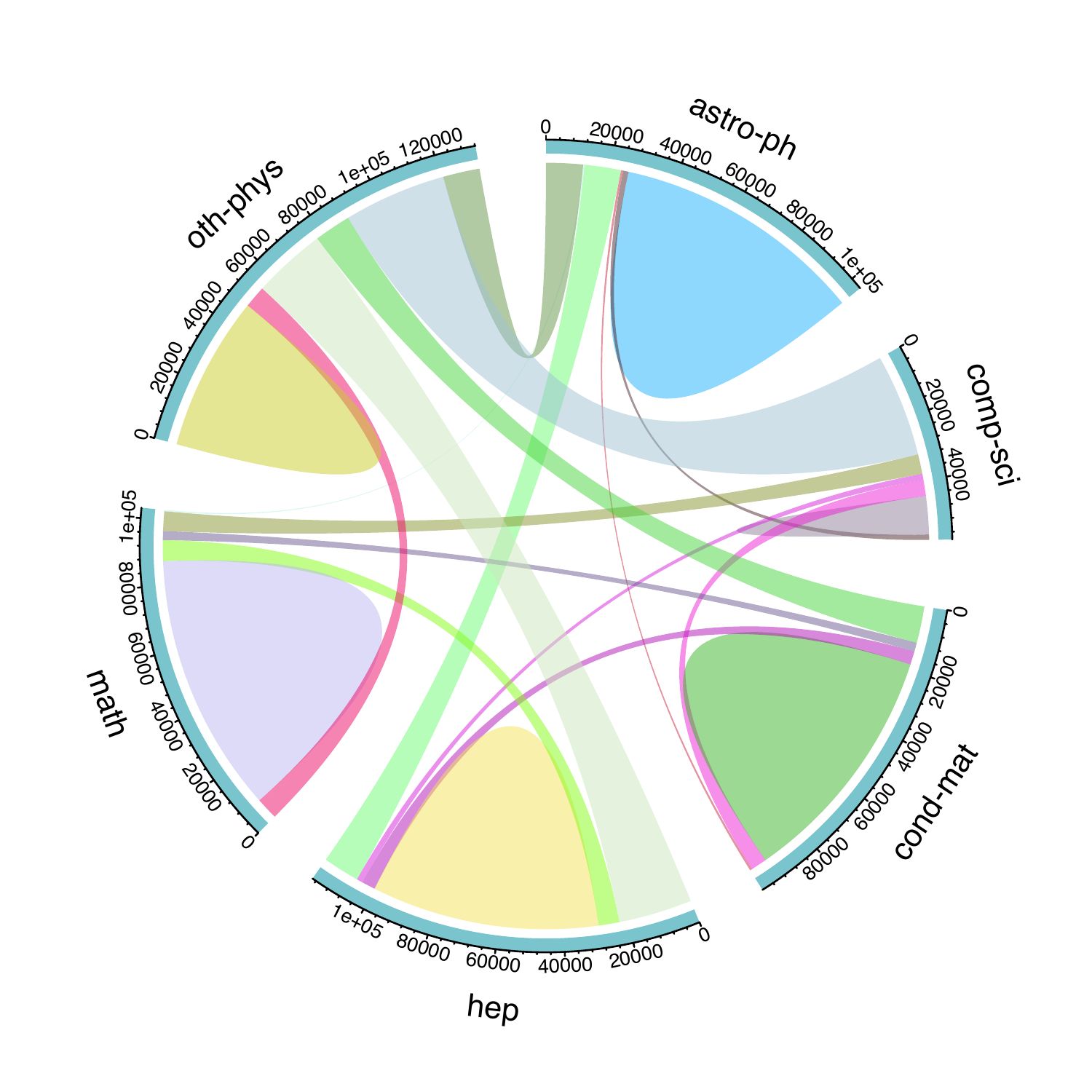}}
\end{flushleft}
      \end{minipage}
\begin{minipage}{0.5\hsize}
\begin{flushleft}
\raisebox{-0.5cm}{\textsf{\textbf{b}\quad {\small 2010-2019}}\quad }
\raisebox{-\height}{\includegraphics[align=c, scale=0.57, vmargin=-5mm]{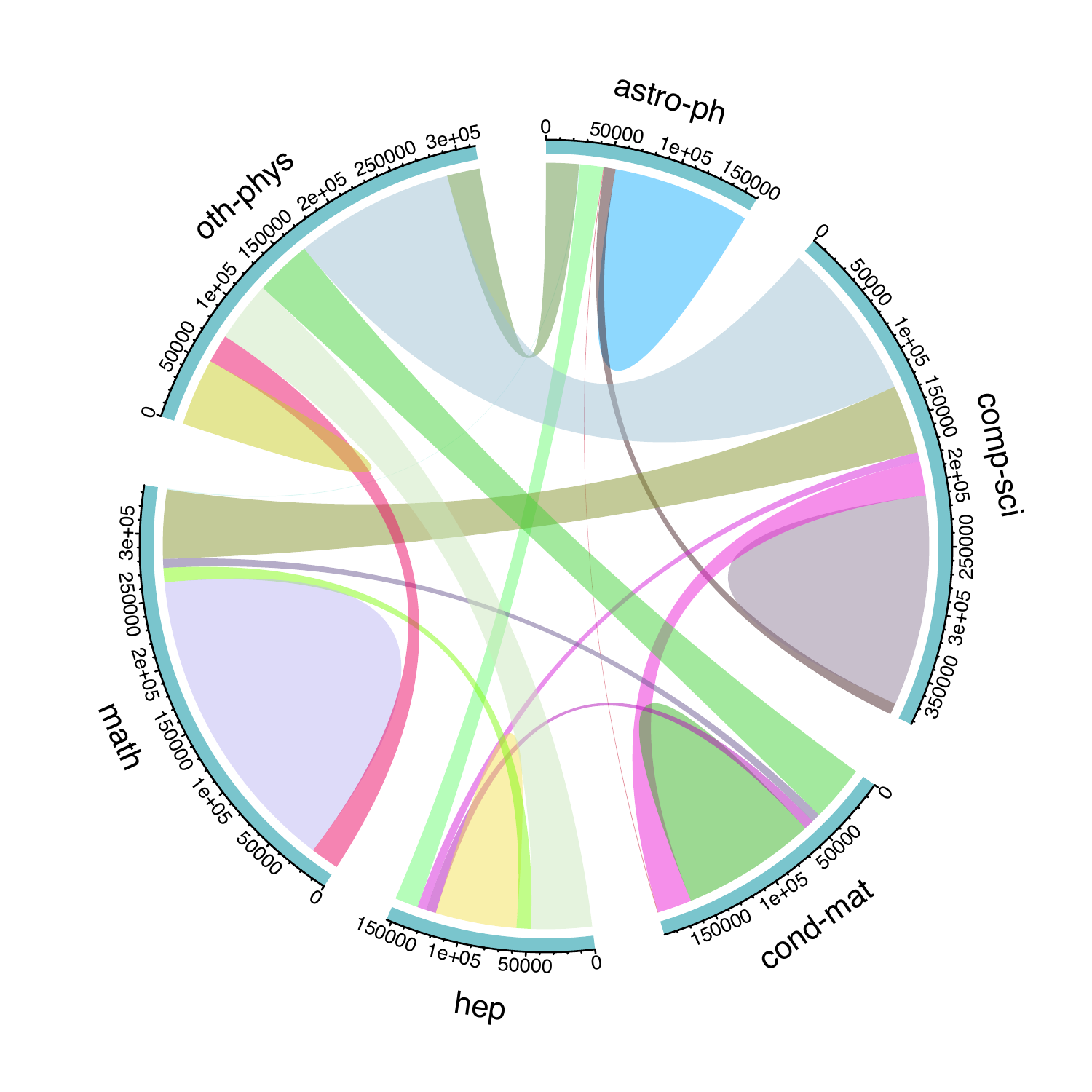}}
\end{flushleft}
          \end{minipage}
    \end{tabular}
\caption{\textbf{Chord diagram representation of the arXiv eprints across disciplines.}
The number of eprints posted on arXiv during the period 2000--2009 (\textbf{a}) and 2010--2019 (\textbf{b}) are represented by the size of each arc.
The arcs (links) connecting two disciplines correspond to the eprints belonging to the two disciplines, while the self-links within each discipline represent the eprints that are not cross-listed to other disciplines.
Those eprints tagged to multiple arXiv disciplines are counted independently for each discipline.
A remarkable change between the two periods can be seen in the distribution of the arXiv eprints and the overlapping feature across disciplines.}
\label{fig:overlap}
\end{figure}
\quad\\
\vfill
}

\afterpage{\clearpage%
\begin{figure}[tp] 
\centering
\vspace{-0.5cm}
\noindent
\includegraphics[align=c, scale=0.9, vmargin=1mm]{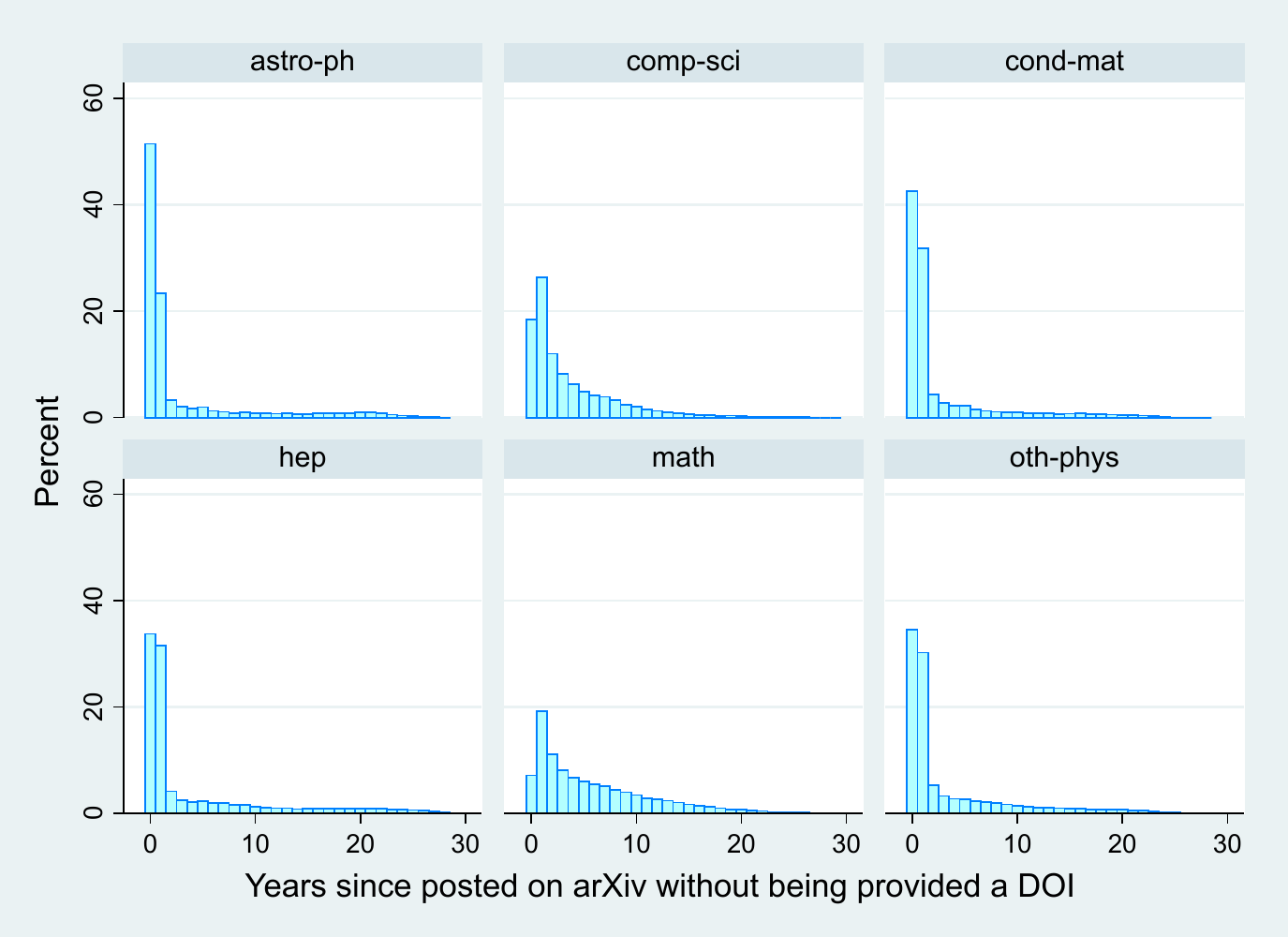}
\caption{\textbf{Elapsed time since posted on arXiv without being assigned a DOI.}
The graph is an excerpt, translated from \citek{Okamura20b}.
The number of eprints for each discipline is provided in Table \ref{tab:arXiv_disciplines} in the main text.
Those eprints tagged to multiple arXiv disciplines are counted independently for each discipline.
A total of 2,045,705 eprints are included, of which 1,102,014 (53.9\% of all articles) eprints are found to be assigned a DOI.}
\label{fig:year_lag_all}
\vspace{5mm}
\end{figure}
\quad\\
\vfill
}

\afterpage{\clearpage%
\begin{figure}[tp] 
\centering
\vspace{-0.5cm}
\noindent
\includegraphics[align=c, scale=0.39, vmargin=1mm]{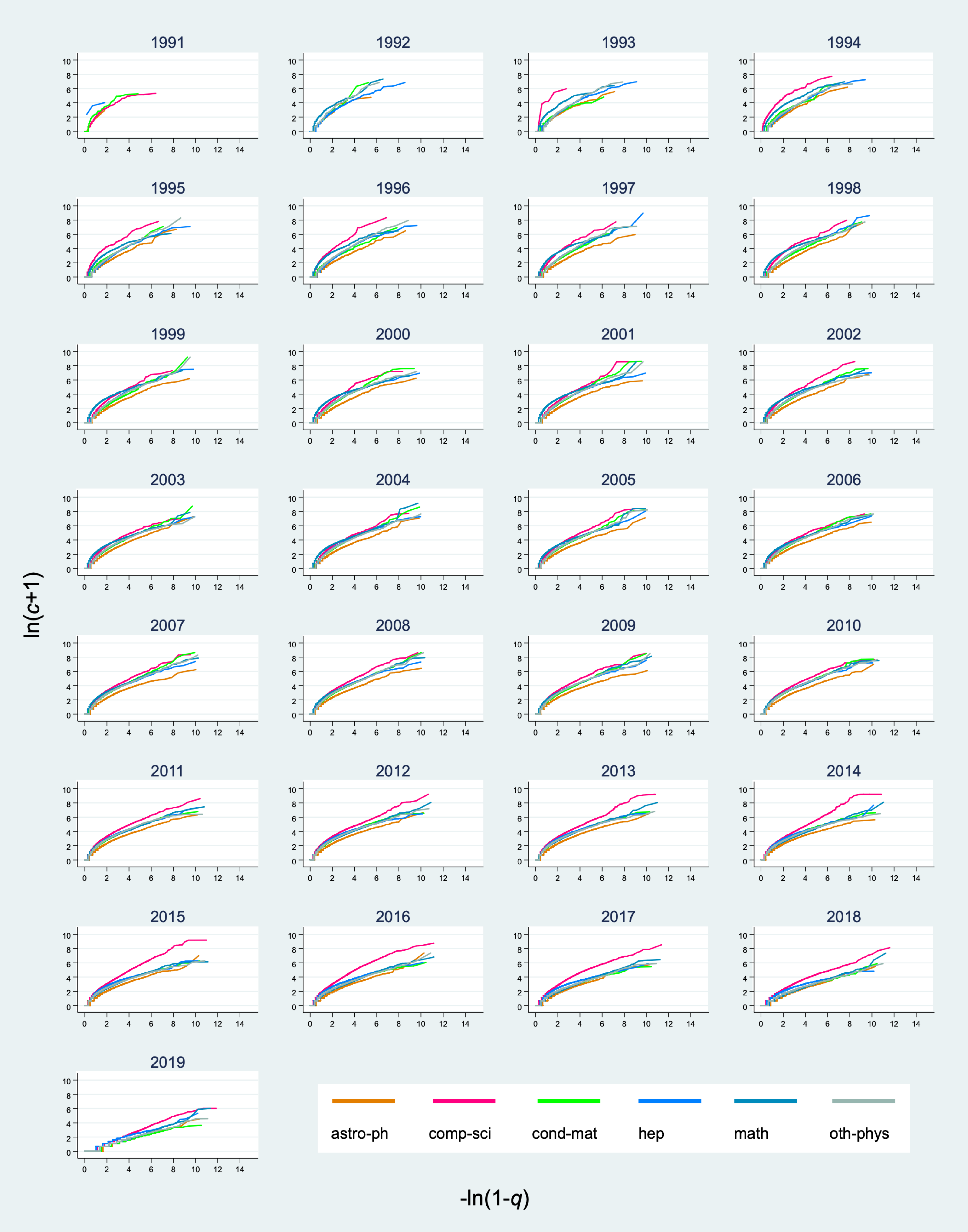}
\caption{\textbf{Power law quantile plot of the arXiv citation data.}
Shown are the plots of $\ln(c_{k}+1)$ against $-\ln(1-q_{k})$, where $c_{k}$ is the cumulative citation of eprint $k$, and $q_{k}$ is the quantile rank of $c_{k}$ in the citation distribution of each discipline.
The plots are depicted by discipline, with each subset of the arXiv eprints classified according to the first submission year (1991--2019).
}
\label{fig:qplot_comb_years_1}
\vspace{5mm}
\end{figure}
}

\afterpage{\clearpage%
\begin{figure}[tp] 
\centering
\vspace{-0.5cm}
\noindent
\includegraphics[align=c, scale=0.39, vmargin=1mm]{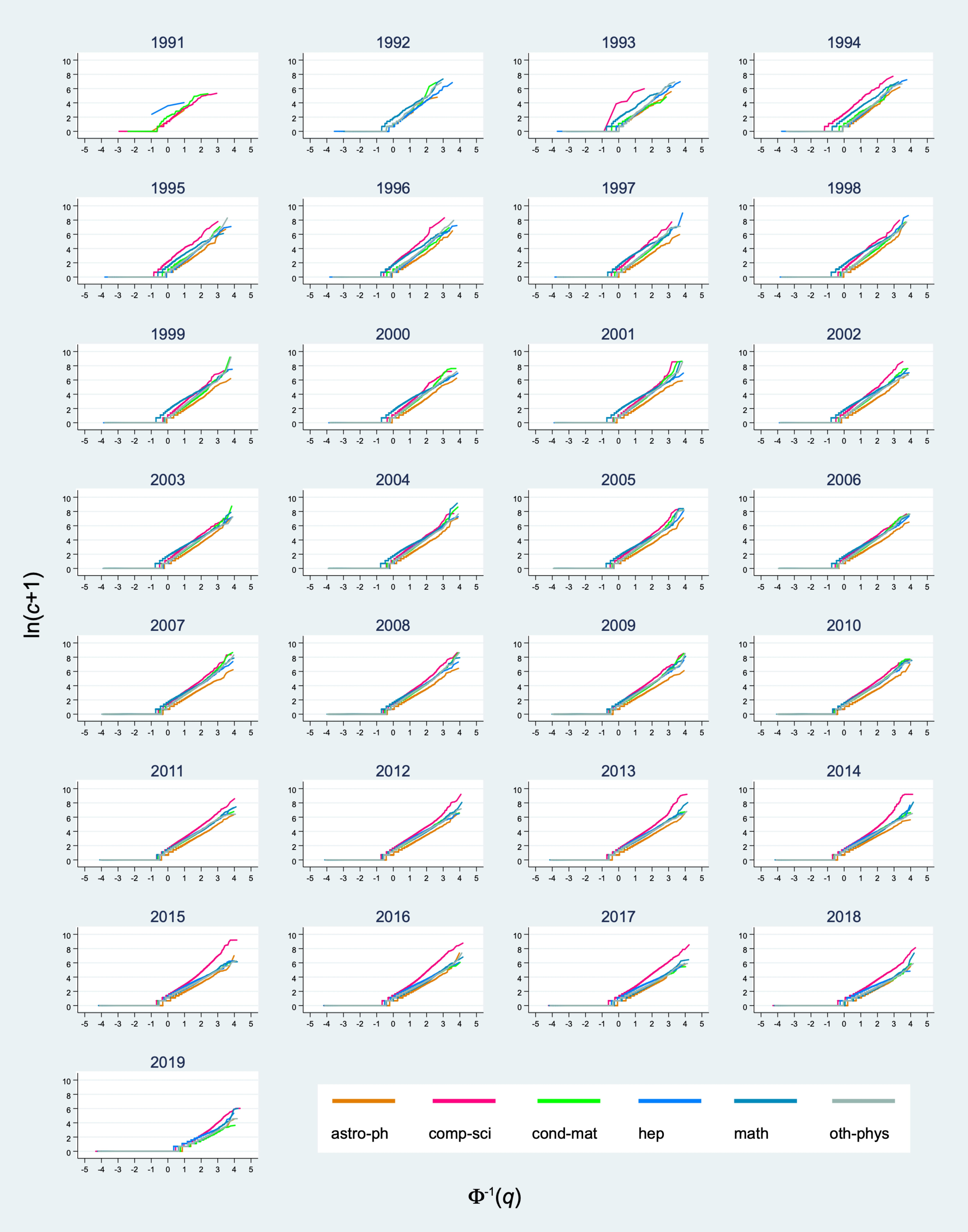}
\caption{\textbf{Lognormal quantile plot of the arXiv citation data.}
Shown are the plots of $\ln(c_{k}+1)$ against $\Phi^{-1}(q_{k})$, where $c_{k}$ is the cumulative citation of eprint $k$, $q_{k}$ is the quantile rank of $c_{k}$ in the citation distribution of each discipline, and $\Phi(\cdot)$ represents the standard normal cumulative density function.
The plots are depicted by discipline, with each subset of the arXiv eprints classified according to the first submission year (1991--2019).
}
\label{fig:qplot_comb_years_2}
\end{figure}
}

\afterpage{\clearpage%
\begin{figure}[tp] 
\centering
\vspace{-1.8cm}
\noindent
\includegraphics[align=c, scale=0.20, vmargin=1mm]{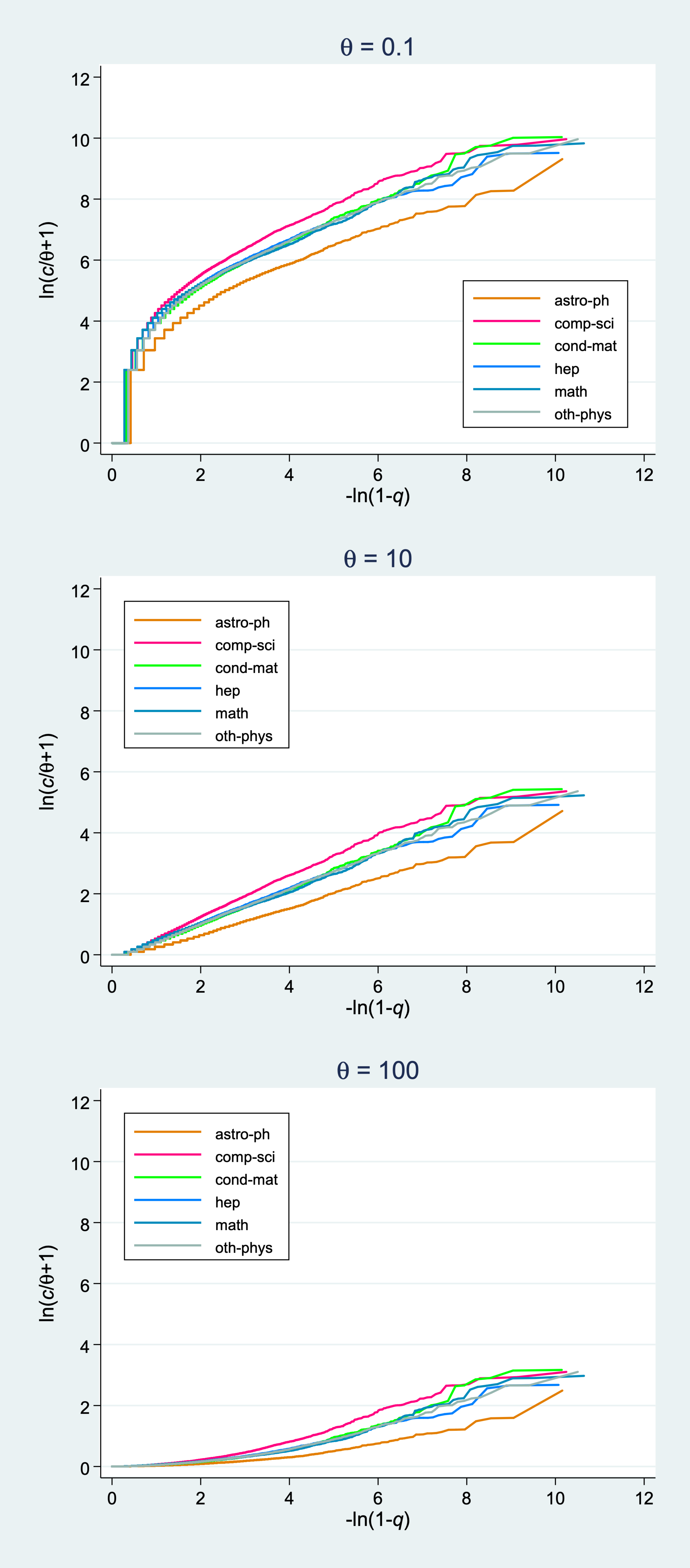}
\caption{\textbf{\q{Shifted power law} quantile plot of the citation data.}
Shown are the plots of $\ln(c_{k}/\theta+1)$ against $-\ln(1-q_{k})$, where $\theta>0$ is the shift parameter, $c_{k}$ is the cumulative citation of eprint $k$, $q_{k}$ is the quantile rank of $c_{k}$ in the citation distribution of each discipline, based on the citation data of eprints posted on arXiv in 2010.}
\label{fig:shiftedpl}
\vspace{5mm}
\end{figure}
}

\afterpage{\clearpage%
\begin{figure}[tp]
 \centering
 \vspace{-0.88cm}
    \begin{tabular}{c}
\raisebox{-0.5cm}{\textsf{\textbf{a}}\quad }
\raisebox{-\height}{\includegraphics[align=c, scale=1.13, vmargin=0mm]{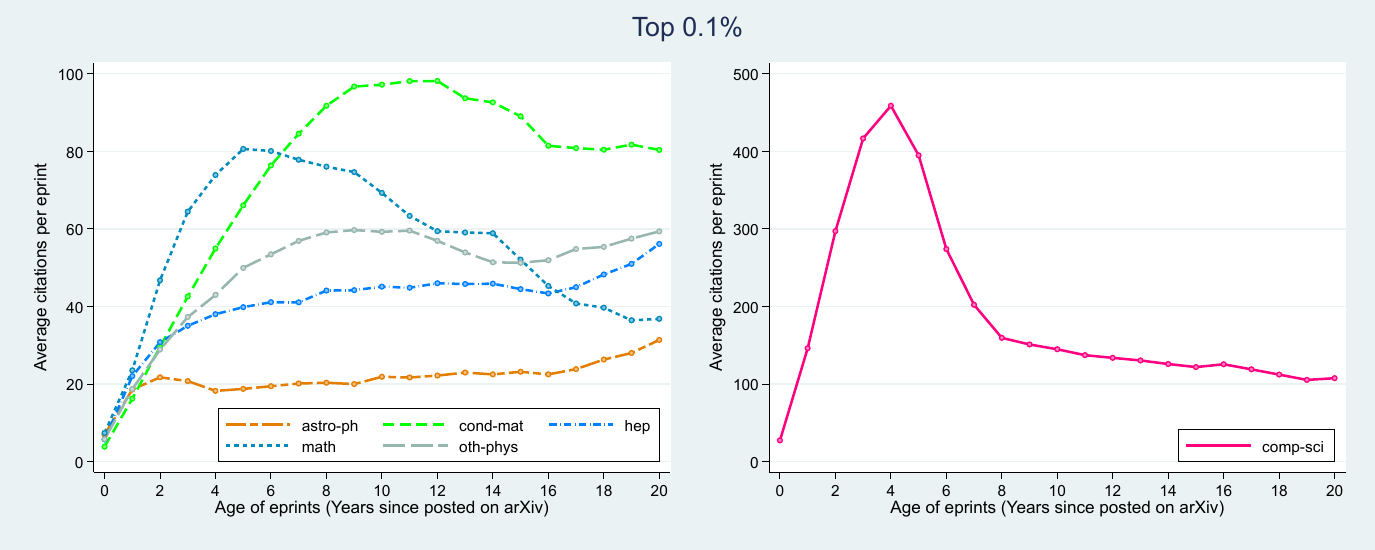}}\\[-2mm]
\raisebox{-0.5cm}{\textsf{\textbf{b}}\quad }
\raisebox{-\height}{\includegraphics[align=c, scale=1.13, vmargin=0mm]{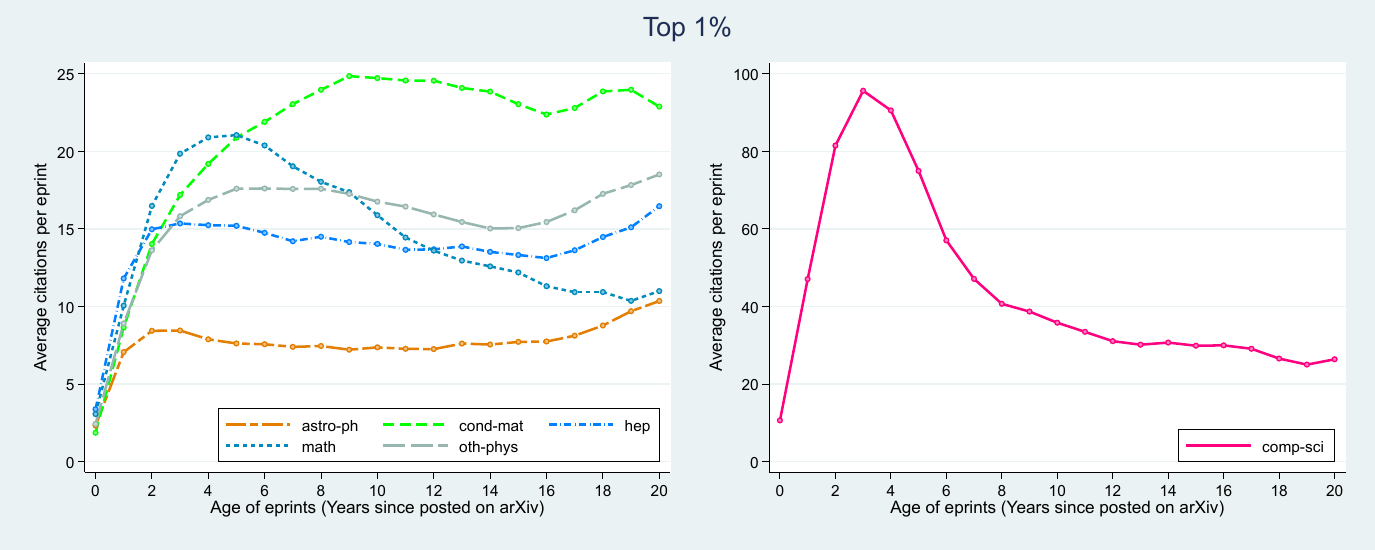}}
    \end{tabular}
\caption{\textbf{Discipline-average citation history curves for highly-cited eprints.}
\textsf{\textbf{a}}, the citation data are based on the top 0.1\% highest cited eprints; 
\textsf{\textbf{b}}, the citation data are based on the top 1\% highest cited eprints.}
\label{fig:lognormfit_ctp_top1}
\end{figure}
\quad\\
\vfill
}

\afterpage{\clearpage%
\begin{figure}[tp] 
\centering
\vspace{-0.5cm}
\noindent
\includegraphics[align=c, scale=0.7, vmargin=1mm]{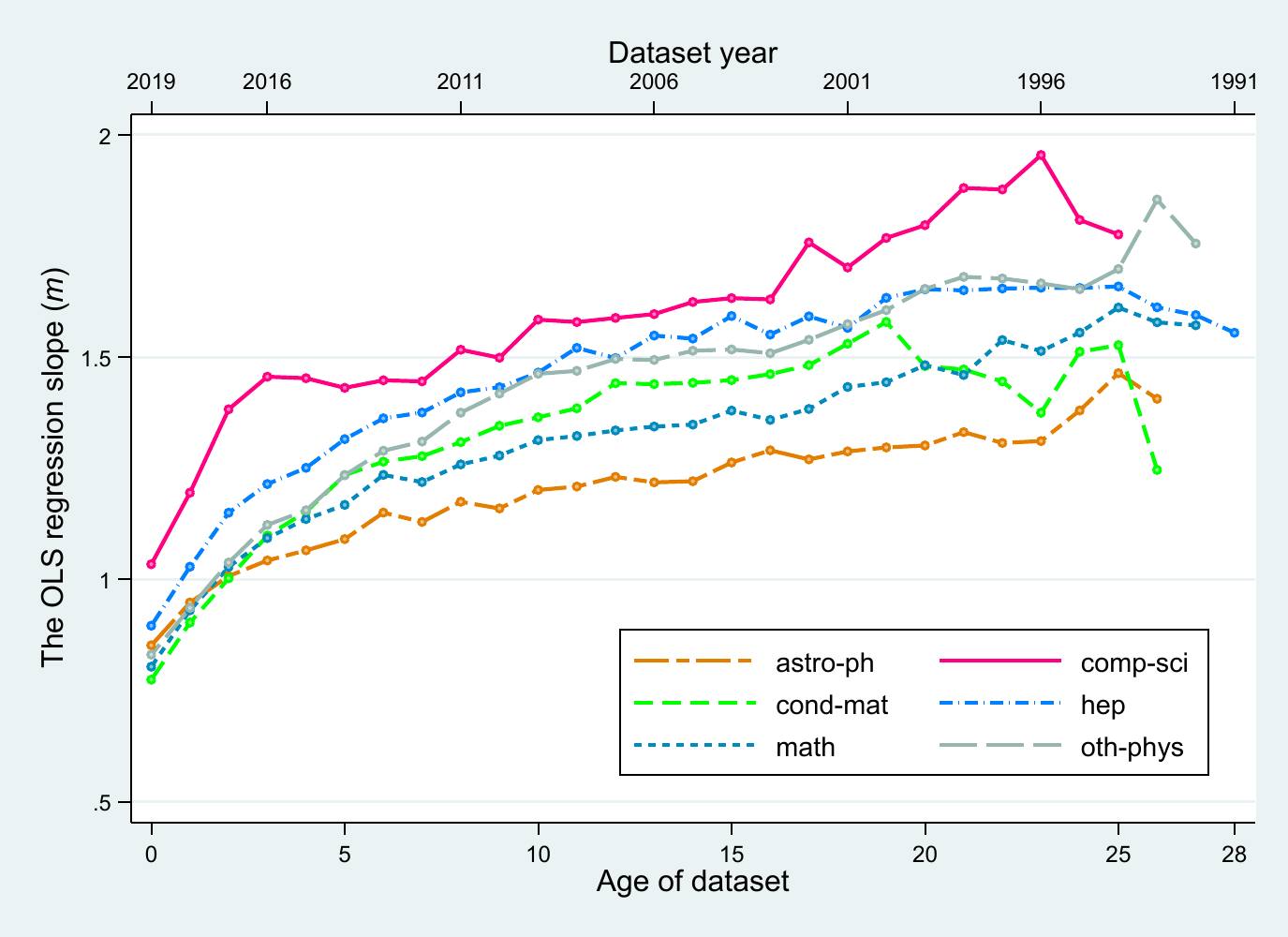}
\caption{\textbf{Trends in the lognormal quantile plot's slope ($\bmt{m}$).}
The citation data are based on the below-99th percentile eprints in the arXiv dataset.
Standard errors are negligibly small for all experimental results and are not shown.}
\label{fig:qplot_slope_lnc_tot_all}
\vspace{5mm}
\end{figure}
\quad\\
\vfill
}

\afterpage{\clearpage%
\begin{figure}[tp] 
\centering
\vspace{-0.5cm}
\noindent
\includegraphics[align=c, scale=1.1, vmargin=1mm]{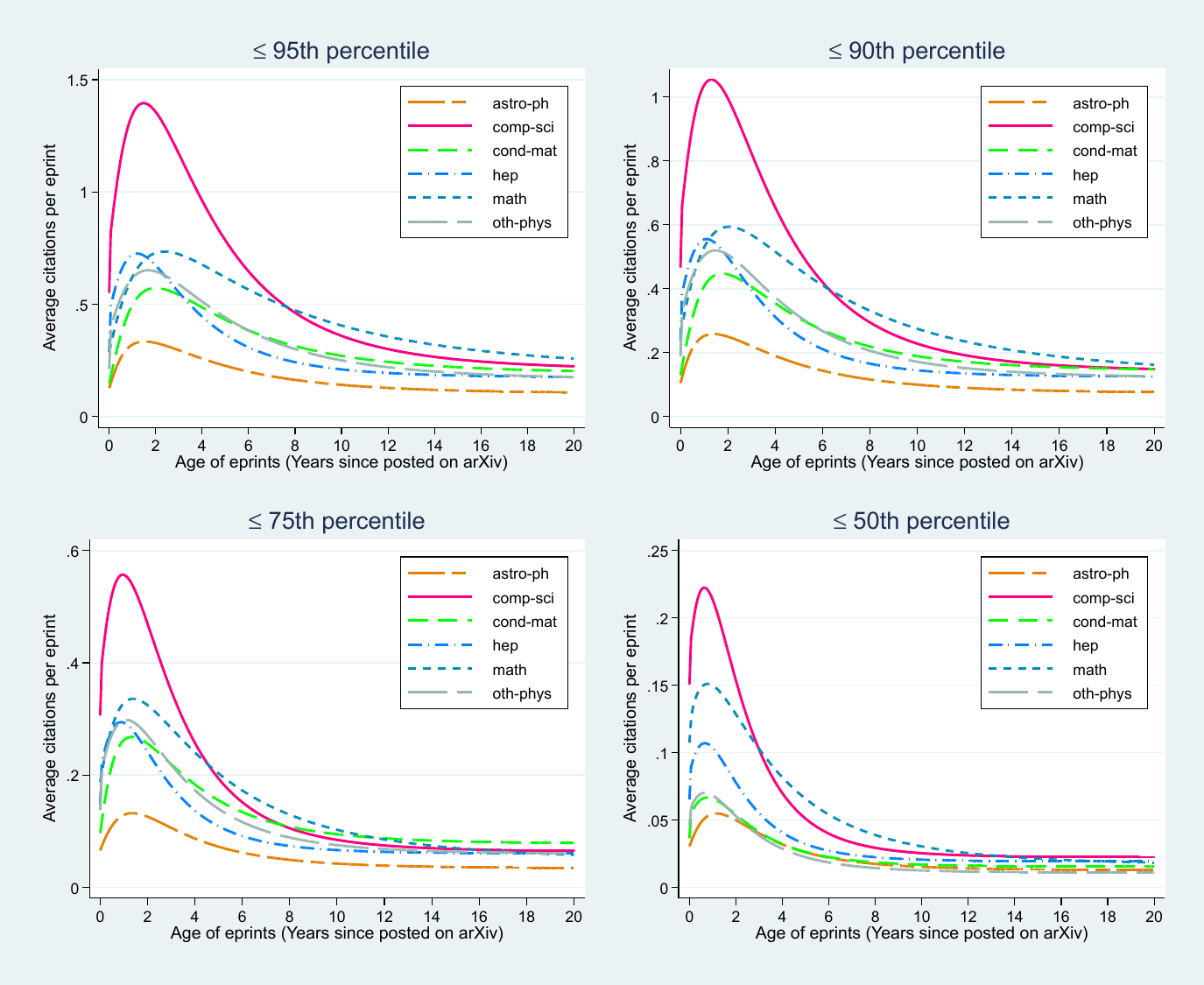}
\caption{\textbf{Discipline-average citation history curves for the arXiv disciplines} (the lower percentile thresholds)\textbf{.}
The estimated results for the below-$p$th percentile eprints ($p=95,\,90,\,75,\,50$) are shown by discipline.
See Suppl.~Table \ref{tab:lognorm_fitting_p50-95} for the statistics.
The results for the below-99th citation data are shown in Fig.\ \ref{fig:citation_curves_p99} and Table \ref{tab:lognorm_fitting_p99} in the main text.}
\label{fig:citation_curves_pctl_rest}
\vspace{5mm}
\end{figure}
\quad\\
\vfill
}

\clearpage%
\begin{figure}[tp] 
\centering
\vspace{-4.5cm}
\noindent
\includegraphics[align=c, scale=0.95, vmargin=1mm]{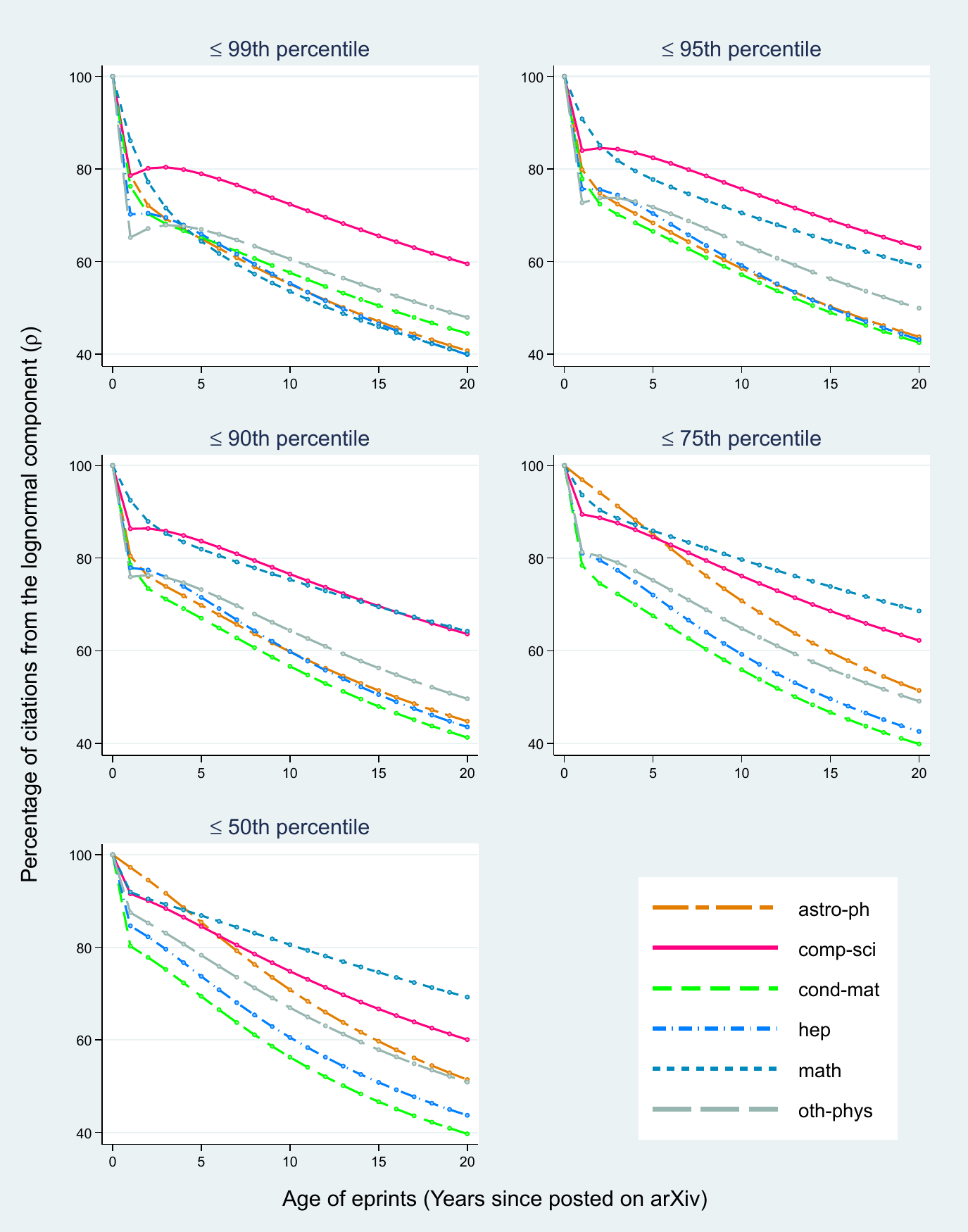}
\caption{\textbf{Trends in the contribution of the lognormal component.}
Percentage of citations from the lognormal component is shown for the below-$p$th percentile eprints ($p=99,\,95,\,90,\,75,\,50$) as a function of the age of eprints.}
\label{fig:fg_balance_by_pctl_rest}
\vspace{5mm}
\end{figure}
\quad\\

\afterpage{\clearpage%
\begin{figure}[tp] 
\centering
\vspace{-0.5cm}
\noindent
\includegraphics[align=c, scale=1.2, vmargin=1mm]{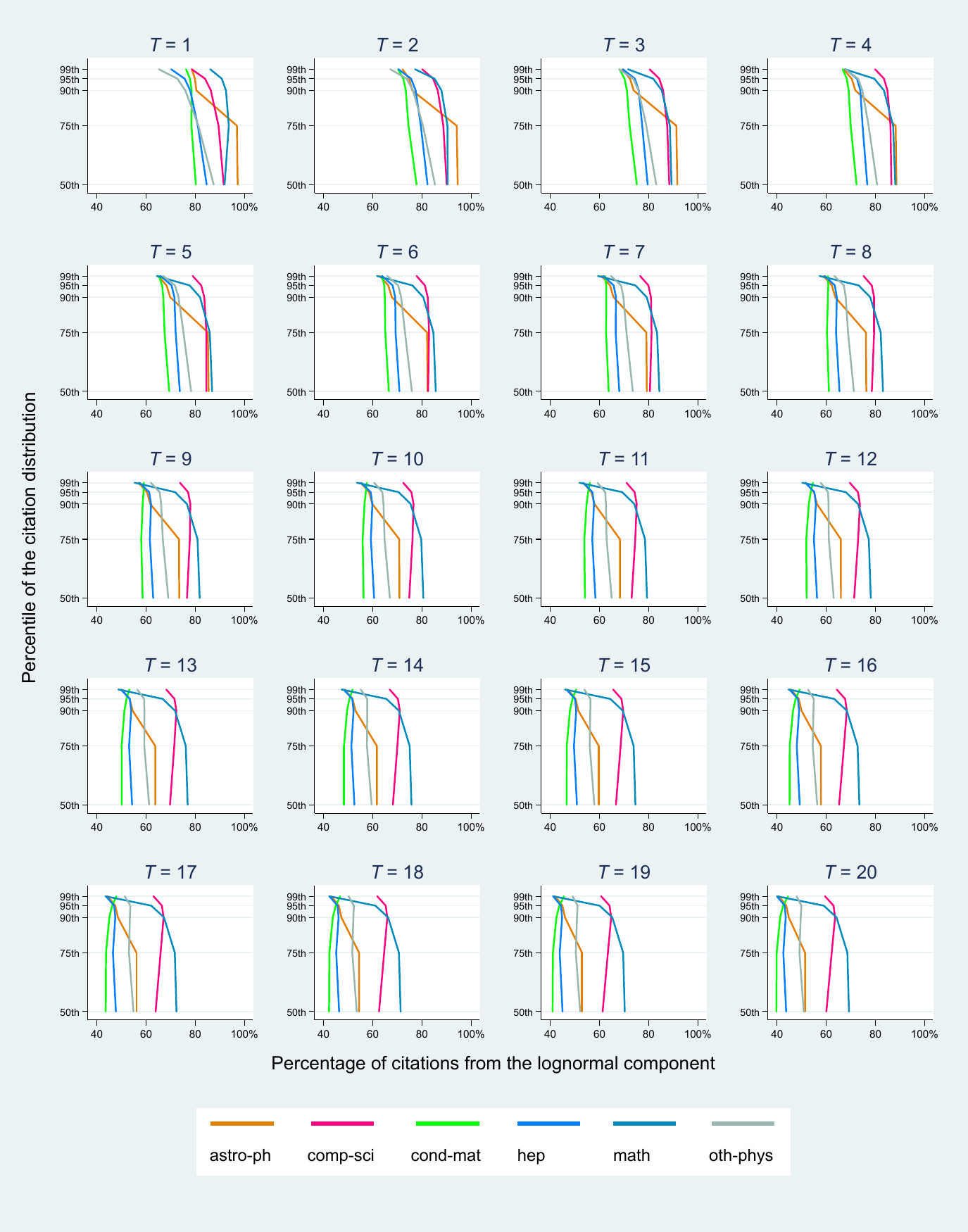}
\caption{\textbf{Contribution of the lognormal component, by age of eprints.}
Percentage of citations from the lognormal component is shown for the below-$p$th percentile eprints ($p=99,\,95,\,90,\,75,\,50$), displayed separately for each age in years ($T=1,\,\dots,\,20$).}
\label{fig:fg_balance_years}
\end{figure}
}

\afterpage{\clearpage%
\begin{figure}[tp] 
\centering
\vspace{-0.5cm}
\noindent
\includegraphics[align=c, scale=0.7, vmargin=1mm]{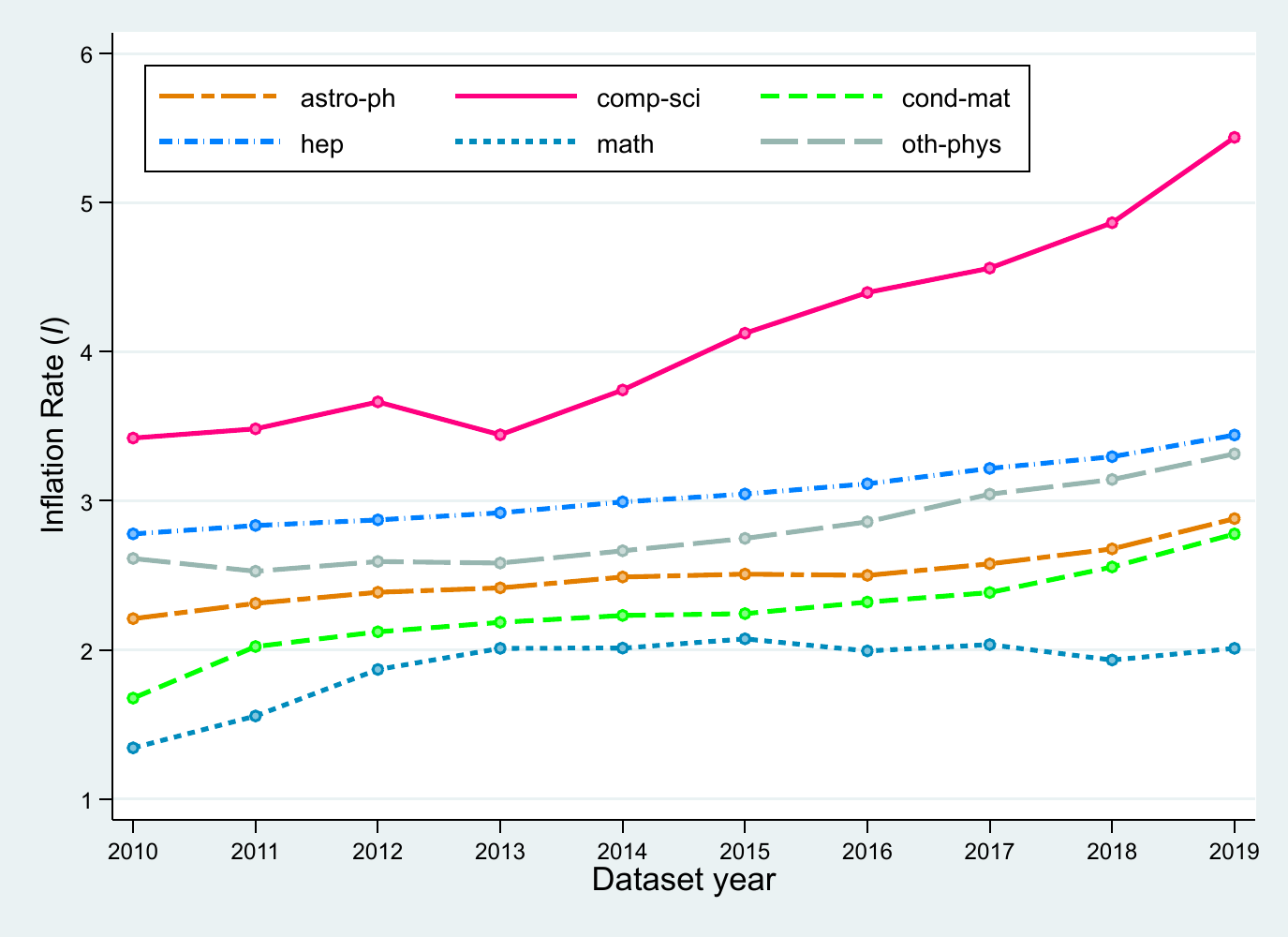}
\caption{\textbf{Trends in the inflation rate ($\bmt{\mathcal{I}}$).}
The inflation rate defined by $\mathcal{I}=\hat{u}_{\mathrm{p}}/\hat{B}$, where $\hat{u}_{\mathrm{p}}$ and $\hat{B}$ respectively represent the peak value and the asymptotic value of the average citation history curve.}
\label{fig:inflation_99}
\end{figure}
\quad\\
\vfill
}

\afterpage{\clearpage%
\begin{figure}[htp]
\centering
\vspace{-0.5cm}
    \begin{tabular}{c}
\begin{minipage}{0.5\hsize}
\begin{flushleft}
\raisebox{-0.5cm}{\textsf{\textbf{a}}\quad }
\raisebox{-\height}{\includegraphics[align=c, scale=0.3685, vmargin=0mm]{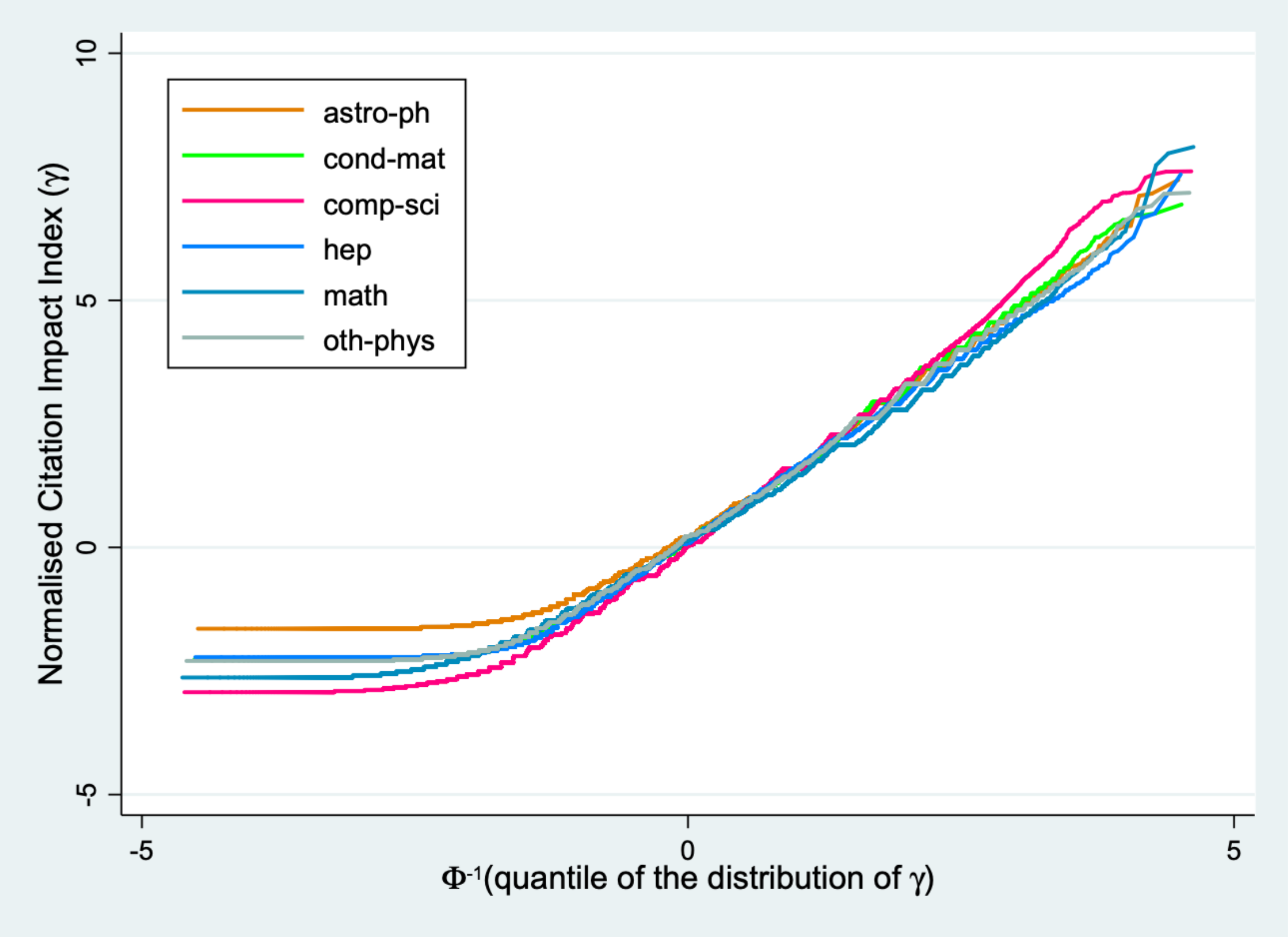}}
\end{flushleft}
      \end{minipage}
\begin{minipage}{0.5\hsize}
\begin{flushleft}
\raisebox{-0.5cm}{\textsf{\textbf{b}}\quad }
\raisebox{-\height}{\includegraphics[align=c, scale=0.6, vmargin=0mm]{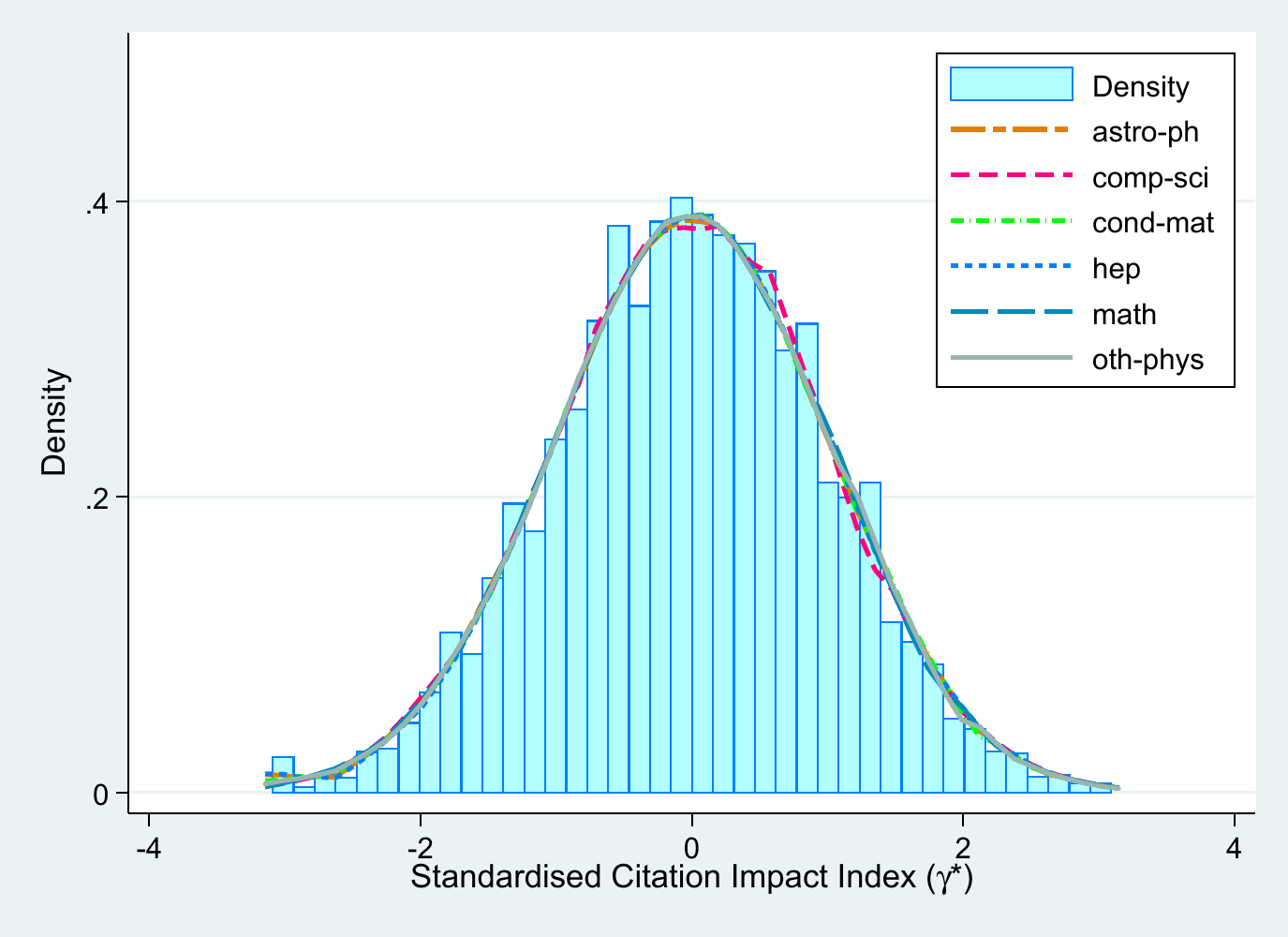}}
\end{flushleft}
          \end{minipage}
    \end{tabular}
\caption{\textbf{Distribution of the normalised ($\bmt{\gamma}$) and the standardised ($\bmt{\gamma^{*}}$) citation indices.}
\textsf{\textbf{a}}, lognormal quantile plot of the $\gamma$-index versus the inverse-normal function of the quantile rank by discipline;
\textsf{\textbf{b}}, frequency distribution of the $\gamma^{*}$-index, overlaid by the graphs of the kernel density estimation ($\textrm{half-width}=0.2$) by discipline.}
\label{fig:gamma_star}
\end{figure}
\quad\\
\vfill
}

\addcontentsline{toc}{subsection}{Supplementary Tables}

\afterpage{\clearpage%
\begin{landscape}
\centering
\vspace{-1cm}
    \captionof{table}{\textbf{Classification of the arXiv disciplines} (details)\textbf{.}
The classification scheme is based on that used on the \citek{arXiv_stats2020} website.}
\label{tab:arXiv_disciplines_de}
{\scriptsize
{\setlength{\tabcolsep}{1.0em}
\renewcommand{\arraystretch}{0.89}
\begin{tabular}{lll}\\[-2mm] \toprule[1pt] \\[-4.8mm]
\multicolumn{1}{c}{\textbf{Discipline}} & \multicolumn{1}{c}{\textbf{arXiv classification}} & \multicolumn{1}{c}{\textbf{Detailed subject categories included}} \\ \midrule
\textbf{astro-ph} & Astrophysics & Astrophysics of Galaxies; Cosmology and Nongalactic Astrophysics; Earth and Planetary Astrophysics; \\
{} & {} & High Energy Astrophysical Phenomena; Instrumentation and Methods for Astrophysics; Solar and Stellar \\
{} & {} & Astrophysics\\[3mm]
\textbf{comp-sci} & Computer Science, Computing Research Repository & Artificial Intelligence; Computation and Language; Computational Complexity; Computational Engineering, \\
{} & {} & Finance, and Science; Computational Geometry; Computer Science and Game Theory; Computer Vision and \\
{} & {} & Pattern Recognition; Computers and Society; Cryptography and Security; Data Structures and Algorithms; \\
{} & {} & Databases;Digital Libraries; Discrete Mathematics; Distributed, Parallel, and Cluster Computing; Emerging\\
{} & {} & Technologies; Formal Languages and Automata Theory; General Literature; Graphics; Hardware Architecture;\\
{} & {} & Human-Computer Interaction; Information Retrieval; Information Theory; Logic in Computer Science; Machine\\
{} & {} & Learning; Mathematical Software; Multiagent Systems; Multimedia; Networking and Internet Architecture; \\
{} & {} & Neural and Evolutionary Computing; Numerical Analysis; Operating Systems; Other Computer Science; \\
{} & {} & Performance; Programming Languages; Robotics; Social and Information Networks; Software Engineering; \\
{} & {} & Sound; Symbolic Computation; Systems and Control \\[3mm]
\textbf{cond-mat} & Condensed Matter Physics & Disordered Systems and Neural Networks; Materials Science; Mesoscale and Nanoscale Physics; \\
{} & {} & Other Condensed Matter; Quantum Gases; Soft Condensed Matter; Statistical Mechanics; Strongly Correlated \\
{} & {} & Electrons; Superconductivity\\[3mm]
\textbf{hep} & High Energy Physics: \q{hep-th} (Theory) + \q{hep-ph} (Phenomenology) &{} \\
{} & {} + \q{hep-lat} (Lattice) + \q{hep-ex} (Experiment) & {} \\[3mm]
\textbf{math}& Mathematics: \q{math} (Mathematics) + \q{math-ph} (Mathematical Physics) & Algebraic Geometry; Algebraic Topology; Analysis of PDEs; Category Theory; Classical Analysis and ODEs; \\
{} & {} & Combinatorics; Commutative Algebra; Complex Variables; Differential Geometry; Dynamical Systems; \\
{} & {} & Functional Analysis; General Mathematics; General Topology; Geometric Topology; Group Theory; History and \\
{} & {} & Overview; Information Theory; K-Theory and Homology; Logic; Mathematical Physics; Metric Geometry; \\
{} & {} & Number Theory; Numerical Analysis; Operator Algebras; Optimization and Control; Probability; Quantum \\
{} & {} & Algebra; Representation Theory; Rings and Algebras; Spectral Theory; Statistics Theory; Symplectic Geometry\\[3mm]
\textbf{oth-phys} & Other Physics: \q{physics} + \q{nucl-th} (Nuclear Theory) + \q{nucl-ex} (Nuclear & Accelerator Physics; Applied Physics; Atmospheric and Oceanic Physics; Atomic and Molecular Clusters; \\
{} & {} Experiment) + \q{gr-qc} (General Relativity and Quantum Cosmology) & Atomic Physics; Biological Physics; Chemical Physics; Classical Physics; Computational Physics; Data Analysis,\\
{} & {} + \q{quant-ph} (Quantum Physics) + \q{nlin} (Nonlinear Sciences) & Statistics and Probability; Fluid Dynamics; General Physics; Geophysics; History and Philosophy of Physics; \\
{} & {} & Instrumentation and Detectors; Medical Physics; Optics; Physics and Society; Physics Education; Plasma Physics; \\
{} & {} & Popular Physics; Space Physics; Adaptation and Self-Organizing Systems; Cellular Automata and Lattice Gases; \\
{} & {} & Chaotic Dynamics; Exactly Solvable and Integrable Systems; Pattern Formation and Solitons\\[0mm] \bottomrule[1pt] 
\end{tabular}}
}
	\end{landscape}
\vspace{5mm}
}

\afterpage{\clearpage%
\begin{table}[h]
\centering
\vspace{-0.5cm}
\caption{\textbf{Percentile values ($\bmt{c_{[p]}}$) of the citation distribution.}
Also shown are the numbers of eprints ($N_{\leq p}$) in each percentile range 0--$p$, i.e.\ $N_{\leq p}=|\{k\,|\,c_{k}\leq c_{[p]}\}|$ with $c_{k}$ the citations of eprint $k$.}
\label{tab:cat_by_pctl}
{\footnotesize
{\setlength{\tabcolsep}{0.5em}
\begin{tabular}{l@{\hspace{1.2cm}}ccccccccccccccc}\\[-2mm] \toprule[1pt] \\[-5.5mm]
         &\multicolumn{2}{c}{$p\leq 50$}  &     &\multicolumn{2}{c}{$p\leq 75$}    &    &\multicolumn{2}{c}{$p\leq 90$}     &   &\multicolumn{2}{c}{$p\leq 95$}   &     &\multicolumn{2}{c}{$p\leq 99$}   \\ \cline{2-3} \cline{5-6} \cline{8-9} \cline{11-12} \cline{14-15}
Category         & $c_{[50]}$  & \multicolumn{1}{c}{$N_{\leq 50}$}    &     & $c_{[75]}$  &  \multicolumn{1}{c}{$N_{\leq 75}$}   &     & $c_{[90]}$  & \multicolumn{1}{c}{$N_{\leq 90}$}    &     & $c_{[95]}$   &  \multicolumn{1}{c}{$N_{\leq 95}$}   &     & $c_{[99]}$    &  \multicolumn{1}{c}{$N_{\leq 99}$}          \\[-0.2mm] \midrule[0.3pt]
\textit{astro-ph} & 1 & 153,818 & & 3 & 195,287 & & 9  & 233,097 & & 16 & 245,346 & & 45  & 255,347  \\
\textit{comp-sci}       & 2 & 218,336 & & 7 & 293,025 & & 22 & 349,189 & & 42 & 367,447 & & 152 & 382,665  \\
\textit{cond-mat} & 1 & 133,510 & & 6 & 205,517 & & 15 & 237,983 & & 27 & 250,555 & & 85  & 260,876  \\
\textit{hep}      & 2 & 166,242 & & 7 & 220,048 & & 19 & 258,507 & & 34 & 272,701 & & 99  & 283,998  \\
\textit{math}     & 2 & 243,052 & & 6 & 321,828 & & 17 & 387,309 & & 29 & 407,414 & & 83  & 424,395  \\
\textit{oth-phys} & 1 & 201,401 & & 6 & 300,939 & & 16 & 351,573 & & 28 & 369,091 & & 84  & 384,072  \\[0mm] \bottomrule[1pt] \\
\end{tabular}}
}
\vspace{5mm}
\end{table}
\quad\\
\vfill
}

\afterpage{\clearpage%
\begin{table}[h]
\renewcommand{\arraystretch}{0.85}
\centering
\vspace{-0.5cm}
\caption{\textbf{Estimated regression coefficients for the lognormal quantile plots.}
The citation data are based on the below-99th percentile eprints posted on arXiv in 2010 with nonzero citations.
The regression function is given by $y_{k}=b+m\Phi^{-1}(q_{k})+\ep_{k}$, where $y_{k}=\ln(c_{k}+1)$ is the \q{(+1)-shifted} and log-transformed citation variable, $q_{k}$ is the quantile rank of $y_{k}$ in the citation distribution of each discipline, $b$ and $m$ are the OLS regression parameters to be estimated, and $\ep_{k}$ is the error term.
Standard errors are in parentheses, which are negligibly small.
The adjusted $R$-squared ($R^{2}_{\mathrm{adj}}$) is also shown, which is nearly one for all disciplines.}
\label{tab:qplot_2010}
{\footnotesize
{\setlength{\tabcolsep}{1.4em}
\begin{tabular}{l@{\hspace{0.4cm}}dddd}\\[-2mm] \toprule[1pt] \\[-5.5mm]
Category & \multicolumn{1}{d}{N}\hspace{-1.1cm}       & \multicolumn{1}{c}{$\hat{b}$} & \multicolumn{1}{c}{$\hat{m}$} & \multicolumn{1}{c}{$R^{2}_{\mathrm{adj}}$} \\[-0.2mm] \midrule[0.3pt]
\textit{astro-ph} & \multicolumn{1}{d}{8,275}\hspace{-1.1cm} & \multicolumn{1}{d}{1.08} & \multicolumn{1}{d}{1.07} & \multicolumn{1}{d}{0.996}  \\[-0.5mm]
{} & {} & \multicolumn{1}{d}{(0.000548)} & \multicolumn{1}{d}{(0.00178)}  & {}   \\[1.2mm]
\textit{comp-sci} & \multicolumn{1}{d}{10,251}\hspace{-1.1cm} & \multicolumn{1}{d}{1.60} & \multicolumn{1}{d}{1.39} & \multicolumn{1}{d}{0.997}  \\[-0.5mm]
{} & {} & \multicolumn{1}{d}{(0.000454)} & \multicolumn{1}{d}{(0.00231)}   & {} \\[1.2mm]
\textit{cond-mat} & \multicolumn{1}{d}{9,002}\hspace{-1.1cm} & \multicolumn{1}{d}{1.39} & \multicolumn{1}{d}{1.23} & \multicolumn{1}{d}{0.996}  \\[-0.5mm]
{} & {} & \multicolumn{1}{d}{(0.000452)} & \multicolumn{1}{d}{(0.00219)}  & {}  \\[1.2mm]
\textit{hep} & \multicolumn{1}{d}{8,041}\hspace{-1.1cm} & \multicolumn{1}{d}{1.39} & \multicolumn{1}{d}{1.33} & \multicolumn{1}{d}{0.994}  \\[-0.5mm]
{} & {} & \multicolumn{1}{d}{(0.000798)} & \multicolumn{1}{d}{(0.00311)}   & {} \\[1.2mm]
\textit{math} & \multicolumn{1}{d}{15,523}\hspace{-1.1cm} & \multicolumn{1}{d}{1.57} & \multicolumn{1}{d}{1.19} & \multicolumn{1}{d}{0.993}  \\[-0.5mm]
{} & {} & \multicolumn{1}{d}{(0.000498)} & \multicolumn{1}{d}{(0.00196)}  & {} \\[1.2mm]
\textit{oth-phys} & \multicolumn{1}{d}{12,312}\hspace{-1.1cm} & \multicolumn{1}{d}{1.35} & \multicolumn{1}{d}{1.30} & \multicolumn{1}{d}{0.993} \\[-0.5mm]
{} & {} & \multicolumn{1}{d}{(0.000634)} & \multicolumn{1}{d}{(0.00256)}  \\[0mm] \bottomrule[1pt] 
\end{tabular}}
}
\end{table}
\quad\\
\vfill
}

\afterpage{\clearpage%
\begin{table}[t]
\centering
\vspace{-1.5cm}
\caption{\textbf{Estimated regression coefficients with the derived metrics} (the lower percentile thresholds)\textbf{.}
Results are shown for the below-$p$th citation data ($p=95,\,90,\,75,\,50$).
The regression function is modelled as Eqs.\ (\ref{reg_model}--\ref{step-like}).
The adjusted $R$-squared was higher than 0.99 for all disciplines.
In addition to the estimated regression coefficients (left half), the peak value ($\hat{u}_{\mathrm{p}}$), the typical time interval of the growth phase ($\delta_{1}$) and the obsolescence phase ($\delta_{2}$), the internal obsolescence rate ($\mathcal{S}=\delta_{1}/\delta_{2}$) and the retention rate ($\mathcal{R}=\hat{B}/\hat{u}_{\mathrm{p}}$) (right half) are also shown by discipline.
The value of $\hat{\lam}$ larger than ten is displayed as \q{$\gg 1$}.
Results for the below-99th citation data are shown in Table \ref{tab:lognorm_fitting_p99} in the main text.}
\label{tab:lognorm_fitting_p50-95}
{\scriptsize
{\setlength{\tabcolsep}{0.6em}
\begin{tabular}{l@{\quad}dddddd@{\hspace{-8mm}}dddddd}\\[-2mm]
\multicolumn{13}{l}{\footnotesize{\hspace{-3mm}\textbf{$\blacktriangledown$ The 95th percentile}}} \\[1mm] \toprule[1pt] \\[-5.5mm]
Category & \multicolumn{1}{c}{$\hat{A}$}        & \multicolumn{1}{c}{$\hat{\mu}$} & \multicolumn{1}{c}{$\hat{\sigma}$} & \multicolumn{1}{c}{$\hat{B}$}  & \multicolumn{1}{c}{$\hat{\lam}$} &{} & \multicolumn{1}{c}{$\hat{u}_{\mathrm{p}}$} & \multicolumn{1}{c}{$\delta_{1}$} & \multicolumn{1}{c}{$\delta_{2}$} & \multicolumn{1}{c}{$\mathcal{S}$} & \multicolumn{1}{c}{$\mathcal{R}$} \\ \cline{1-6}\cline{8-13}\\[-5.0mm]
\textit{astro-ph} & \multicolumn{1}{d}{1.58} & \multicolumn{1}{d}{1.53} & \multicolumn{1}{d}{0.801} & \multicolumn{1}{d}{0.101} & \multicolumn{1}{d}{1.55} &{} & \multicolumn{1}{d}{0.334} & \multicolumn{1}{d}{2.44} & \multicolumn{1}{d}{2.19} & \multicolumn{1}{d}{1.11} & \multicolumn{1}{d}{0.303} \\
\textit{comp-sci} & \multicolumn{1}{d}{7.13} & \multicolumn{1}{d}{1.45} & \multicolumn{1}{d}{0.735} & \multicolumn{1}{d}{0.206} & \multicolumn{1}{d}{\gg 1}\hspace{-8mm} &{} & \multicolumn{1}{d}{1.40} & \multicolumn{1}{d}{2.48} & \multicolumn{1}{d}{1.78} & \multicolumn{1}{d}{1.39} & \multicolumn{1}{d}{0.148} \\
\textit{cond-mat} & \multicolumn{1}{d}{2.81} & \multicolumn{1}{d}{1.63} & \multicolumn{1}{d}{0.757} & \multicolumn{1}{d}{0.190} & \multicolumn{1}{d}{1.07} &{} & \multicolumn{1}{d}{0.571} & \multicolumn{1}{d}{2.87} & \multicolumn{1}{d}{2.22} & \multicolumn{1}{d}{1.29} & \multicolumn{1}{d}{0.333} \\
\textit{hep} & \multicolumn{1}{d}{2.65} & \multicolumn{1}{d}{1.26} & \multicolumn{1}{d}{0.688} & \multicolumn{1}{d}{0.174} & \multicolumn{1}{d}{\gg 1}\hspace{-8mm} &{} & \multicolumn{1}{d}{0.726} & \multicolumn{1}{d}{2.20} & \multicolumn{1}{d}{1.33} & \multicolumn{1}{d}{1.65} & \multicolumn{1}{d}{0.240} \\
\textit{math} & \multicolumn{1}{d}{6.02} & \multicolumn{1}{d}{1.95} & \multicolumn{1}{d}{0.930} & \multicolumn{1}{d}{0.196} & \multicolumn{1}{d}{0.581} &{} & \multicolumn{1}{d}{0.727} & \multicolumn{1}{d}{2.96} & \multicolumn{1}{d}{4.06} & \multicolumn{1}{d}{0.728} & \multicolumn{1}{d}{0.270} \\
\textit{oth-phys} & \multicolumn{1}{d}{3.33} & \multicolumn{1}{d}{1.56} & \multicolumn{1}{d}{0.761} & \multicolumn{1}{d}{0.163} & \multicolumn{1}{d}{\gg 1}\hspace{-8mm} &{} & \multicolumn{1}{d}{0.651} & \multicolumn{1}{d}{2.68} & \multicolumn{1}{d}{2.10} & \multicolumn{1}{d}{1.28} & \multicolumn{1}{d}{0.250} \\[0mm] \bottomrule[1pt] \\
\end{tabular}}
{\setlength{\tabcolsep}{0.6em}
\begin{tabular}{l@{\quad}dddddd@{\hspace{-8mm}}dddddd}\\[-5mm]
\multicolumn{13}{l}{\footnotesize{\hspace{-3mm}\textbf{$\blacktriangledown$ The 90th percentile}}} \\[1mm] \toprule[1pt] \\[-5.5mm]
Category & \multicolumn{1}{c}{$\hat{A}$}        & \multicolumn{1}{c}{$\hat{\mu}$} & \multicolumn{1}{c}{$\hat{\sigma}$} & \multicolumn{1}{c}{$\hat{B}$}  & \multicolumn{1}{c}{$\hat{\lam}$} &{} & \multicolumn{1}{c}{$\hat{u}_{\mathrm{p}}$} & \multicolumn{1}{c}{$\delta_{1}$} & \multicolumn{1}{c}{$\delta_{2}$} & \multicolumn{1}{c}{$\mathcal{S}$} & \multicolumn{1}{c}{$\mathcal{R}$} \\ \cline{1-6}\cline{8-13}\\[-5.0mm]
\textit{astro-ph} & \multicolumn{1}{d}{1.18} & \multicolumn{1}{d}{1.47} & \multicolumn{1}{d}{0.790} & \multicolumn{1}{d}{0.0725} & \multicolumn{1}{d}{1.86} &{} & \multicolumn{1}{d}{0.258} & \multicolumn{1}{d}{2.34} & \multicolumn{1}{d}{2.02} & \multicolumn{1}{d}{1.15} & \multicolumn{1}{d}{0.281} \\
\textit{comp-sci} & \multicolumn{1}{d}{4.95} & \multicolumn{1}{d}{1.36} & \multicolumn{1}{d}{0.722} & \multicolumn{1}{d}{0.140} & \multicolumn{1}{d}{\gg 1}\hspace{-8mm} &{} & \multicolumn{1}{d}{1.05} & \multicolumn{1}{d}{2.31} & \multicolumn{1}{d}{1.58} & \multicolumn{1}{d}{1.46} & \multicolumn{1}{d}{0.133} \\
\textit{cond-mat} & \multicolumn{1}{d}{1.98} & \multicolumn{1}{d}{1.51} & \multicolumn{1}{d}{0.735} & \multicolumn{1}{d}{0.142} & \multicolumn{1}{d}{1.23} &{} & \multicolumn{1}{d}{0.447} & \multicolumn{1}{d}{2.64} & \multicolumn{1}{d}{1.89} & \multicolumn{1}{d}{1.40} & \multicolumn{1}{d}{0.318} \\
\textit{hep} & \multicolumn{1}{d}{1.92} & \multicolumn{1}{d}{1.19} & \multicolumn{1}{d}{0.673} & \multicolumn{1}{d}{0.124} & \multicolumn{1}{d}{\gg 1}\hspace{-8mm} &{} & \multicolumn{1}{d}{0.556} & \multicolumn{1}{d}{2.10} & \multicolumn{1}{d}{1.20} & \multicolumn{1}{d}{1.74} & \multicolumn{1}{d}{0.223} \\
\textit{math} & \multicolumn{1}{d}{4.60} & \multicolumn{1}{d}{1.83} & \multicolumn{1}{d}{0.902} & \multicolumn{1}{d}{0.123} & \multicolumn{1}{d}{0.672} &{} & \multicolumn{1}{d}{0.591} & \multicolumn{1}{d}{2.77} & \multicolumn{1}{d}{3.48} & \multicolumn{1}{d}{0.796} & \multicolumn{1}{d}{0.208} \\
\textit{oth-phys} & \multicolumn{1}{d}{2.40} & \multicolumn{1}{d}{1.45} & \multicolumn{1}{d}{0.737} & \multicolumn{1}{d}{0.120} & \multicolumn{1}{d}{\gg 1}\hspace{-8mm} &{} & \multicolumn{1}{d}{0.520} & \multicolumn{1}{d}{2.47} & \multicolumn{1}{d}{1.78} & \multicolumn{1}{d}{1.38} & \multicolumn{1}{d}{0.231} \\[0mm] \bottomrule[1pt] \\
\end{tabular}}
{\setlength{\tabcolsep}{0.6em}
\begin{tabular}{l@{\quad}dddddd@{\hspace{-8mm}}dddddd}\\[-5mm]
\multicolumn{13}{l}{\footnotesize{\hspace{-3mm}\textbf{$\blacktriangledown$ The 75th percentile}}} \\[1mm] \toprule[1pt] \\[-5.5mm]
Category & \multicolumn{1}{c}{$\hat{A}$}        & \multicolumn{1}{c}{$\hat{\mu}$} & \multicolumn{1}{c}{$\hat{\sigma}$} & \multicolumn{1}{c}{$\hat{B}$}  & \multicolumn{1}{c}{$\hat{\lam}$} &{} & \multicolumn{1}{c}{$\hat{u}_{\mathrm{p}}$} & \multicolumn{1}{c}{$\delta_{1}$} & \multicolumn{1}{c}{$\delta_{2}$} & \multicolumn{1}{c}{$\mathcal{S}$} & \multicolumn{1}{c}{$\mathcal{R}$} \\ \cline{1-6}\cline{8-13}\\[-5.0mm]
\textit{astro-ph} & \multicolumn{1}{d}{0.613} & \multicolumn{1}{d}{1.29} & \multicolumn{1}{d}{0.707} & \multicolumn{1}{d}{0.0339} & \multicolumn{1}{d}{0.231} &{} & \multicolumn{1}{d}{0.132} & \multicolumn{1}{d}{2.20} & \multicolumn{1}{d}{1.42} & \multicolumn{1}{d}{1.54} & \multicolumn{1}{d}{0.257} \\
\textit{comp-sci} & \multicolumn{1}{d}{2.10} & \multicolumn{1}{d}{1.14} & \multicolumn{1}{d}{0.688} & \multicolumn{1}{d}{0.0636} & \multicolumn{1}{d}{\gg 1}\hspace{-8mm} &{} & \multicolumn{1}{d}{0.557} & \multicolumn{1}{d}{1.95} & \multicolumn{1}{d}{1.18} & \multicolumn{1}{d}{1.65} & \multicolumn{1}{d}{0.114} \\
\textit{cond-mat} & \multicolumn{1}{d}{1.02} & \multicolumn{1}{d}{1.34} & \multicolumn{1}{d}{0.713} & \multicolumn{1}{d}{0.0774} & \multicolumn{1}{d}{1.95} &{} & \multicolumn{1}{d}{0.268} & \multicolumn{1}{d}{2.30} & \multicolumn{1}{d}{1.52} & \multicolumn{1}{d}{1.51} & \multicolumn{1}{d}{0.289} \\
\textit{hep} & \multicolumn{1}{d}{0.894} & \multicolumn{1}{d}{1.06} & \multicolumn{1}{d}{0.655} & \multicolumn{1}{d}{0.0603} & \multicolumn{1}{d}{\gg 1}\hspace{-8mm} &{} & \multicolumn{1}{d}{0.294} & \multicolumn{1}{d}{1.88} & \multicolumn{1}{d}{1.01} & \multicolumn{1}{d}{1.86} & \multicolumn{1}{d}{0.205} \\
\textit{math} & \multicolumn{1}{d}{2.09} & \multicolumn{1}{d}{1.57} & \multicolumn{1}{d}{0.865} & \multicolumn{1}{d}{0.0473} & \multicolumn{1}{d}{1.12} &{} & \multicolumn{1}{d}{0.335} & \multicolumn{1}{d}{2.27} & \multicolumn{1}{d}{2.52} & \multicolumn{1}{d}{0.900} & \multicolumn{1}{d}{0.141} \\
\textit{oth-phys} & \multicolumn{1}{d}{1.15} & \multicolumn{1}{d}{1.25} & \multicolumn{1}{d}{0.710} & \multicolumn{1}{d}{0.0594} & \multicolumn{1}{d}{\gg 1}\hspace{-8mm} &{} & \multicolumn{1}{d}{0.298} & \multicolumn{1}{d}{2.10} & \multicolumn{1}{d}{1.38} & \multicolumn{1}{d}{1.53} & \multicolumn{1}{d}{0.199} \\[0mm] \bottomrule[1pt] \\
\end{tabular}}
{\setlength{\tabcolsep}{0.6em}
\begin{tabular}{l@{\quad}dddddd@{\hspace{-8mm}}dddddd}\\[-5mm]
\multicolumn{13}{l}{\footnotesize{\hspace{-3mm}\textbf{$\blacktriangledown$ The 50th percentile}}} \\[1mm] \toprule[1pt] \\[-5.5mm]
Category & \multicolumn{1}{c}{$\hat{A}$}        & \multicolumn{1}{c}{$\hat{\mu}$} & \multicolumn{1}{c}{$\hat{\sigma}$} & \multicolumn{1}{c}{$\hat{B}$}  & \multicolumn{1}{c}{$\hat{\lam}$} &{} & \multicolumn{1}{c}{$\hat{u}_{\mathrm{p}}$} & \multicolumn{1}{c}{$\delta_{1}$} & \multicolumn{1}{c}{$\delta_{2}$} & \multicolumn{1}{c}{$\mathcal{S}$} & \multicolumn{1}{c}{$\mathcal{R}$} \\ \cline{1-6}\cline{8-13}\\[-5.0mm]
\textit{astro-ph} & \multicolumn{1}{d}{0.232} & \multicolumn{1}{d}{1.19} & \multicolumn{1}{d}{0.694} & \multicolumn{1}{d}{0.0127} & \multicolumn{1}{d}{0.245} &{} & \multicolumn{1}{d}{0.0547} & \multicolumn{1}{d}{2.04} & \multicolumn{1}{d}{1.26} & \multicolumn{1}{d}{1.62} & \multicolumn{1}{d}{0.233} \\
\textit{comp-sci} & \multicolumn{1}{d}{0.672} & \multicolumn{1}{d}{0.927} & \multicolumn{1}{d}{0.658} & \multicolumn{1}{d}{0.0223} & \multicolumn{1}{d}{\gg 1}\hspace{-8mm} &{} & \multicolumn{1}{d}{0.222} & \multicolumn{1}{d}{1.64} & \multicolumn{1}{d}{0.888} & \multicolumn{1}{d}{1.85} & \multicolumn{1}{d}{0.100} \\
\textit{cond-mat} & \multicolumn{1}{d}{0.201} & \multicolumn{1}{d}{1.05} & \multicolumn{1}{d}{0.692} & \multicolumn{1}{d}{0.0153} & \multicolumn{1}{d}{\gg 1}\hspace{-8mm} &{} & \multicolumn{1}{d}{0.0667} & \multicolumn{1}{d}{1.78} & \multicolumn{1}{d}{1.09} & \multicolumn{1}{d}{1.63} & \multicolumn{1}{d}{0.229} \\
\textit{hep} & \multicolumn{1}{d}{0.298} & \multicolumn{1}{d}{0.938} & \multicolumn{1}{d}{0.657} & \multicolumn{1}{d}{0.0192} & \multicolumn{1}{d}{\gg 1}\hspace{-8mm} &{} & \multicolumn{1}{d}{0.107} & \multicolumn{1}{d}{1.66} & \multicolumn{1}{d}{0.895} & \multicolumn{1}{d}{1.85} & \multicolumn{1}{d}{0.179} \\
\textit{math} & \multicolumn{1}{d}{0.745} & \multicolumn{1}{d}{1.31} & \multicolumn{1}{d}{0.857} & \multicolumn{1}{d}{0.0162} & \multicolumn{1}{d}{\gg 1}\hspace{-8mm} &{} & \multicolumn{1}{d}{0.151} & \multicolumn{1}{d}{1.78} & \multicolumn{1}{d}{1.93} & \multicolumn{1}{d}{0.923} & \multicolumn{1}{d}{0.107} \\
\textit{oth-phys} & \multicolumn{1}{d}{0.225} & \multicolumn{1}{d}{1.01} & \multicolumn{1}{d}{0.708} & \multicolumn{1}{d}{0.0109} & \multicolumn{1}{d}{\gg 1}\hspace{-8mm} &{} & \multicolumn{1}{d}{0.0702} & \multicolumn{1}{d}{1.66} & \multicolumn{1}{d}{1.08} & \multicolumn{1}{d}{1.54} & \multicolumn{1}{d}{0.155} \\[0mm] \bottomrule[1pt] \\
\end{tabular}}
}
\vspace{5mm}
\end{table}
}

\afterpage{\clearpage%
\begin{table}[h]
\renewcommand{\arraystretch}{0.85}
\centering
\vspace{-0.5cm}
\caption{\textbf{Summary statistics of the distribution of the normalised citation index ($\bmt{\gamma}$).}
The citation data are based on the below-99th percentile eprints in the arXiv dataset.
See Fig.\ \ref{fig:impact} for the corresponding graphs of the kernel density estimation by discipline.}
\label{tab:impact_nlx}
{\footnotesize
{\setlength{\tabcolsep}{0.5em}
\begin{tabular}{l@{\hspace{1.5cm}}dddddd}\\[-2mm] \toprule[1pt] \\[-5.5mm]
Category & \multicolumn{1}{d}{N}        & \multicolumn{1}{c}{Mean} & \multicolumn{1}{c}{S.D.} & \multicolumn{1}{c}{Min}  & \multicolumn{1}{c}{Max} \\[-0.2mm] \midrule[0.3pt]
all & \multicolumn{1}{d}{1,206,997} & \multicolumn{1}{d}{0.183} & \multicolumn{1}{d}{1.38} & \multicolumn{1}{d}{-2.93} & \multicolumn{1}{d}{8.10}   \\ \hdashline
\textit{astro-ph} & \multicolumn{1}{d}{138,776} & \multicolumn{1}{d}{0.352} & \multicolumn{1}{d}{1.25} & \multicolumn{1}{d}{-1.64} & \multicolumn{1}{d}{7.44}   \\
\textit{comp-sci} & \multicolumn{1}{d}{249,666} & \multicolumn{1}{d}{0.0884} & \multicolumn{1}{d}{1.53} & \multicolumn{1}{d}{-2.93} & \multicolumn{1}{d}{7.61}  \\
\textit{cond-mat} & \multicolumn{1}{d}{161,661} & \multicolumn{1}{d}{0.191} & \multicolumn{1}{d}{1.40} & \multicolumn{1}{d}{-2.27} & \multicolumn{1}{d}{6.94}  \\
\textit{hep} & \multicolumn{1}{d}{155,410} & \multicolumn{1}{d}{0.179} & \multicolumn{1}{d}{1.40} & \multicolumn{1}{d}{-2.22} & \multicolumn{1}{d}{7.55}  \\
\textit{math} & \multicolumn{1}{d}{273,731} & \multicolumn{1}{d}{0.152} & \multicolumn{1}{d}{1.28} & \multicolumn{1}{d}{-2.63} & \multicolumn{1}{d}{8.10}  \\
\textit{oth-phys} & \multicolumn{1}{d}{227,753} & \multicolumn{1}{d}{0.217} & \multicolumn{1}{d}{1.37} & \multicolumn{1}{d}{-2.30} & \multicolumn{1}{d}{7.18}  \\[0mm] \bottomrule[1pt] 
\end{tabular}}
}
\end{table}
\vspace{0.5cm}

\begin{table}[h]
\renewcommand{\arraystretch}{0.85}
\centering
\caption{\textbf{Summary statistics of the distribution of the standardised citation index ($\bmt{\gamma^{*}}$).}
The citation data are based on the below-99th percentile eprints in the arXiv dataset.
See Suppl.~Fig.\ \ref{fig:gamma_star}b for the corresponding graphs of the kernel density estimation by discipline.}
\label{tab:impact_normal}
{\footnotesize
{\setlength{\tabcolsep}{0.5em}
\begin{tabular}{l@{\hspace{1.5cm}}dddddd}\\[-2mm] \toprule[1pt] \\[-5.5mm]
Category & \multicolumn{1}{d}{N}        & \multicolumn{1}{c}{Mean} & \multicolumn{1}{c}{S.D.} & \multicolumn{1}{c}{Min}  & \multicolumn{1}{c}{Max} \\[-0.2mm] \midrule[0.3pt]
all & \multicolumn{1}{d}{1,206,997} & \multicolumn{1}{d}{-0.0165} & \multicolumn{1}{d}{1.00} & \multicolumn{1}{d}{-3.09} & \multicolumn{1}{d}{3.09}   \\ \hdashline
\textit{astro-ph} & \multicolumn{1}{d}{138,776} & \multicolumn{1}{d}{-0.0201} & \multicolumn{1}{d}{1.01} & \multicolumn{1}{d}{-3.09} & \multicolumn{1}{d}{3.09}   \\
\textit{comp-sci} & \multicolumn{1}{d}{249,666} & \multicolumn{1}{d}{-0.0187} & \multicolumn{1}{d}{1.00} & \multicolumn{1}{d}{-3.09} & \multicolumn{1}{d}{3.09}  \\
\textit{cond-mat} & \multicolumn{1}{d}{161,661} & \multicolumn{1}{d}{-0.0154} & \multicolumn{1}{d}{1.01} & \multicolumn{1}{d}{-3.09} & \multicolumn{1}{d}{3.09}  \\
\textit{hep} & \multicolumn{1}{d}{155,410} & \multicolumn{1}{d}{-0.0140} & \multicolumn{1}{d}{1.01} & \multicolumn{1}{d}{-3.09} & \multicolumn{1}{d}{3.09}  \\
\textit{math} & \multicolumn{1}{d}{273,731} & \multicolumn{1}{d}{-0.0155} & \multicolumn{1}{d}{1.00} & \multicolumn{1}{d}{-3.09} & \multicolumn{1}{d}{3.09}  \\
\textit{oth-phys} & \multicolumn{1}{d}{227,753} & \multicolumn{1}{d}{-0.0155} & \multicolumn{1}{d}{1.00} & \multicolumn{1}{d}{-3.09} & \multicolumn{1}{d}{3.09}  \\[0mm] \bottomrule[1pt] 
\end{tabular}}
}
\end{table}
\vspace{0.5cm}

\begin{table}[h]
\renewcommand{\arraystretch}{0.85}
\centering
\caption{\textbf{The Pearson's correlation coefficient ($\bmt{r}$) between the $\bmt{\gamma}$-index and the $\bmt{\gamma^{*}}$-index.}
The citation data are based on the below-99th percentile eprints in the arXiv dataset.
All coefficients are significant at the 0.1\% level.}
\label{tab:pwcorr}
{\footnotesize
{\setlength{\tabcolsep}{0.5em}
\begin{tabular}{l@{\hspace{1.5cm}}dddddd}\\[-2mm] \toprule[1pt] \\[-5.5mm]
Category & \multicolumn{1}{d}{N}        & \multicolumn{1}{c}{$r$} \\[-0.2mm] \midrule[0.3pt]
all & \multicolumn{1}{d}{1,205,870} & \multicolumn{1}{d}{0.987}  \\ \hdashline
\textit{astro-ph} & \multicolumn{1}{d}{138,647} & \multicolumn{1}{d}{0.980}  \\
\textit{comp-sci} & \multicolumn{1}{d}{249,418} & \multicolumn{1}{d}{0.993}  \\
\textit{cond-mat} & \multicolumn{1}{d}{161,527} & \multicolumn{1}{d}{0.988}  \\
\textit{hep} & \multicolumn{1}{d}{155,283} & \multicolumn{1}{d}{0.988}  \\
\textit{math} & \multicolumn{1}{d}{273,466} & \multicolumn{1}{d}{0.995}  \\
\textit{oth-phys} & \multicolumn{1}{d}{227,529} & \multicolumn{1}{d}{0.991}  \\[0mm] \bottomrule[1pt] \\
\end{tabular}}
}
\vspace{-3.5cm}
\end{table}
}

\afterpage{\clearpage%
\begin{table}[h]
\renewcommand{\arraystretch}{0.85}
\centering
\vspace{-0.5cm}
\caption{\textbf{Estimated characteristics of the quantile plot's slope ($\bmt{m}$).}
The citation data are based on the below-99th percentile eprints in the arXiv dataset.
The regression function is given by $\hat{m}_{i}=\sqrt{s_{2}\ln(t_{i}/s_{1}+1)}+\ep_{i}$ with $s_{1}$ and $s_{2}$ the regression parameters to be estimated and $\ep_{k}$ the error term.
Standard errors are in parentheses.
The adjusted $R$-squared ($R^{2}_{\mathrm{adj}}$) is also shown, which is nearly one for all disciplines.}
\label{tab:reg_beta}
{\footnotesize
{\setlength{\tabcolsep}{1.0em}
\begin{tabular}{l@{\hspace{0.2cm}}dddd}\\[-2mm] \toprule[1pt] \\[-5.5mm]
Category & \multicolumn{1}{d}{N}\hspace{-1.1cm}       & \multicolumn{1}{c}{$\hat{s}_{1}$} & \multicolumn{1}{c}{$\hat{s}_{2}$} & \multicolumn{1}{c}{$R^{2}_{\mathrm{adj}}$} \\[-0.2mm] \midrule[0.3pt]
\textit{astro-ph} & \multicolumn{1}{d}{24}\hspace{-1.1cm} & \multicolumn{1}{d}{0.0281} & \multicolumn{1}{d}{0.200} & \multicolumn{1}{d}{1.00}  \\[-0.5mm]
{} & {} & \multicolumn{1}{d}{(0.00422)} & \multicolumn{1}{d}{(0.00515)}  & {}   \\[1.2mm]
\textit{comp-sci} & \multicolumn{1}{d}{20}\hspace{-1.1cm} & \multicolumn{1}{d}{0.0842} & \multicolumn{1}{d}{0.423} & \multicolumn{1}{d}{1.00}  \\[-0.5mm]
{} & {} & \multicolumn{1}{d}{(0.0211)} & \multicolumn{1}{d}{(0.0230)}   & {} \\[1.2mm]
\textit{cond-mat} & \multicolumn{1}{d}{23}\hspace{-1.1cm} & \multicolumn{1}{d}{0.320} & \multicolumn{1}{d}{0.432} & \multicolumn{1}{d}{1.00}  \\[-0.5mm]
{} & {} & \multicolumn{1}{d}{(0.0427)} & \multicolumn{1}{d}{(0.0163)}  & {}  \\[1.2mm]
\textit{hep} & \multicolumn{1}{d}{28}\hspace{-1.1cm} & \multicolumn{1}{d}{0.191} & \multicolumn{1}{d}{0.455} & \multicolumn{1}{d}{1.00}  \\[-0.5mm]
{} & {} & \multicolumn{1}{d}{(0.0315)} & \multicolumn{1}{d}{(0.0181)}   & {} \\[1.2mm]
\textit{math} & \multicolumn{1}{d}{24}\hspace{-1.1cm} & \multicolumn{1}{d}{0.222} & \multicolumn{1}{d}{0.371} & \multicolumn{1}{d}{1.00}  \\[-0.5mm]
{} & {} & \multicolumn{1}{d}{(0.0286)} & \multicolumn{1}{d}{(0.0124)}  & {} \\[1.2mm]
\textit{oth-phys} & \multicolumn{1}{d}{26}\hspace{-1.1cm} & \multicolumn{1}{d}{0.434} & \multicolumn{1}{d}{0.522} & \multicolumn{1}{d}{1.00} \\[-0.5mm]
{} & {} & \multicolumn{1}{d}{(0.0386)} & \multicolumn{1}{d}{(0.0138)}  \\[0mm] \bottomrule[1pt] 
\end{tabular}}
}
\end{table}
\quad\\
\vfill
}

\afterpage{\clearpage%
\begin{table}[t]
\renewcommand{\arraystretch}{0.7}
\centering
\vspace{-0.5cm}
\caption{\textbf{\q{Ready reckoner} for the normalised citation index ($\bmt{\gamma}$)} (the below-90th citation data)\textbf{.}
Pre-calculated values are shown by discipline.
The equivalent for the below-99th citation data is shown in Table \ref{tab:hayamihyou99} in the main text.
Negative $\gamma$-values are omitted and shown as a dash (--).}
\label{tab:hayamihyou90}
{\scriptsize
{\setlength{\tabcolsep}{0.7em}
\begin{tabular}{ld@{\hspace{0cm}}d@{\hspace{0cm}}ddddddddd}\\[-2mm] 
\toprule[1pt] \\[-3.5mm]
{} & {} &{} & \multicolumn{1}{c}{\hspace{-4mm}$T=2$}    & \multicolumn{1}{c}{\hspace{-4mm}$T=3$}     & \multicolumn{1}{c}{\hspace{-4mm}$T=4$}     & \multicolumn{1}{c}{\hspace{-4mm}$T=5$}     & \multicolumn{1}{c}{\hspace{-4mm}$T=6$}     & \multicolumn{1}{c}{\hspace{-4mm}$T=7$}     & \multicolumn{1}{c}{\hspace{-4mm}$T=8$}     & \multicolumn{1}{c}{\hspace{-4mm}$T=9$}     & \multicolumn{1}{c}{\hspace{-4mm}$T=10$}    \\ \cline{4-12}\\[-2.5mm]
{\footnotesize{\textit{astro-ph}}} & \multicolumn{1}{l}{~$c={}$} & 5   & 3.05 & 2.32 & 1.93 & 1.68 & 1.50  & 1.37  & 1.27  & 1.19  & 1.12  \\
{} & \multicolumn{1}{l}{~$c={}$} & 10       & 3.74 & 3.01 & 2.62 & 2.37 & 2.20  & 2.07  & 1.97  & 1.88  & 1.81  \\
{} & \multicolumn{1}{l}{~$c={}$} & 50       & 5.35 & 4.62 & 4.23 & 3.98 & 3.81  & 3.68  & 3.58  & 3.49  & 3.42  \\
{} & \multicolumn{1}{l}{~$c={}$} & 100      & 6.04 & 5.31 & 4.92 & 4.67 & 4.50  & 4.37  & 4.27  & 4.19  & 4.12  \\[0mm] \hline \\[-2.5mm]
{\footnotesize{\textit{comp-sci}}} & \multicolumn{1}{l}{~$c={}$} & 5   & 1.58  & 0.88  & 0.52  & 0.30  & 0.15  & 0.05  & \multicolumn{1}{c}{---} & \multicolumn{1}{c}{---} & \multicolumn{1}{c}{---} \\
{} & \multicolumn{1}{l}{~$c={}$} & 10       & 2.28  & 1.58  & 1.21  & 0.99  & 0.84  & 0.74  & 0.66  & 0.60  & 0.55  \\
{} & \multicolumn{1}{l}{~$c={}$} & 50       & 3.89  & 3.19  & 2.82  & 2.60  & 2.45  & 2.35  & 2.27  & 2.21  & 2.16  \\
{} & \multicolumn{1}{l}{~$c={}$} & 100      & 4.58  & 3.88  & 3.51  & 3.29  & 3.15  & 3.04  & 2.97  & 2.91  & 2.86  \\[0mm] \hline \\[-2.5mm]
{\footnotesize{\textit{cond-mat}}} & \multicolumn{1}{l}{~$c={}$} & 5   & 2.71 & 1.87 & 1.43  & 1.15  & 0.96  & 0.82  & 0.71  & 0.62  & 0.55  \\
{} & \multicolumn{1}{l}{~$c={}$} & 10       & 3.40 & 2.56 & 2.12  & 1.85  & 1.66  & 1.51  & 1.40  & 1.31  & 1.24  \\
{} & \multicolumn{1}{l}{~$c={}$} & 50       & 5.01 & 4.17 & 3.73  & 3.45  & 3.26  & 3.12  & 3.01  & 2.92  & 2.85  \\
{} & \multicolumn{1}{l}{~$c={}$} & 100      & 5.70 & 4.86 & 4.42  & 4.15  & 3.96  & 3.82  & 3.71  & 3.62  & 3.54  \\[0mm] \hline \\[-2.5mm]
{\footnotesize{\textit{hep}}} & \multicolumn{1}{l}{~$c={}$} & 5     & 2.19 & 1.52  & 1.17  & 0.97  & 0.83  & 0.73  & 0.65  & 0.59  & 0.53  \\
{} & \multicolumn{1}{l}{~$c={}$} & 10        & 2.88 & 2.21  & 1.87  & 1.66  & 1.52  & 1.42  & 1.35  & 1.28  & 1.23  \\
{} & \multicolumn{1}{l}{~$c={}$} & 50        & 4.49 & 3.82  & 3.48  & 3.27  & 3.13  & 3.03  & 2.95  & 2.89  & 2.83  \\
{} & \multicolumn{1}{l}{~$c={}$} & 100       & 5.18 & 4.51  & 4.17  & 3.96  & 3.83  & 3.73  & 3.65  & 3.58  & 3.53  \\[0mm] \hline \\[-2.5mm]
{\footnotesize{\textit{math}}} & \multicolumn{1}{l}{~$c={}$} & 5     & 2.28 & 1.53  & 1.10  & 0.81  & 0.61  & 0.47  & 0.35  & 0.25  & 0.18  \\
{} & \multicolumn{1}{l}{~$c={}$} & 10       & 2.97 & 2.22  & 1.79  & 1.51  & 1.31  & 1.16  & 1.04  & 0.95  & 0.87  \\
{} & \multicolumn{1}{l}{~$c={}$} & 50       & 4.58 & 3.83  & 3.40  & 3.12  & 2.92  & 2.77  & 2.65  & 2.56  & 2.48  \\
{} & \multicolumn{1}{l}{~$c={}$} & 100      & 5.27 & 4.52  & 4.09  & 3.81  & 3.61  & 3.46  & 3.34  & 3.25  & 3.17  \\[0mm] \hline \\[-2.5mm]
{\footnotesize{\textit{oth-phys}}} & \multicolumn{1}{l}{~$c={}$} & 5    & 2.34 & 1.61 & 1.22  & 0.98  & 0.81  & 0.69  & 0.59  & 0.52  & 0.45  \\
{} & \multicolumn{1}{l}{~$c={}$} & 10       & 3.03 & 2.30 & 1.91  & 1.67  & 1.50  & 1.38  & 1.29  & 1.21  & 1.15  \\
{} & \multicolumn{1}{l}{~$c={}$} & 50       & 4.64 & 3.91 & 3.52  & 3.28  & 3.11  & 2.99  & 2.90  & 2.82  & 2.76  \\
{} & \multicolumn{1}{l}{~$c={}$} & 100      & 5.33 & 4.61 & 4.22  & 3.97  & 3.81  & 3.68  & 3.59  & 3.51  & 3.45  \\[-0.3mm] \bottomrule[1pt] \\
\end{tabular}}
}
\vspace{5mm}
\end{table}
\quad\\
\vfill
}

\afterpage{\clearpage%
\renewcommand\refname{\fontsize{14}{15}\selectfont References for Supplementary Materials}

}

\end{document}